\documentclass{article}

\usepackage{authblk}
\usepackage{array,epsfig,fancyhdr,rotating}
\usepackage[]{hyperref}  
\usepackage{sectsty, secdot}
\sectionfont{\fontsize{12}{14pt plus.8pt minus .6pt}\selectfont}

\subsectionfont{\fontsize{12}{14pt plus.8pt minus .6pt}\selectfont}
\usepackage{algorithm}
\usepackage{datatool}
\usepackage{tcolorbox}
\usepackage{graphicx}
\usepackage{caption}
\usepackage{subcaption}
\usepackage{upgreek}

\usepackage[left=3cm,right=3cm,top=3cm,bottom=3cm]{geometry}

\usepackage{enumitem}
\usepackage{float}
\floatstyle{plaintop}
\restylefloat{table}
\usepackage{amsmath}
\usepackage[linesnumbered,ruled,vlined, algo2e]{algorithm2e}
\usepackage{amssymb}
\usepackage{amsfonts}
\usepackage{multirow}
\usepackage{amsthm}
\usepackage{mathtools}
\usepackage{bbm}

\newtheorem{theorem}{Theorem}
\newtheorem{lemma}{Lemma}
\newtheorem{corollary}{Corollary}
\newtheorem{proposition}{Proposition}
\theoremstyle{definition}

\newtheorem{remark}{Remark}
\newtheorem{assumption}{\textbf{A}}[section]

\usepackage{ifthen}
\usepackage{stmaryrd}
\usepackage{xargs}
\usepackage{accents}
\usepackage{cleveref}
\usepackage{todonotes}
\SetKwInput{KwInput}{Input}                
\SetKwInput{KwOutput}{Output}   

\crefname{assumption}{{\textbf{A}}\!}{{\textbf{A}}}
\Crefname{assumption}{{\textbf{A}}\!}{{\textbf{A}}}
\Crefname{algorithm}{{Algorithm}}{Algorithm}
\def\ie{\emph{i.e.}}
\newcommand{\aind}[2]{\alpha_{#1}^{#2}}
\newcommand{\af}[1]{h_{#1}}
\newcommand{\afterm}[1]{\tilde{h}_{#1}}
\newcommand{\bcdot}{\cdot}
\newcommand{\bdpart}[1]{\mathbb{B}_{#1}}
\newcommand{\bklip}[1]{L^{\overleftarrow{Q}}_{#1}}
\newcommand{\udlip}[1]{L^q_{#1}}
\newcommandx{\bk}[2][1=]{
\ifthenelse{\equal{#1}{}}
{\overleftarrow{Q}_{#2}}
{\overleftarrow{Q}_{#2}^{#1}}
}
\newcommand{\lipsmu}[1]{L^{\mathbb{S}\mu} _#1}
\newcommand{\idx}[2]{\mathrm{J}_{#1} ^{#2}}
\newcommand{\scorelip}[1]{L^s_{#1}}
\def\cstcondparisc{\mathsf{c}}
\def\cstcondparisd{\mathsf{d}}
\newcommand{\supscore}[1]{s^{\infty} _{#1}}

\newcommand{\bdpartlip}[1]{L^{\mathbb{B}}_{#1}}
\newcommand{\bkwidx}[2]{\mathcal{B}_{\paramopt{#2}, #1}}
\newcommand{\bkklip}[1]{L^{BK}_{#1}}
\newcommand{\maxcoupling}[2]{\mathbbm{M} \left[{#1}, {#2}\right]}
\newcommand{\ckjtidx}[2]{\widetilde{\mathbb{S}}_{#1, \paramopt{#2}}}
\newcommand{\ckidx}[2]{\widetilde{\mbletter{S}}_{#1, \paramopt{#2}}}
\newcommand{\weight}[2]{\omega_{#1} ^{#2}}
\newcommand{\paramopt}[1]{\theta_{#1}}
\newcommand{\bmf}{\mathsf{F}}
\newcommand{\borel}{\mathcal{B}}
\newcommand{\bpart}[2]{\upsilon_{#1}^{#2}}
\newcommand{\bpartmb}[1]{\boldsymbol{\upsilon}_{#1}}
\newcommandx{\canlaw}[2][1=]{\prob^{#1}_{#2}}
\newcommandx{\canlawexp}[2][1=]{\E^{#1}_{#2}}
\newcommand{\ck}[1]{\mbletter{S}_{#1}}
\newcommand{\ckjt}[1]{\mathbb{S}_{#1}}
\newcommand{\condparis}{\mathsf{CondPaRIS}}
\newcommand{\E}{\mathbb{E}}
\newcommand{\efd}[1]{\boldsymbol{\mathcal{E}}_{#1}}
\newcommand{\eg}{\emph{e.g.}}
\newcommand{\epart}[2]{\xi_{#1}^{#2}}
\newcommand{\eparttd}[2]{\tilde{\xi}_{#1}^{#2}}
\newcommand{\epartmb}[1]{\boldsymbol{\xi}_{#1}}
\newcommand{\eqsp}{\;}

\newcommand{\eqdef}{\coloneqq}
\newcommand{\esp}[1]{\boldsymbol{\mathsf{E}}_{#1}}

\newcommand{\gibbs}[1]{K_{#1}}
\newcommand{\init}{\eta_0}
\newcommandx{\initmb}[1][1=]{\ifthenelse{\equal{#1}{}}{\boldsymbol{\eta}_0}{\boldsymbol{\eta}_{0}\langle #1 \rangle}}
\newcommand{\intvect}[2]{\llbracket #1, #2 \rrbracket}
\newcommand{\koskimat}[1]{\mathbf{J}}

\newcommand{\M}{M}
\newcommandx{\mbjt}[2][2=]{
 \ifthenelse{\equal{#2}{}}
 {\mathbb{C}_{#1}}
 {\mathbb{C}_{{#1}, {#2}}}}
\newcommand{\mbjtlip}[1]{L^\mathbb{C}_{#1}}
\newcommand{\mkmblip}[1]{L^{\mbletter{M}}_{#1}}
\newcommand{\mbletter}[1]{\boldsymbol{#1}}
\newcommand{\meas}{\mathsf{M}}
\newcommand{\md}[1]{m_{#1}}
\newcommand{\mdlow}[1]{\ubar{\sigma}_{#1}}
\newcommand{\mdhigh}[1]{\bar{\sigma}_{#1}}
\newcommand{\mixrate}[2]{\kappa_{#2,#1}}
\newcommand{\mk}[1]{M_{#1}}

\newcommandx{\mkmb}[2][2=]{
\ifthenelse{\equal{#2}{}}
{{\mbletter{M}}_{#1}}
{{\mbletter{M}}_{#1} \langle #2 \rangle}
}

\newcommand{\epartvaridx}[2]{x^{\idx{#1}{#2}} _{#1}}
\newcommand{\epartvar}[2]{x^{#2} _{#1}}
\newcommand{\eparttdvar}[2]{\tilde{x}^{#2} _{#1}}
\newcommand{\stattdvar}[2]{\tilde{b}^{#2} _{#1}}
\newcommand{\statvaridx}[2]{{b^{\idx{#1}{#2}} _{#1}}}
\newcommand{\statvar}[2]{b^{#2} _{#1}}
\newcommand{\PE}[2]{{\mathbb E}_{#2}\left[#1\right]}

\newcommand{\N}{N}
\newcommand{\noisekerhom}[1]{\mathbb{P}_{\paramopt{#1}}}
\newcommand{\noiseker}[2]{\mathbb{P}_{\paramopt{#2}, #1}}
\newcommand{\noisekerlip}[1]{L^P_{#1}}

\newcommand{\nset}{\mathbb{N}}
\newcommand{\nsets}{\mathbb{N}^*}
\newcommand{\nsetpos}{\mathbb{N}^\ast}
\newcommand{\occm}{\mu}
\def\1{\mathbbm{1}}
\newcommand{\zpath}{z}
\newcommand{\parisgibbs}[1]{\mathbb{K}_{#1}}
\newcommand{\parisgibbslip}[1]{L^{\mathbb{K}}_t}

\newcommand{\partfilt}[1]{\mathcal{F}_{#1}}
\newcommand{\partfiltbar}[1]{\tilde{\mathcal{F}}_{#1}}
\newcommand{\pd}[1]{\Phi_{#1}}

\newcommand{\pinit}{\mbletter{\psi}_0}
\newcommand{\pkl}{\mbletter{P}}
\newcommandx{\pk}[2][2=]{
\ifthenelse{\equal{#2}{}}
{\pkl_{#1}}
{\pkl_{#1} \langle #2 \rangle}
}
\newcommandx{\pkjt}[2][1=]{
\ifthenelse{\equal{#1}{}}
{\mbletter{\psi}_{#2}}
{\mbletter{\psi}_{#2} \langle #1 \rangle}
}
\def\paramspace{\Theta}
\newcommand{\pot}[1]{g_{#1}}
\newcommand{\pothigh}[1]{\bar{\tau}_{#1}}
\newcommand{\potlow}[1]{\ubar{\tau}_{#1}}
\newcommand{\potmb}[1]{\mbletter{g}_{#1}}
\newcommand{\prob}{\mathbb{P}}
\newcommand{\score}[2]{s_{#2, \paramopt{#1}}}
\def\gradest{H}

\newcommand{\gradestP}[1]{\widehat{H}_{\paramopt{#1}}}
\newcommand{\probmeas}{\mathsf{M}_1}
\newcommand{\refm}[1]{\lambda_{#1}}
\newcommandx{\rk}[2][1=]{
\ifthenelse{\equal{#1}{}}
{B_{#2}}
{B_{#2} \langle #1 \rangle}
}
\newcommand{\rmd}{\mathrm{d}} 
\newcommand{\rset}{\mathbb{R}}
\newcommand{\rsetpos}{\mathbb{R}_{\geq 0}}
\newcommand{\stat}[2]{\beta_{#1}^{#2}}
\newcommand{\statl}{b}
\newcommand{\statlmb}[1]{\mbletter{b}_{#1}}
\newcommand{\statmb}[1]{\boldsymbol{\beta}_{#1}}
\newcommand{\statmbtd}[1]{\tilde{\boldsymbol{\beta}}_{#1}}
\newcommand{\stattd}[2]{\tilde{\beta}_{#1}^{#2}}

\newcommandx{\targ}[2][1=]{
\ifthenelse{\equal{#1}{}}
{\eta_{#2}}
{\eta_{#2} \langle #1 \rangle}
}
\newcommandx{\targp}[3][1=]{
\ifthenelse{\equal{#1}{}}
{\eta_{#2, \paramopt{#3}}}
{\eta_{ #2, \paramopt{#3}} \langle #1 \rangle}
}
\newcommand{\targmb}[1]{\boldsymbol{\eta}_{#1}}

\newcommand{\tensprod}{\varotimes}
\newcommand{\ubar}[1]{\underaccent{\bar}{#1}}
\newcommand{\ud}[1]{q_{#1}}
\newcommandx{\uk}[2][1=]{
\ifthenelse{\equal{#1}{}}
{Q_{#2}}
{Q_{#2} \langle #1 \rangle}
}
\newcommand{\ukmb}[1]{\boldsymbol{Q}_{#1}}
\newcommandx{\ukp}[2][1=]{
\ifthenelse{\equal{#1}{}}
{\mathcal{Q}_{#2}}
{\mathcal{Q}_{#2} \langle #1 \rangle}
}
\newcommandx{\untarg}[2][1=]{
\ifthenelse{\equal{#1}{}}
{\gamma_{{#2}}}
{\gamma_{{#2}} \langle #1 \rangle}
}

\def\PARIS{\texttt{PARIS}}
\def\PPG{\texttt{PPG}}
\def\PGAS{\texttt{PGAS}}

\newcommand{\untargmb}[1]{\boldsymbol{\gamma}_{#1}}
\newcommand{\xsp}[1]{\mathsf{X}_{#1}}
\newcommand{\xfd}[1]{\mathcal{X}_{#1}}
\newcommand{\zsp}[1]{\mathsf{Z}_{#1}}
\newcommand{\zfd}[1]{\mathcal{Z}_{#1}}
\newcommand{\tv}[2]{\left\| #1 - #2 \right\|_{\mathrm{TV}}}
\newcommand{\xfdmb}[1]{\boldsymbol{\mathcal{X}}_{#1}}
\newcommand{\xmb}[1]{\mbletter{x}_{#1}}
\newcommand{\xmbtd}[1]{\tilde{\mbletter{x}}_{#1}}
\newcommand{\xpfd}[2]{\xfd{#1:#2}}
\newcommand{\xpsp}[2]{\xsp{#1:#2}}
\newcommand{\xspmb}[1]{\boldsymbol{\mathsf{X}}_{#1}}
\newcommand{\yfd}[1]{\mathcal{Y}_{#1}}
\newcommand{\yfdmb}[1]{\boldsymbol{\mathcal{Y}}_{#1}}
\newcommand{\ymb}[1] {{\mbletter{y}}_{#1}}
\newcommand{\ymbtd}[1]{\tilde{\mbletter{y}}_{#1}}
\newcommand{\ysp}[1]{\mathsf{Y}_{#1}}
\newcommand{\yspmb}[1]{\boldsymbol{\mathsf{Y}}_{#1}}
\newcommand{\indi}[1]{\1_{{#1}}}
\newcommand{\invar}[2]{\pi_{\paramopt{#2}, #1}}
\newcommand{\indin}[1]{\1\left\{#1\right\}}
\newcommand{\chunk}[3]{#1_{{#2:#3}}}

\newcommandx{\CPE}[4][1=,2=]{{\mathbb E}^{#2}_{#1}\left[ #3 \mid #4 \right]}
\newcommandx{\cPE}[4][1=,2=]{{\mathbb E}^{#2}_{#1}[ #3 \mid #4 ]} 
\newcommandx{\CPP}[3][1=]{{\mathbb P}_{#1}\left(\left. #2 \, \right| #3 \right)} 
\newcommandx{\cPP}[3][1=]{{\mathbb P}_{#1}[ #2 | #3 ]} 
\newcommand{\gsupbound}[1]{\pothigh{#1}}
\newcommand{\ginfbound}[1]{\potlow{#1}}
\newcommand{\constpdist}{\mathsf{c}}
\newcommandx{\mserunconst}[2][1=n]{\upzeta^{\scriptsize{\mbox{{\it mse}}}}_{#1,#2}}
\newcommandx{\biasrunconst}[2][1=n]{\upzeta^{\scriptsize{\mbox{{\it bias}}}}_{#1,#2}}
\newcommand{\cstcondparisbias}[1]{\bar{\mathsf{c}}_{#1}^{\scriptsize{\mbox{{\it bias}}}}}
\newcommand{\cstparisbias}[1]{\mathsf{c}_{#1}^{\scriptsize{\mbox{{\it bias}}}}}
\newcommand{\cstparismse}[1]{\mathsf{c}_{#1}^{\scriptsize{\mbox{{\it mse}}}}}

\newcommand{\cstpariscov}[1]{\mathsf{c}_{#1}^{\scriptsize{\mbox{{\it cov}}}}}
\newcommand{\gibbslip}[1]{L^{K}_{#1}}
\newcommand{\ki}{{k}}
\def\burningloss{\upsilon}
\newcommandx{\rollingestim}[4][1=N,2=K_0,3=K,4=f]{\Pi_{(#1,#2),#3}(#4)}
\def\totalbudget{C} 
\def\Id{\operatorname{Id}}

\def\lyapunovlip{L^V}
\def\kergradestplip{L^{\mathbb{P}\widehat{H}}}
\def\kergradestpbound{L_{0}^{\mathbb{P}\widehat{H}}}
\def\boundmse{\sigma_{\scriptsize{\mbox{{\it mse}}}}}
\newcommand{\lr}[1]{\gamma_{#1}}
\def\boundgradestp{L^{\widehat{H}}}
\def\targlip{L^{\eta}}

\newcommandx{\ckjttd}[3][3=]{\ifthenelse{\equal{#3}{}}{\widetilde{\mathbb{S}}_{\paramopt{#2}, #1} \langle \score{#2}{0:#1} \rangle}{\widetilde{\mathbb{S}}_{\paramopt{#2}, #1} \langle #3 \rangle}}

\newcommand{\constmsealg}{\sigma_{\scriptsize{\mbox{{\it mse}}}}}
\newcommand{\constbiasalg}{\sigma_{\scriptsize{\mbox{{\it bias}}}}}

\begin{document}

\title{State and parameter learning with {\PARIS} particle Gibbs}

\author[$\dag$]{Gabriel Cardoso}
\author[$\ddag$]{Yazid Janati El Idrissi}
\author[$\star$]{Sylvain Le Corff}
\author[$\dag$]{\'Eric Moulines}
\author[$\top$]{Jimmy Olsson}

\affil[$\dag$]{{\small CMAP, \'Ecole Polytechnique, Institut Polytechnique de Paris, Palaiseau.}}
\affil[$\ddag$]{{\small Samovar, T\'el\'ecom SudParis, d\'epartement CITI, TIPIC, Institut Polytechnique de Paris, Palaiseau.}}
\affil[$\star$]{{\small LPSM, Sorbonne Universit\'e, UMR CNRS 8001, 4 Place Jussieu, 75005 Paris.}}
\affil[$\top$]{{\small Department of Mathematics, KTH Royal Institute of Technology, Stockholm, Sweden.}}

\date{}
\maketitle 

\begin{abstract}
Non-linear state-space models, also known as general hidden Markov models, are ubiquitous in statistical machine learning, being the most classical generative models for serial data and sequences in general. 
The particle-based, rapid incremental smoother (\PARIS) is a sequential Monte Carlo (SMC) technique allowing for efficient online approximation of expectations of additive functionals under the smoothing distribution in these models.
Such expectations appear naturally in several learning contexts, such as likelihood estimation (MLE) and Markov score climbing (MSC). {\PARIS} has linear computational complexity, limited memory requirements and comes with non-asymptotic bounds, convergence results and stability guarantees.
Still, being based on self-normalised importance sampling, the {\PARIS} estimator is biased.
Our first contribution is to design a novel additive smoothing algorithm, the Parisian particle Gibbs (\PPG) sampler, which can be viewed as a {\PARIS} algorithm driven by conditional SMC moves, resulting in bias-reduced estimates of the targeted quantities. We substantiate the {\PPG} algorithm with theoretical results, including new bounds on bias and variance as well as deviation inequalities.  
Our second contribution is to apply {\PPG} in a learning framework, covering MLE and MSC as special examples. In this context, we establish, under standard assumptions, non-asymptotic bounds highlighting the value of bias reduction and the implicit Rao--Blackwellization of {\PPG}. These are the first non-asymptotic results of this kind in this setting.
We illustrate our theoretical results with numerical experiments supporting our claims.
\end{abstract}

\section{Introduction}

\emph{Sequential Monte Carlo} (SMC) \emph{methods}, or \emph{particle filters}, are simulation-based approaches used for the online approximation of posterior distributions in the context of Bayesian inference in state space models. In nonlinear \emph{hidden Markov models} (HMM), they have been successfully applied for approximating online the typically intractable posterior distributions of sequences of unobserved states $(X_{s_1},\ldots,X_{s_2})$ given observations $(Y_{t_1},\ldots,Y_{t_2})$ for $0\leq s_1\leq s_2$ and $0\leq t_1\leq t_2$. Standard SMC methods use Monte
Carlo samples generated recursively by means of sequential importance sampling
and resampling steps. A particle filter approximates the flow of marginal posteriors by a sequence of occupation measures associated with a sequence $\{\epart{t}{i}\}_{i=1}^{N}$, $t \in \nset$, of Monte Carlo samples, each \emph{particle} $\epart{t}{i}$ being a random draw in the state space of the hidden process. Particle filters revolve around two operations: a \emph{selection step} duplicating/discarding particles with large/small importance weights, respectively, and a \emph{mutation step} evolving randomly the selected particles in the state space. 
Applying alternatingly and iteratively selection and mutation results in swarms of particles being both temporally and spatially dependent. The joint state posteriors of an HMM can also be interpreted as laws associated with a certain kind of Markovian backward dynamics; this interpretation is useful, for instance, when designing backward-sampling-based particle algorithms for nonlinear smoothing \cite{douc:garivier:moulines:olsson:2009,delmoral:doucet:singh:2010}.

Throughout the years, several convergence results as the number $\N$ of particles tends to infinity have been established; see, \emph{e.g.}, \cite{delmoral:2004,douc:moulines:2008,cappe:moulines:ryden:2005} and the references therein. In addition, a number of non-asymptotic results have been established, including time-uniform bounds on the SMC $\mathsf{L}_p$ error and bias as well as bounds describing the propagation of chaos among the particles. Extensions to the backward-sampling-based particle algorithms can also be found for instance in \cite{douc:garivier:moulines:olsson:2009,delmoral:doucet:singh:2010,dubarry:lecorff:2013}.

In this paper, we focus on the problem of recursively computing smoothed expectations $\targ{0:t}h_t = \mathbb{E}[h_t(X_{0:t}) \mid Y_{0:t}]$ for additive functionals $h_t$ in the form
\begin{equation} \label{eq:add:functional}
h_{t}(\chunk{x}{0}{t}) \eqdef \sum_{s=0}^{t-1} \tilde{h}_s(\chunk{x}{s}{s + 1}), 
\end{equation}
where $X_{0:n}$ and $Y_{0:n}$ denote vectors of states and observations (see below for precise definitions). Such expectations appear frequently in the context of maximum-likelihood parameter estimation in nonlinear HMMs, for instance, when computing the score function (the gradient of the log-likelihood function) or the Expectation Maximization intermediate quantity; see \cite{cappe:2001,andrieu:doucet:2003,poyiadjis:doucet:singh:2005,cappe:2009,poyiadjis:doucet:singh:2011}.
The \emph{particle-based, rapid incremental smoother} (\PARIS) proposed in \cite{olsson:westerborn:2017} is tailored for solving online this additive smoothing problem. When the transition density of the latent states is lower and upper bounded, this algorithm can  be shown to have a linear computational complexity in the number $N$ of particles and limited memory requirements. 
An interesting feature of the {\PARIS}, which samples on-the-fly from the backward dynamics induced by the particle filter, is that it requires two or more backward draws per particle to cope with the degeneracy of the sampled trajectories and remain numerically stable in the long run, with an asymptotic variance that grows only linearly with time.

In this paper, we introduce a method to reduce the bias of the {\PARIS} estimator of $\targ{0:t} h_t$. The idea is to mix---by introducing a \emph{conditional {\PARIS} algorithm}---the {\PARIS} algorithm with a backward-sampling-based version of the \emph{particle Gibbs sampler}  \cite{andrieu:doucet:holenstein:2010,lindsten:jordan:schoen:2014,chopin:singh:2015,del2016particle,del2018sharp}.
This leads to a batch mode \emph{{\PARIS} particle Gibbs} (\PPG) \emph{sampler}, which we furnish with an upper bound of the bias that decreases inversely proportionally to the number $N$ of particles and exponentially fast with the particle Gibbs iteration index (under the assumption that the particle Gibbs sampler is uniformly ergodic). 

As an application we consider the problem of likelihood maximization with stochastic gradient. In this specific context, where the smoothing estimator is employed repeatedly to produce mean-field estimates, controlling the bias becomes critical. Thus, it is natural to aim at minimizing the bias for a fixed computational budget, provided that the variance does not explode. For this reason, bias reduction in stochastic simulation has been the subject of extensive research during the last decades \cite{https://doi.org/10.1111/rssb.12336, glynn_rhee_2014}. The present paper contributes to this line of research. In particular, we show that stochastic approximation (SA) with {\PPG} achieves a $\mathcal{O}(\log(n) / \sqrt{n})$ rate, where $n$ is the number of SA steps. This improves on a previous result of \cite{lindholm2018learning}, which establishes the almost sure convergence (to a stationary point of the likelihood) of an SA \emph{Expectation Maximization} (EM) algorithm based on particle Gibbs with \emph{ancestor sampling} (\PGAS).

The paper is structured as follows. In \Cref{sec:background}, we recall the hidden Markov model framework, the particle filter and the \PARIS\ algorithm. In \Cref{sec:PPG}, we lay out the \PPG\  algorithm and present the first central result of this paper, an upper bound on the bias of our estimator as a function of the number of particles and the iteration index of the Gibbs algorithm. In addition, we provide an upper bound on the mean-squared error (MSE). In \Cref{sec:learning}, we undertake the learning problem and present the second result of this paper, a $\mathcal{O}(\log(n) / \sqrt{n})$ non-asymptotic bound on the expectation of the squared gradient norm taken at a random index $K$. In \Cref{sec:numerics}, we illustrate our results through numerical experiments. All the proofs are collected in the supplementary material.

\paragraph{Notation.}
For a given measurable space $(\xsp{}, \xfd{})$, where $\xfd{}$ is a countably generated $\sigma$-algebra, we denote by $\bmf(\xfd{})$ the set of bounded $\xfd{} / \mathcal{B}(\rset)$-measurable functions on $\xsp{}$. For any $h \in \bmf(\xfd{})$, we let
$\| h \|_\infty \eqdef \sup _{x \in \xsp{}}|h(x)|$ and $\operatorname{osc}(h) \eqdef \sup _{(x, x') \in \xsp{}^2}| h(x) - h(x')|$ denote the supremum and oscillator norms of $h$, respectively. Let $\meas(\xfd{})$ be the set of $\sigma$-finite measures on $(\xsp{}, \xfd{})$ and $\probmeas(\xfd{}) \subset \meas(\xfd{})$ the probability measures. For any $h \in \bmf(\xfd{})$ and $\mu \in \meas(\xfd{})$ we write $\mu(f) = \int h(x) \mu(\rmd x)$. For a Markov kernel $K$ from $(\xsp{}, \xfd{})$ to another measurable space $(\ysp{}, \yfd{})$, we define the measurable function $Kh : \xsp{} \ni x \mapsto \int h(y) K(x,\rmd y)$. 
The composition $\mu K$ is a probability measure on $(\ysp{}, \yfd{})$ such that $\mu K: \xfd{} \ni A \mapsto \int \mu(\rmd x) K(x,\rmd y) \indi{A}(y)$. For all sequences
 $\{a_u\}_{u\in\mathbb{Z}}$ and $\{b^u\}_{u\in\mathbb{Z}}$, and all $s\leq t$ we write $a_{s:t} = \{a_s,\ldots,a_t\}$ and $b^{s:t} = \{b^s,\ldots,b^t\}$.

\section{Background}
\label{sec:background}
\subsection{Hidden Markov models}
\textit{Hidden Markov models} consist of an unobserved state process $\{X_t\}_{t \in \nset}$ and observations $\{Y_t\}_{t \in \nset}$, where, at each time $t \in \nset$, the unobserved state $X_t$ and the observation $Y_t$ are assumed to take values in some general measurable spaces $(\xsp{t}, \xfd{t})$ and $(\ysp{t}, \yfd{t})$, respectively. It is assumed that $\{ X_t\}_{t \in \nset}$ is a Markov chain with transition kernels $\{\mk{t+1}\}_{t \in \nset}$ and initial distribution $\init$. Given the states $\{X_t\}_{t \in \nset}$, the observations $\{Y_t\}_{t \in \N}$ are assumed to be independent and such that for all $t \in \nset$, the conditional distribution of the observation $Y_t$ depends only on the current state $X_t$. This distribution is assumed to admit 
a density $\pot{t}(X_t, \cdot)$ with respect to some reference measure. In the following we assume that we are given a fixed sequence $\{y_t\}_{t \in \nset}$ of observations and define, abusing notations, $\pot{t}(\cdot) = \pot{t}(\cdot, y_t)$ for each $t \in \nset$. 
We denote, for $0 \leq s \leq t$, $\xsp{s:t} \eqdef \prod_{u=s}^t\xsp{u}$ and $\xfd{s:t} \eqdef \bigotimes_{u=s}^t\xfd{u}$. Consider the unnormalized transition kernel
\begin{equation} \label{eq:unnormalised:kernel}
    \uk{s} : \xsp{s} \times \xfd{s + 1} \ni (x, A) \mapsto \pot{s}(x) \mk{s}(x, A)
    \end{equation}
and let 
\begin{equation}
    \label{eq:unnormalized:F-K:measures} \untarg{0:t} : \xpfd{0}{t} \ni A 
    \mapsto \int \indi{A}(\chunk{x}{0}{t}) \, \targ{0}(\rmd x_0) \prod_{s = 0}^{t - 1} \uk{s}(x_s, \rmd x_{s + 1}).
\end{equation}
Using these quantities, we may define the \emph{joint-smoothing} and \emph{predictor distributions} at time $t \in \nset$ as
\begin{align}
    \label{eq:normalized:F-K:measures} 
    \targ{0:t}  : \xpfd{0}{t} \ni A & \mapsto \frac{\untarg{0:t}(A)}{\untarg{0:t}(\xpsp{0}{t})}, \\
    \targ{t} : \xfd{t} \ni A & \mapsto \targ{0:t}(\xsp{0:t - 1} \times A), 
\end{align}
respectively. It can be shown (see \cite[Section~3]{cappe:moulines:ryden:2005}) that $\targ{0:t}$ and $\targ{t}$ are the conditional distributions of $X_{0:t}$ and $X_t$ given $Y_{0:t - 1} $ respectively, evaluated at  $y_{0:t - 1}$. Unfortunately, these distributions, which are vital in Bayesian smoothing and filtering as they enable the estimation of hidden states through the observed data stream, are available in a closed form only in the cases of linear Gaussian models or models with finite state spaces; see \cite{infhidden} for a comprehensive coverage. 

\subsection{Particle filters}
For most models of interest in practice, the joint smoothing and predictor distributions are intractable, and so are also any expectation associated with these distributions. Still, such expectations can typically be efficiently estimated using \emph{particle methods}, which are based on the predictor recursion $\targ{t+1} = \targ{t} \uk{t} / \targ{t} \pot{t}$. At time $t$, if we assume that we  have at hand a consistent particle approximation of $\targ{t}$, formed by $N$ random draws $\{\epart{t}{i}\}_{i = 1}^N$, so-called \emph{particles}, in $\xsp{t}$ and given by $\targ{t}^N = N^{-1} \sum_{i = 1}^N \delta_{\epart{t}{i}}$, plugging $\targ{t}^N$ into the recursion tying $\targ{t+1}$ and $\targ{t}$ yields the mixture $\targ{t}^N \uk{t}$, from which a sample of $N$ new particles can be drawn in order to construct $\targ{t+1}^N$. To do so, we sample, for all $1 \leq i \leq N$, ancestor indices $\aind{t}{i} \sim \mbox{Categorical}(\{ \pot{t}(\epart{t}{\ell}) \}_{\ell = 1}^N)$ and then propagate $\epart{t+1}{i} \sim \mk{t}(\epart{t}{\aind{t}{i}}, \cdot)$.
 This procedure, which is initialized by sampling the initial particles $\{\epart{0}{i}\}_{i = 1}^N$ independently from $\init$, describes the particle filter with multinomial resampling and produces consistent estimators such that for every $h \in \bmf(\xsp{t})$, $\targ{t}^N (h)$ converges almost surely to $\targ{t}(h)$ as the number $N$ of particles tends to infinity. 

This procedure can also be extended to produce particle approximations of the joint-smoothing distributions $\{\targ{0:t}\}_{t \in \nset}$. Note that the successive ancestor selection steps described previously generates an ancestor line for each terminal particle $\epart{t}{i}$, which we denote by $\epart{0:t}{i}$. It can then be easily shown that  $\targ{0:t}^N = N^{-1} \sum_{i = 1}^N \delta_{\epart{0:t}{i}}$ forms a particle approximation of the joint-smoothing distribution $\targ{0:t}$. However, it is well known that the same selection operation also depletes the ancestor lines, since, at each step, two different particles are likely to originate from the same parent in the previous generation. Thus, eventually, all the particles end up having a large portion of their initial ancestry in common. This means that in practice, this naive approach, which we refer to as the \emph{poor man's smoother}, suffers generally from high variance when used for estimating joint-smoothing expectations of objective functionals depending on the whole state trajectory.

\subsection{Backward smoothing and the {\PARIS} algorithm}
\label{sec:std:paris}
We now discuss how to avoid the problem of particle degeneracy relative to the smoothing problem by means of so-called \emph{backward sampling}. While this line of research has broader applicability, we restrict ourselves for the sake of simplicity to the case of \emph{additive state functionals} in the form 
\begin{equation} \label{eq:add:functional}
    h_{t}(\chunk{x}{0}{t}) \eqdef \sum_{s=0}^{t-1} \tilde{h}_s(\chunk{x}{s}{s + 1}), \quad \chunk{x}{0}{t} \in \xpsp{0}{t}.
\end{equation}
Appealingly, using the poor man's smoother  described in the previous section, smoothing of additive functionals can be performed online alongside the particle filter by letting, for each $s$, 
\begin{equation} \label{eq:def:PaRIS:est}
\targ{0:s}^N h_s \eqdef N^{-1} \sum_{i = 1}^N \stat{s}{i},
\end{equation}
where the statistics $\{\stat{s}{i}\}_{i = 1}^N$ satisfy the recursion 
\begin{equation}
    \label{eq:pms:online}
  \stat{s + 1}{i} = \stat{s}{\aind{s}{i}} + \tilde{h}_s(\epart{s}{\aind{s}{i}}, \epart{s + 1}{i}),
\end{equation}
where $\aind{s}{i}$ is, as described, the ancestor at time $s$ of particle $\epart{s + 1}{i}$. 

As mentioned above, the previous estimator suffers from high variance when $s$ is  relatively large with respect to  $\N$. However, assume now that the model is \emph{fully dominated} in the sense that each state process kernel $\mk{s}$ has a transition density $\md{s}$ with respect to some reference measure;  then, interestingly, it is easily seen that the conditional probability that $\aind{s}{i} = j$ given the offspring $\epart{s + 1}{i}$ and the ancestors $\{\epart{s}{\ell}\}_{\ell = 1}^N$ is given by 
\begin{equation} \label{eq:def:bk}
    \mathbf{\Lambda}_s(i, j) \eqdef  \frac{\weight{s}{j} \md{s}(\epart{s}{j}, \epart{s + 1}{i})}{\sum_{\ell = 1}^N \weight{s}{\ell}\md{s}(\epart{s}{\ell}, \epart{s + 1}{i})}.  
\end{equation}
Here $\mathbf{\Lambda}_s$ forms a backward Markov transition kernel on $\intvect{1}{N} \times \intvect{1}{N}$. 
Using this observation, we may avoid completely the particle-path degeneracy of the poor man's smoother by simply replacing the naive update \eqref{eq:pms:online} by the Rao--Blackwellized counterpart  
\begin{equation}
    \label{eq:ffbs:bwupdate}
    \stat{s + 1}{i} = \sum_{j = 1}^N \mathbf{\Lambda}_s(i,j) \{ \stat{s}{j} + \tilde{h}_s(\epart{s}{j}, \epart{s + 1}{i})\}.
\end{equation}
This approach, proposed in \cite{delmoral:doucet:singh:2010}, avoids elegantly the path degeneracy as is eliminates the ancestral connection between the particles by means of averaging. Furthermore, it is entirely online since at step $s$ only the particle populations $\epart{s}{1:N}$ and $\epart{s + 1}{1:N}$ are needed to perform the update. Still, a significant drawback is the overall $\mathcal{O}(N^2)$ complexity for the computation of $\stat{t}{1:N}$, since the calculation of each $\stat{s + 1}{i}$ in \eqref{eq:ffbs:bwupdate} involves the computation of $N^2$ terms, which can be prohibitive when the number $N$ of particles is large. Thus, in \cite{olsson:westerborn:2017}, the authors propose to sample $M \ll N$ conditionally independent indices $\{ J^{i,j}_s \}_{j = 1}^M$ from the distribution $\mathbf{\Lambda}_s(i, \cdot)$ and to update the statistics according to 
\begin{equation} 
    \label{eq:ffbs:paris}
    \stat{s + 1}{i} = M^{-1} \sum_{j = 1}^M \left(\stat{s}{J^{i,j}_s} + \tilde{h}_s(\epart{s}{J^{i,j}_s}, \epart{s + 1}{i})\right).
\end{equation}  
If the transition density $\md{s}$ is uniformly bounded from above and below, an accept-reject approach allows the sampling-based update \eqref{eq:ffbs:paris} to be performed for $i \in \intvect{1}{N}$ at an $\mathcal{O}(N(M+1))$ overall complexity if a pre-initialized multinomial sampler is used. A key aspect of this approach is that the number $M$ of sampled indices at each step can be very small; indeed, for any fixed $M \geq 2$, the algorithm, which is referred to as the \PARIS, can be shown to be stochastically stable with an $\mathcal{O}(t)$ variance (see \cite[Section~1]{olsson:westerborn:2017} for details), and setting $M$ to $2$ or $3$ yields typically fully satisfying results. 

The {\PARIS} estimator can be viewed as an alternative to the FFBSm, rather than the FFBSi. Even if the {\PARIS} and FFBSi are both randomised versions of the FFBSm estimator, the {\PARIS} is of a fundamentally different nature than the FFBSi. The {\PARIS} approximates the forward-only FFBSm online in the context of additive functionals by approximating each updating step  by additional Monte Carlo sampling. The sample size $M$ is an accuracy parameter that determines the precision of this approximation, and by increasing $M$ the statistical properties of the {\PARIS} approaches those of the  forward-only FFBSm. 
On the other hand, as shown in  \cite[Corollary~9]{douc:garivier:moulines:olsson:2009}, the asymptotic variance of FFBSi is always larger than that of the FFBSm, with a gap given by the variance of the state functional under the joint-smoothing distribution. Thus, we expect, especially in the case of a low signal-to-noise ratio, the {\PARIS} to be more accurate than the FFBSi for a given computational budget. 
Another important reason to focus on the {\PARIS} estimator rather than the FFBSi is the appealing online properties of the latter, whose interplay with and relevance to the particle MCMC methodology is to be explored. Our results can be naturally extended to the FFBSi and PGAS but since the {\PARIS} has a practical edge, 
we chose to center our contribution around it although the main idea behind our paper is more general. 

\section{{\PARIS} particle Gibbs}
\label{sec:PPG}
\subsection{Particle Gibbs methods}
The \emph{conditional particle filter} (CPF) introduced in \cite{andrieu2010particle} serves the basis of a particle-based MCMC algorithm targeting the joint-smoothing distribution $\targ{0:t}$. Let $\ell \in \nset^*$ be an iteration index and $\zeta_{0:t}[\ell]$ a conditional path used at iteration $\ell$ of the CPF to construct a particle approximation of $\targ{0:t}$ as follows. At step $s \in \intvect{1}{t}$ of the CPF, a randomly selected particle, with uniform probability $1/N$, is set to $\zeta_s[\ell]$, whereas the remaining $N-1$ particles are all drawn from the mixture $\targ{s-1}^N \uk{s-1}$. At the final step, a new particle path $\zeta_{0:t}[\ell + 1]$ is drawn either:
\begin{itemize}
    \item by selecting randomly, again with uniform probability $1/N$, a genealogical trace from the ancestral tree of the particles $\{\epart{s}{1:N}\}_{s = 0}^t$ produced by the CPF, as in the vanilla particle Gibbs sampler;
    \item or by generating the path by means of backward sampling, \ie, by drawing  indices $J_{0:t}$ backwards in time according to $J_t \sim \mbox{Categorical}(\{1/N \}_{i = 1}^N )$ and, conditionally to $J_{s + 1}$, $J_s \sim \mathbf{\Lambda}_s(J_{s + 1}, \cdot)$, $s \in \intvect{0}{t - 1}$, and letting $\zeta_{0:t}[\ell + 1] = (\epart{0}{J_0}, \ldots \epart{t}{J_t})$ (where the transition kernels $\{\mathbf{\Lambda}_s\}_{s = 0}^t$, defined by \eqref{eq:def:bk}, are induced by the particles produced by the CPF), as proposed in \cite{Whiteley:2010}.
\end{itemize} 
The theoretical properties of the different versions of the particle Gibbs sampler are well studied \cite{singh2017blocking, chopin2015particle,andrieu2018uniform}. In short, the produced conditional paths $(\zeta_{0:t}[\ell])_{\ell \in \nset}$ form a Markov chain whose marginal law converges geometrically fast in total variation to the target distribution $\targ{0:t}$. As it is the case for smoothing algorithms, the vanilla particle Gibbs sampler suffers from bad mixing due to particle path degeneracy while its backward-sampling counterpart exhibits superior performance as $t$ increases \cite{lee2020coupled}.

\subsection{The {\PPG} algorithm}
Remarkably, in order for the standard particle Gibbs samplers to output a single conditional path, a whole particle filter is run and then discarded, resulting in significant waste of computational work. Thus, we now introduce a variant of the {\PARIS} algorithm, coined the {\PARIS} particle Gibbs ({\PPG}), 
in which the conditional path of particle Gibbs with backward sampling is merged with the intermediate particles, ensuring less computational waste and reduced bias with respect to the vanilla {\PARIS}. 

In the following we let $t \in \nset$ be a fixed time horizon, and describe in detail how the {\PPG} approximates iteratively $\targ{0:t} \af{t}$, where $\af{t}$ is an additive functional in the form \eqref{eq:add:functional}. Using a given conditional path $\zeta_{0:t}[\ell - 1]$ as input, the $\ell$-th iteration of the {\PPG} outputs a many-body system $\bpartmb{t}[\ell] = ((\epart{0:t}{1}, \stat{t}{1}), \ldots, (\epart{0:t}{N}, \stat{t}{N}))$ 
comprising $N$ backward particle paths $\{\epart{0:t}{i}\}_{i = 1}^N$ with associated {\PARIS} statistics $\{\stat{t}{i}\}_{i = 1}^N$. This is the so-called \emph{conditional {\PARIS} update} detailed in \Cref{alg:parisian:Gibbs}. After this, an updated conditional path is selected with probability $1/N$ among the $N$ particle paths $\{\epart{0:t}{i}\}_{i = 1}^N$ and used as input in the next conditional {\PARIS} operation. At each iteration, the produced statistics $\{ \stat{t}{i} \}_{i = 1}^N$ provide an approximation of $\targ{0:t}h_t$ according to \eqref{eq:def:PaRIS:est}. The overall algorithm is summarized in Algorithm~\ref{alg:parisian:particle:Gibbs}. The function $\mathsf{CPF}_s$ describes one step of the conditional particle filter and is given in the supplementary material. In addition, the \PPG\ algorithm defines a Markov chain with Markov transition kernel denoted by $\parisgibbs{t}{}$ and detailed in \eqref{eq:ppg_ker_def}.

\begin{algorithm}
  \KwInput{$\{(\epart{0:s}{i}, \stat{s}{i})\}_{i = 1}^\N$, $\zeta_{s + 1}$, $\tilde{h}_{s-1}$}
  \KwResult{$\{(\epart{0:s+1}{i}, \stat{s+1}{i})\}_{i = 1}^\N$}
  draw $\epart{s + 1}{1:N} \sim \mathsf{CPF}_s(\zeta_{s+1}, \epart{s}{1:N})$\;\\
  \For{$i \gets 1$ \KwTo $\N$}{
  draw $ \{ J^{i, \ell}_s \}_{\ell = 1}^\M \sim \mathbf{\Lambda}(i, \cdot)^{\varotimes M}$\;
  
  set $\stat{s + 1}{i} \gets M^{-1} \sum_{\ell = 1}^\M \left( \stat{s}{i, J^{i, \ell}_s} + \afterm{s}(\epart{s}{i, J^{i,\ell}_s}, \epart{s + 1}{i}) \right)$\;
  
  set $\epart{0:s+1}{i} \gets (\epart{0:s}{i, J^{i,1}_s}, \epart{s + 1}{i})$\;  
  
  }
  \caption{One conditional {\PARIS}  update ($\mathsf{CondPaRIS}$)} \label{alg:parisian:Gibbs}
  \label{alg:paris}
  \end{algorithm}
  
  \begin{algorithm}
  \KwInput{Initial path $\chunk{\zeta}{0}{t}$, $\{\tilde{h}_s\}_{s = 0} ^{t-1}$}
  \KwResult{$\{\stat{t}{i}\}_{i = 1}^N$, $\chunk{\zeta'}{0}{t}$}
  draw $\epart{0}{1:N} \sim \mathsf{CPF}_0(\zeta_0)$\;
  
  set $\stat{0}{i} \gets 0$ for $i \in \intvect{1}{N}$\;
  
  \For{$s \gets 0$ \KwTo $t - 1$}{
  set $\{(\epart{0:s+1}{i}, \stat{s+1}{i})\}_{i = 1}^N \gets
  \condparis(\{(\epart{0:s}{i}, \stat{s}{i})\}_{i = 1}^N, \zeta_{s + 1}, \tilde{h}_{s})$\;
  }

  draw $\zeta_{0:t}' \sim N^{-1} \sum_{i = 1}^N \delta_{\epart{0:t}{i}}$\;
  \caption{One iteration of \PPG} \label{alg:parisian:particle:Gibbs}
  \label{alg:ppg}
  \end{algorithm}

As performing $k$ steps of the {\PPG} results in $k$ many-body systems, it is natural to consider the following \emph{roll-out estimator} which combines the backward statistics from step $k_0 < k$ to $k$:
\begin{equation}
\label{eq:rolling-estimator}
\rollingestim[\ki_0][\ki][N][\af{t}] = \left[\N(\ki-\ki_0)\right]^{-1} \sum_{\ell=\ki_{0}+1}^\ki \sum_{j=1}^\N \stat{t}{j}[\ell].
\end{equation}
The total number of particles used in this estimator is $\totalbudget = (\N - 1) \ki$ per time step.
We denote by $\burningloss= (\ki-\ki_0)/\ki$ the ratio of the number of particles used in the estimator to the total number of sampled particles.

We now state the first main results of the present paper, in the form of theoretical bounds on the bias and mean-squared error (MSE) of the roll-out estimator \eqref{eq:rolling-estimator}. These results are obtained under the following \emph{strong mixing} assumptions, which are now standard in the literature (see \cite{delmoral:2004,douc:moulines:2008,delmoral:2013,del2016particle}). It is crucial for obtaining quantitative bounds for particle smoothing algorithms, see \cite{olsson:westerborn:2017} or \cite{gloaguen:lecorff:olsson:2022} but also for the coupled conditional backward sampling particle filter \cite{lee2020coupled}. 
\begin{assumption}[strong mixing] 
  \label{assumption:strong_mixing}
For every $s \in \nset$ there exist $\potlow{s}$, $\pothigh{s}$, $\mdlow{s}$, and $\mdhigh{s}$ in $\rset_+^\ast$ such that
 \begin{enumerate}[label=(\roman*),nosep]
 \item  $\potlow{s} \leq \pot{s}(x_s) \leq \pothigh{s}$ for every $x_s \in \xsp{s}$,
 \item  $\mdlow{s} \leq \md{s}(x_s, x_{s + 1}) \leq \mdhigh{s}$ for every $(x_s, x_{s + 1}) \in \xsp{s:s+1}$.
 \end{enumerate}
\end{assumption}

Under \Cref{assumption:strong_mixing}, define, for every $s \in \nset$,
\begin{equation} \label{eq:def:rho:n}
\rho_s \eqdef \max_{m \in \intvect{0}{s}} \frac{\pothigh{m} \mdhigh{m}}{\potlow{m} \mdlow{m}}
\end{equation}
and, for every $\N \in \nsetpos$ and $t \in \nset$ such that $\N > \N_t \eqdef (1+ 5 \rho^2 _t / 2) \vee 2 t (1 + \rho_t^2)$,
\begin{equation} \label{eq:def:kappa}
\kappa_{N, t} \eqdef 1 - \frac{1 - (1 + 5 t \rho_t^2 / 2)/ N}{1 + 4 t (1+2 \rho_t^2) / N}.
\end{equation}
Note that $\kappa_{N, t} \in (0, 1)$ for all $\N$ and $t$ as above. 
\begin{theorem}
\label{theo:bias-mse-rolling}
Assume \Cref{assumption:strong_mixing}. Then for every $t \in \nset$,  $\M \in \nsetpos$, $\upxi \in \probmeas(\xpfd{0}{t})$,
$k_0 \in \nsetpos$, $k >k_0$ and $\N \in \nsetpos$ such that $\N > \N_t$,
\begin{align}
\left| \E_{\upxi}[\rollingestim[{\ki}_0][\ki][N][\af{t}]] - \targ{0:t} \af{t} \right| 
&\leq \constbiasalg \label{eq:theo:bias-mse-rolling:bias} \\
\E_{\upxi}\left[ \left(\rollingestim[{\ki}_0][\ki][N][\af{t}] - \targ{0:t} \af{t} \right)^2 \right] \nonumber &\leq \constmsealg^2, \label{eq:theo:bias-mse-rolling:mse}
\end{align}
where 
\begin{align*}
    \constbiasalg &\eqdef \frac{ \cstparisbias{t} \kappa_{t, N}^{\ki_0} \sum_{m=0}^{t-1} \| \afterm{m} \|_\infty}{(\ki-\ki_0) (1 - \kappa_{t,\N})\N} \eqsp ,\\
    \constmsealg^2 &\eqdef \frac{(\sum_{m=0}^{t - 1} \| \afterm{m} \|_\infty)^2}{\N (\ki- \ki_0)} \left(\cstparismse{t} + \frac{2 \cstpariscov{t}}{\N^{1/2} (1-\kappa_{t,N})}\right) 
\end{align*}
and $\cstparisbias{t}, \cstparismse{t}$ and $\cstpariscov{t}$ are constants that do not depend on $N$ and $\E_{\upxi}$ denotes the expectation under the law of the Markov chain formed by the {\PPG} when initialized according to $\upxi$. 
\end{theorem} 
The proof is provided in the supplementary material. Importantly, \eqref{eq:theo:bias-mse-rolling:bias} provides a bound on the bias of the roll-out estimator that decreases exponentially with the burn-in period $k_0$ and is inversely proportional to the number $N$ of particles. This means that we can improve the bias of the {\PARIS} estimator with a better allocation of the computational resources.

\section{Parameter learning with $\PPG$}
\label{sec:learning}
We now turn to parameter learning using {\PPG} and gradient-based methods.
We set the focus on learning the parameter $\paramopt{}$ of a function $V(\paramopt{})$ whose gradient is the smoothed expectation of an additive functional $\score{}{0:t}$ in the form \eqref{eq:add:functional}.
\Cref{alg:scoreasc:full} defines a stochastic approximation (SA) scheme where the noise forms a parameter dependent Markov chain with associated invariant measure $\pi_{\paramopt{}}$.
We follow the approach of \cite{pmlr-v99-karimi19a} to establish a non-asymptotic bound over the mean field $h(\paramopt{}) \eqdef \pi_{\paramopt{}} \score{}{0:t}$.
Such a setting encompasses for instance  the following estimation procedures.
\begin{enumerate}[leftmargin = *]
    \item[\textbf{(1)}] \textit{Score ascent. } In the case of fully dominated HMMs, we are often interested in optimizing the log-likelihood of the observations given by $V(\paramopt{}) = \log \int \untarg{0:t,\paramopt{}}{} (\rmd x_{0:t})$. By applying \textit{Fisher's identity}, we may express its gradient as a smoothed expectation of an additive functional according to   
    \begin{align*}
        \nabla_{\paramopt{}} V(\paramopt{}) & = \int \nabla_{\paramopt{}} \log \untarg{0:t}(x_{0:t}) \,  \targp{0:t}{}(\rmd x_{0:t}), \\
        & = \int \sum_{\ell = 0}^{t-1} \score{}{\ell}(x_{\ell}, x_{\ell +1}) \, \targp{0:t}{}(\rmd x_{0:t}),
    \end{align*}
    where $\score{}{\ell}: \xsp{\ell:\ell+1} \ni (x, x') \mapsto \nabla_{\paramopt{}} \log \{\pot{\ell,\theta}(x)\md{\ell,\theta}(x,x')\}$ and $\score{}{0:t} \eqdef \sum_{\ell = 0}^{t-1} \score{}{\ell}$. 
    \item[\textbf{(2)}]\textit{Inclusive KL surrogates. } 
    Inspired by \cite{naesseth:2O2O}, we may consider the problem of learning a surrogate model for $\targp{0:t}{}$ in the form $q_{\phi}(x_{0:t})=q_{\phi}(x_0) \prod_{\ell=0}^{t-1} q_{\phi}(x_{\ell + 1}, x_{\ell})$ by minimizing $V(\phi) = \mbox{KL}(\targp{0:t}{}, {q_{\phi}})$.
\end{enumerate}
\begin{algorithm}
    \KwInput{$\paramopt{},\chunk{\zeta}{0}{t}[0]$, $\{ s_{\ell, \paramopt{}} \}_{\ell = 0} ^{t-1}$, number $k$ of {\PPG} iterations, burn-in $k_0$.}
    \KwResult{$\stat{t}{1:N}[k_0:k], \zeta_{0:t}[k]$}
    
    \For{$\ell \gets 0$ \KwTo $k-1$}{
    run $(\stattd{t}{1:N}[\ell + 1], \zeta_{0:t}[\ell + 1]) \gets \mathsf{PPG}( \paramopt{};\zeta_{0:t}[\ell], \{ s_{\ell, \paramopt{}}\}_{\ell = 0}^{t-1})$\;

    \If{$\ell \geq k_0 - 1$}{
        set $\stat{t}{1:N}[\ell + 1] = \stattd{t}{1:N}[\ell+1] $\;
    }
    }
    \caption{Gradient estimation with roll-out \PPG\  ($\mathsf{GdEst}$)} \label{alg:scoreasc:onestep}
\end{algorithm}
\begin{algorithm}
        \KwInput{$\paramopt{0}$, $\chunk{\zeta}{0}{t}[0]$, number $k$ of {\PPG} iterations, burn-in $k_0$, number of SA iterations $n$, learning-rate sequence $\{\gamma_{\ell}\}_{\ell \in \nset}$.}
        \KwResult{$\paramopt{n}$}
        \For{$i \gets 0$ \KwTo $n-1$}{
        run $(\stat{t}{1:N}[k_0:k], \zeta_{0:t}[i+1]) \gets \mathsf{GdEst}(\paramopt{i}, \zeta_{0:t}[i], \{ s_{\ell, \paramopt{i}} \}_{\ell = 0} ^{t-1}, k, k_0)$ \;
        
        set $\rollingestim[\ki_0][\ki][N][s_{0:t, \paramopt{i}}]=(N(k-k_0))^{-1} \sum_{\ell=k_0}^{k-1} \sum_{j=1}^{N} \stat{t}{j}[\ell]$
        
        set $\paramopt{i+1} \gets \paramopt{i} + \gamma_{i+1} \rollingestim[\ki_0][\ki][N][s_{0:t, \paramopt{i}}]$\;}
        \caption{Score ascent with {\PPG}.} \label{alg:scoreasc:full}
\end{algorithm}
Note that \Cref{alg:scoreasc:onestep} defines a (collapsed) Markov kernel $\noiseker{t}{}$ defining for each path $\zeta_{0:t}$ a measure  $\noiseker{t}{}(\zeta_{0:t}, \rmd(\tilde{\zeta}_{0:t}, \stattd{t}{1:N}[k_0:k]))$ over the extended space of paths and sufficient statistics. Note that  by evaluating   
the function $b_{t}^{1:N}[k_0:k] \mapsto \left[\N(\ki-\ki_0)\right]^{-1} \sum_{\ell=\ki_{0}+1}^\ki \sum_{j=1}^\N b_{t}^{j}[\ell]$ at a realisation of this kernel gives the roll-out estimator whose properties are analysed in \Cref{theo:bias-mse-rolling}. The Markov kernel $\noiseker{t}{}$ is detailed in \eqref{eq:def:bbP}.

The following assumptions,  
 are vital when analysing the convergence of \Cref{alg:scoreasc:full}.
\begin{assumption}
    \label{assumption:lyapunov_smoothness}
    \begin{itemize}[label=(\roman*),nosep]
        \item[(i)] The function $\paramopt{} \mapsto V(\paramopt{})$ is $\lyapunovlip$-smooth.
        \item[(ii)] The function $\paramopt{} \mapsto \targp{0:t}{}$ is $\targlip$-Lipschitz in total variation distance.
        \item[(iii)] For each path $\zeta_{0:t} \in \xsp{0:t}$, the function \begin{equation}
            \paramopt{} \mapsto  \gibbs{\paramopt{},t}{}(\zeta_{0:t}, \rmd \tilde{\zeta}_{0:t}) 
        \end{equation}
        is $L^P_1$-Lipschitz in total variation distance, where $\gibbs{\paramopt{},t}{}$ is path-marginalized Markov transition kernel associated with the \PPG\ algorithm when the model is parameterized by $\paramopt{}$, see \eqref{eq:ppg_ker_def}. 
        \item[(iv)] For each path $\zeta_{0:t} \in \xsp{0:t}$, the function
        \begin{equation}
            \paramopt{} \mapsto \noiseker{t}{}\Pi_{k_0 - 1, k, N}(s_{0:t, \paramopt{}})(\zeta_{0:t})
        \end{equation}
        is $L^P_2$-Lipschitz in total variation distance. 
    \end{itemize}
\end{assumption} 
In the case of score ascent we check, in \Cref{sec:appendix:learning_with_ppg}, that these assumptions hold if the strong mixing assumption  \cref{assumption:strong_mixing} is satisfied uniformly in $\paramopt{}$, and with additional assumptions on the model.
We are now ready to state a bound on the mean field $h(\paramopt{})$ for  \Cref{alg:scoreasc:full}.
 \begin{theorem}
    \label{theo:bound_grad}
      Assume \cref{assumption:strong_mixing} uniformly in $\paramopt{}$ and \Cref{assumption:lyapunov_smoothness} and suppose that the stepsizes $\{\lr{\ell+1}\}_{\ell \in \intvect{0}{n-1}}$ satisfy
    $ \lr{\ell+1} \leq \lr{\ell}$, $\lr{\ell} < a\lr{\ell + 1}$, $\lr{\ell} - \lr{\ell + 1} < a' \lr{\ell}^2$ and $\lr{1} \leq 0.5 (\lyapunovlip + C_h)$
    for some $a > 0$, $a' > 0$ and all $n \in \nset$. Then, 

    \begin{equation}
       \PE{\|h(\paramopt{\varpi})\|^2}{} \leq 2\frac{V_{0,n} + C_{0,n} + C_{0,\gamma}\sum_{k=0}^{n} \lr{k+1}^2}{\sum_{k=0}^{n} \lr{k+1}} \eqsp,
    \end{equation}
    where $V_{0, n} = \PE{V(\paramopt{}) - V(\paramopt{n})}{}$ and
    \begin{align}
        C_{0,n} &\eqdef \lr{1} h(\paramopt{0}) C_0 + \constbiasalg (\lr{1} - \lr{n+1} + 1)\delta_{k,N,t}^{-1}\eqsp, \label{eq:non_ass:c_0}\\
        C_{0,\gamma}& \eqdef  \boundmse^2 \lyapunovlip + \boundmse C_1 + \boundmse \constbiasalg \left(\lyapunovlip + \frac{C_2}{1 - \mixrate{t}{\N}}\right)\delta_{k,N,t}^{-1}
    + \constbiasalg\lyapunovlip\delta_{k,N,t}^{-1}\eqsp, \label{eq:non_ass:c_gamma}\\
        C_{h} &\eqdef \left( C_1 + \constbiasalg \frac{C_2}{(1 - \mixrate{t}{\N})\delta_{k,N,t}} \right)\left[(a+1)/2 + a \boundmse \right] + (\lyapunovlip + a' + 1) \constbiasalg\delta_{k,N,t}^{-1} \eqsp,\label{eq:non_ass:c_h} \\
        C_1 &= L_2^P\left[1 + \mixrate{t}{\N}^{k}\delta_{k,N,t}^{-1}\right] + \lyapunovlip \\
    C_2 &= L_1^P \delta_{k,N,t}^{-1} + \targlip\mixrate{t}{\N}^{k} \eqsp .      
    \end{align}
     where $C_0$ is independent of $\constbiasalg, \constmsealg, \N$ and where $\delta_{k,N,t} = 1-\mixrate{t}{\N}^{\ki} $. 
 \end{theorem}
 
\Cref{theo:bound_grad} establishes not only the convergence of \Cref{alg:scoreasc:full}, but also illustrates the impact of the bias and the variance of the {\PPG} on the convergence rate. 
\begin{remark}
    \label{corollary:bound_grad}
    Under additional assumptions on the model (cf \Cref{sec:appendix:learning_with_ppg}), if we consider $\lr{1} \leq 0.5 (\lyapunovlip + C_h)$, $\lr{\ell} = \lr{1} {\ell}^{-1/2}$ for all $\ell \in \intvect{1}{n}$, then $\sum_{k=0}^{n} \lr{k+1}^2/\sum_{k=0}^{n} \lr{k+1} \sim  \log n / \sqrt{n} $, showing that $\PE{\|h(\paramopt{\varpi})\|^2}{}$ is $\mathcal{O}(\log n / \sqrt{n})$, where the leading constant depends on $\constbiasalg$ and $\constmsealg$.
\end{remark}
\Cref{corollary:bound_grad} establishes the rate of convergence of \Cref{alg:scoreasc:full}. In principle we could try to optimize the parameters $\ki, \ki_0$ and $\N$ of the algorithm using these bounds, but one of the main challenges with this approach is the determination of the mixing rate, which is underestimated by $\kappa_{N, t}$. Still, our bound provides interesting information of the role of both bias and MSE.

\section{Numerics}
In this section, we focus on the numerical analysis of the two main results of the paper, namely the bias and MSE bounds of the roll-out estimator established in \Cref{theo:bias-mse-rolling} and the efficiency of using {\PPG} for learning in the framework developed in \Cref{sec:learning}. For the latter, we will restrict ourselves to the case of parameter learning via score ascent. In this setting, the competing method that corresponds most closely to the one presented here consists of using, as presented in  \Cref{alg:scoreasc:pgas}, a standard particle Gibbs sampler $\Pi_{\paramopt{}}$ instead of the {\PPG}. One of the most common such samplers is the \emph{particle Gibbs with ancestor sampling} (\PGAS) presented in \cite{lindsten14a}. In \cite{lindholm2018}, the {\PGAS} is used for parameter learning in HMMs via the Expectation Maximization (EM) algorithm. 
\begin{algorithm}
        \KwData{$\chunk{\zeta}{0}{t}[0]$, $\paramopt{0}$, number $k$ of paths per trajectory, burn-in $k_0$, number $n$ of SA iterations, learning-rate sequence $\{\gamma_{\ell}\}_{\ell \in \nset}$, $ \Pi_{\paramopt{}}(\zeta_{0:t}, \rmd \tilde{\zeta}_{0:t})$ a Markov kernel targeting $\targ{0:t}$.}
        \KwResult{$\paramopt{n}$}
        \For{$i \gets 0$ \KwTo $n-1$}{
            
            \For{$j \gets 0$ \KwTo $k-1$}{
                sample $\tilde{\zeta}_{0:t}[j+1] \sim \Pi_{\paramopt{}}(\tilde{\zeta}_{0:t}[j], \cdot)$ \;
            }
            set $\paramopt{i+1} \gets \paramopt{i} + \frac{\gamma_{i+1}}{k - k_0} \sum_{\ell=k_0+1}^{k} \score{i}{0:t}(\tilde{\zeta}_{0:t}[\ell])$\;
            
            set $\zeta_{0:t}[i+1] = \tilde{\zeta}_{0:t}[k]$\;
        }
        \caption{Score ascent with particle Gibbs kernel.} \label{alg:scoreasc:pgas}
\end{algorithm}

\subsection{\PPG\ }
\label{sec:numerics}
\paragraph{Linear Gaussian state-space model (LGSSM).}
We first consider a linear Gaussian HMM
\begin{equation}
X_{m + 1} = A X_m + Q \epsilon_{m + 1}, \quad Y_m = B X_m + R \zeta_m, \quad m \in \nset,
\end{equation}
where $\{\epsilon_m\}_{m \in \nsetpos}$ and $\{\zeta_m\}_{m \in \nset}$ are sequences of independent standard normally distributed random variables, independent of $X_0$.
The coefficients $A$, $Q$, $B$, and $R$ are assumed to be known and equal to $0.97$, $0.60$, $0.54$, and $0.33$, respectively.
Using this parameterisation, we generate, by simulation, a record of $t = 999$ observations.

\begin{figure}[h]
    \centering
    \begin{subfigure}{0.4\textwidth}
        \includegraphics[width=\textwidth]{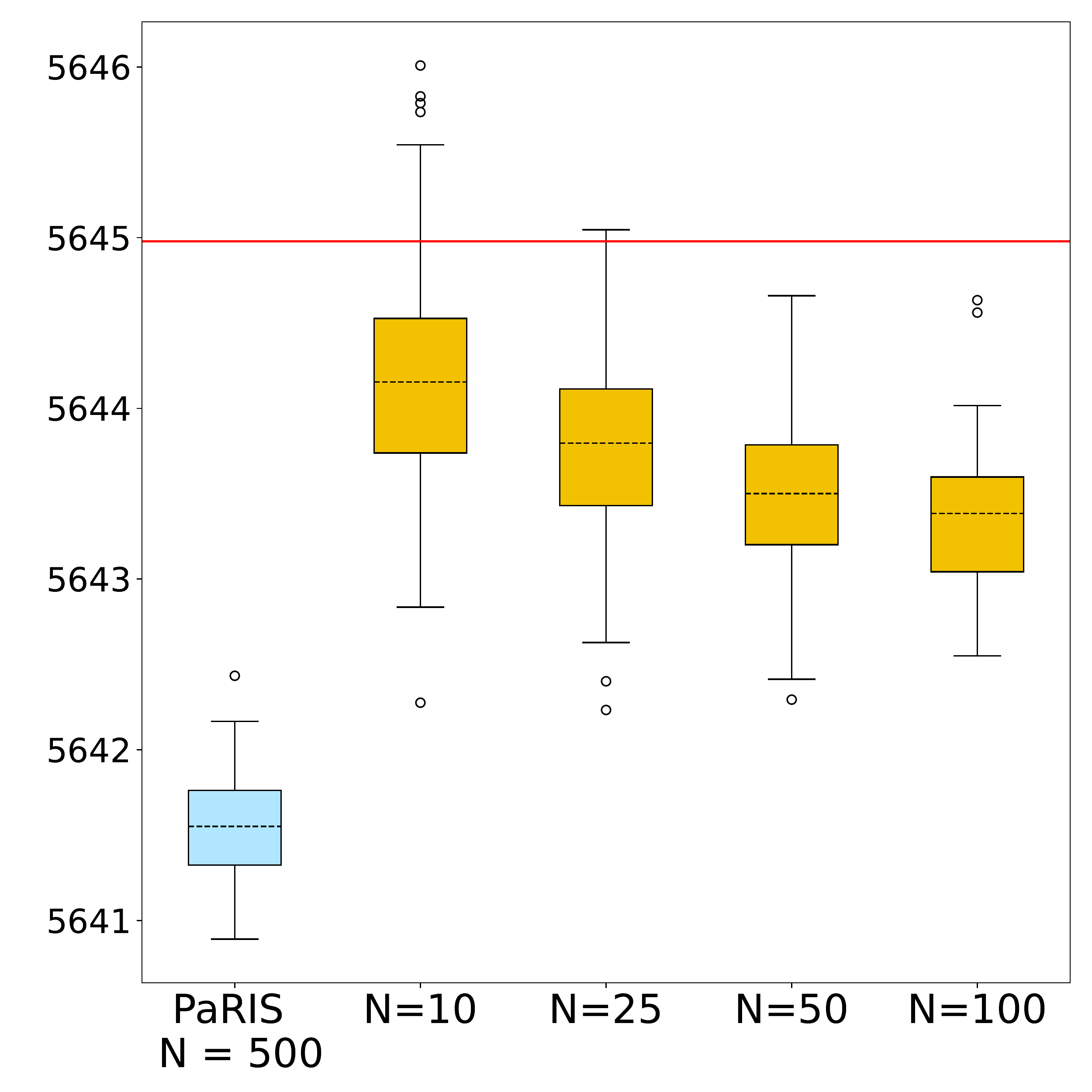}
    \end{subfigure}
    \begin{subfigure}{0.4\textwidth}
        \includegraphics[width=\textwidth]{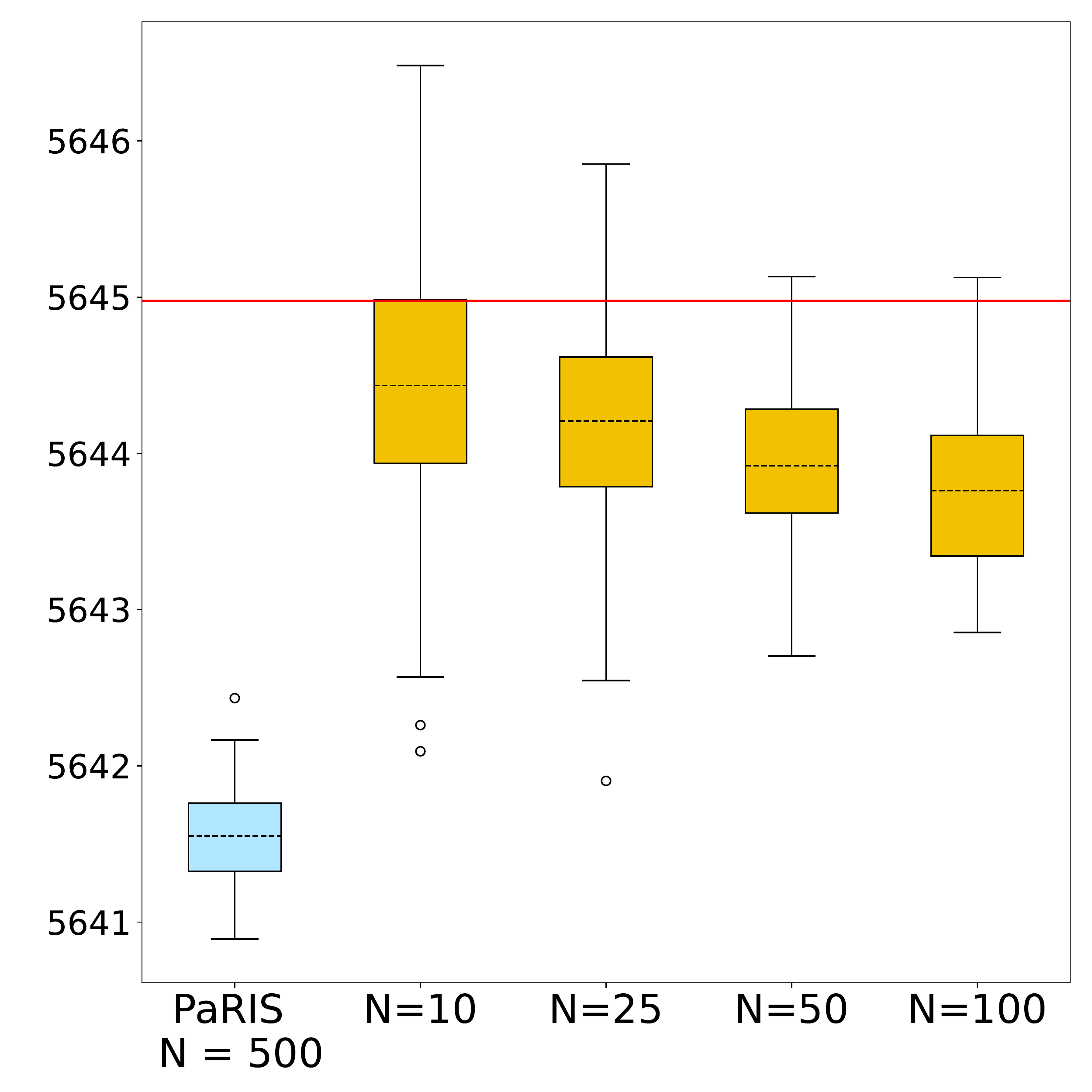}
    \end{subfigure}
    \caption{{\PARIS} and {\PPG} outputs for the LGSSM for $\totalbudget = 500$, yellow boxes correspond to {\PPG} outputs produced using $\ki \in \{50, 20, 10, 5\}$ iterations and $\N \in \{C/50, C/20, C/10, C/5\}$ particles. The image on the left corresponds to taking $\ki_0 = \ki / 2$ and the one on the right to $\ki_0 = \ki / 4$.}
    \label{fig:LGSSM:comparison_component_2}
\end{figure}

In this setting, we aim at computing smoothed expectations of the state one-lag covariance
$\af{t}(\chunk{x}{0}{t}) \eqdef \sum_{m=0}^{t-1} x_m x_{m + 1}$. In the linear Gaussian case, the \emph{disturbance smoother} (see \cite[Algorithm 5.2.15]{cappe:moulines:ryden:2005}) provides the exact values of the smoothed sufficient statistics, which allows us to study the bias of the estimator for a given computational budget $\totalbudget$.
\Cref{fig:LGSSM:comparison_component_2}
displays, for three different total budgets $\totalbudget$, the distribution of estimates of $\targ{0:n} h_{n}$ using the {\PARIS} as well as three different configurations of the {\PPG} corresponding to $\ki \in \{2, 4, 10\}$ (and $N = \totalbudget / \ki$) with $\ki_0 = \ki / 2$ and $\ki_0 = \ki / 4$.
The reference value is shown as a red-dashed line and the mean value of each distribution is shown as a black-dashed line.
Each boxplot is based on $1000$ independent replicates of the corresponding estimator. We observe that in this example, all configurations of the {\PPG} are less biased than the equivalent {\PARIS} estimator.
The illustration of the bounds from \Cref{theo:bias-mse-rolling} is postponed to  \Cref{sec:appendix:numerics:ppg}.

\subsection{Score ascent}

\paragraph{LGSSM.}

\begin{figure}[h]
     \centering
     \begin{subfigure}{0.65\textwidth}
         \includegraphics[width=\textwidth]{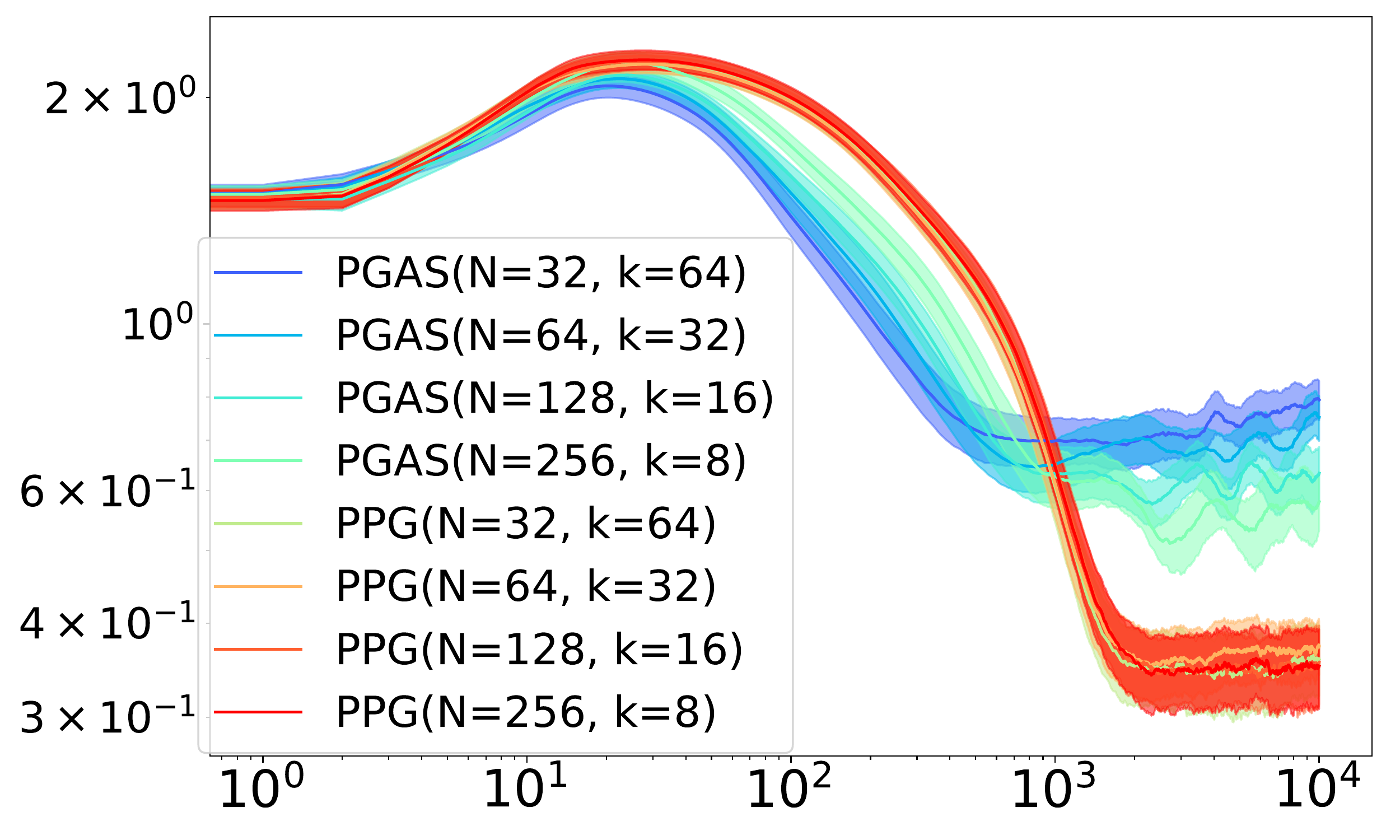}
     \end{subfigure}
     \caption{Distance to the MLE estimator as a function of the iteration step for PGAS and \PPG\ with different parameters while keeping the particle budget fixed for LGSSM for $25$ different seeds.}
     \label{fig:LGSSM:dist_mle}
 \end{figure}

We consider the LGSSM with state and observation spaces being $\rset^{5}$. We assume that the parameters $R$ and $Q$ are known and consider the inference of $\paramopt{} = (A, B)$ on the basis of a simulated sequence of $n = 999$ observations.
In this setting, the M-step of the EM algorithm can be solved exactly with the disturbance smoother \cite[Chapter 11]{cappe:moulines:ryden:2005}. The parameter obtained by this procedure (denoted $\paramopt{\scriptsize{\mbox{{\it mle}}}}$) is the reference value for any likelihood maximization algorithm.
 \Cref{table:comparison LGSSM} shows the $\mathrm{L}_2$ distance between the singular values of $\paramopt{\scriptsize{\mbox{{\it mle}}}}$ and those of the parameters obtained by \Cref{alg:scoreasc:full} and \Cref{alg:scoreasc:pgas}.
The CLT confidence intervals were obtained on the basis of $25$ replicates. The configurations respect a given particle budget $kN = \totalbudget = 1024$.
The choice of keeping $k_0 = k/2$ is a heuristic rule to achieve a good bias--variance trade-off, but other combinations of $k_0$ and $k$ may lead to better performance for different problems. We analyse this for the LGSMM in \Cref{sec:appendix:numerics:learning}.
 All settings are the same for both algorithms and are described in \Cref{sec:appendix:numerics:learning}. The {\PPG} achieves consistently a smaller distance to $\paramopt{\scriptsize{\mbox{{\it mle}}}}$.  \Cref{fig:LGSSM:dist_mle} displays, for each estimator and configuration, the evolution of the distance to the MLE estimator as a function of the iteration index. 

\begin{filecontents*}{6_comparison_fixed_budget.csv}
Algorithm,Distance,NLL,lr,N,k,k_0,gradNorm
PGAS,0.793 ± 0.048,-2832.256 +/- 1.207,32.0,32,64,32,2.591 +/- 0.258
PGAS,0.751 ± 0.052,-2851.789 +/- 2.256,16.0,64,32,16,2.051 +/- 0.202
PGAS,0.633 ± 0.054,-2864.262 +/- 3.748,8.0,128,16,8,2.232 +/- 0.167
PGAS,0.580 ± 0.049,-2884.268 +/- 2.245,4.0,256,8,4,2.353 +/- 0.180
PPG,0.358 ± 0.038,-2943.734 +/- 0.395,32.0,32,64,32,2.728 +/- 0.262
PPG,0.373 ± 0.031,-2942.622 +/- 0.446,16.0,64,32,16,2.647 +/- 0.230
PPG,0.355 ± 0.043,-2942.859 +/- 0.543,8.0,128,16,8,2.672 +/- 0.313
PPG,0.351 ± 0.042,-2942.507 +/- 0.753,4.0,256,8,4,2.417 +/- 0.240
\end{filecontents*}
\DTLloaddb{comparison_score_ascent}{6_comparison_fixed_budget.csv}
\begin{table}
    \centering
    \resizebox{.45\textwidth}{!}{%
    \begin{tabular}{|c|c|c|c|c|}%
          \hline
          Algorithm & $N$ & $k_0$ & $k$ & $D_{\scriptsize{\mbox{{\it mle}}}}$\\
          \hline
          \DTLforeach*{comparison_score_ascent}{\Alg=Algorithm,\D=Distance, \kze=k_0, \k=k, \Npart=N, \NLL=NLL, \grad=gradNorm}{
            \Alg & \Npart & \kze & \k &\D \DTLiflastrow{}{\\
          }}
        \\\hline
    \end{tabular}
    }
    \caption{Distance to $\paramopt{\operatorname{MLE}}$ for each configuration in the LGSSM case.}
    \label{table:comparison LGSSM}
\end{table}

\paragraph{CRNN.}

We consider now the problem of inference in a non-linear HMM and in particular the chaotic recurrent neural network
introduced by \cite{zhao2021streaming}. We use the same setting as in the original paper.
The state and observation equations are
\begin{align*}
X_{m + 1} &= X_m + \tau^{-1}\Delta \left(-X_{m} + \gamma W \tanh(X_{m})\right) + \epsilon_{m + 1}, \\
Y_m &= B X_m + \zeta_m, \quad m \in \nset,
\end{align*}
where $\{\epsilon_m\}_{m \in \nsetpos}$ is a sequence of $20$-dimensional independent multivariate Gaussian random variables with zero mean
and covariance $0.01 \mathbf{I}$ and $\{\zeta_m\}_{m \in \nset}$ is a sequence of independent random variables where each component is distributed independently
according to a Student's t-distribution with scale $0.1$ and $2$ degrees of freedom.

In this case, the natural metric used to evaluate the different estimators is the negative log likelihood (NLL).
We use the unbiased estimator of the likelihood given by the mean of the log weights produced by a particle filter \cite[Section 12.1]{douc2014nonlinear} using $N = 10^4$ particles.
\Cref{table:comparison_CRNN} shows the results obtained for $25$ different replications
for several different configurations of {\PPG} and {\PGAS}, while keeping total budget of particles fixed. Further numerical details are given in \Cref{sec:appendix:numerics:learning}.
We observe that {\PPG} achieves the a considerably lower NLL than {\PGAS} in all configurations.

\begin{filecontents*}{2_comparison_fixed_budget.csv}
Algorithm,lr,N,k,k_0,Distance,NLL,gradNorm
PGAS,16,32,32,16,15.485 ± 0.196,31364.932 ± 173.708,10.231 ± 1.374
PGAS,8,64,16,8,10.939 ± 0.383,31083.408 ± 380.527,18.660 ± 3.407
PGAS,4,128,8,4,7.292 ± 0.217,30264.836 ± 265.880,32.930 ± 3.606
PPG,16,32,32,16,16.013 ± 0.711,22291.971 ± 47.683,3.959 ± 0.242
PPG,8,64,16,8,15.878 ± 0.384,22314.537 ± 25.028,3.893 ± 0.285
PPG,4,128,8,4,15.659 ± 0.762,22353.416 ± 39.443,3.521 ± 0.159
\end{filecontents*}
\DTLloaddb{comparison_score_ascent_crnn}{2_comparison_fixed_budget.csv}
\begin{table}
    \centering
    \resizebox{.45\textwidth}{!}{%
    \begin{tabular}{|c|c|c|c|c|}%
          \hline
          Algorithm & N & $k_0$ & $k$ & NLL\\
          \hline
          \DTLforeach*{comparison_score_ascent_crnn}{\Alg=Algorithm, \kze=k_0, \k=k, \Npart=N, \NLL=NLL}{
            \Alg & \Npart & \kze & \k & \NLL \DTLiflastrow{}{\\
          }}
        \\\hline
    \end{tabular}
    }
    \caption{Per configuration negative loglikelihood for the CRNN model.}
    \label{table:comparison_CRNN}
\end{table}
\section{Conclusion}
We have presented a new algorithm, referred to as \PPG\, as well as bounds on its bias and MSE in \Cref{theo:bias-mse-rolling}. We then propose a way of using \PPG\ in a learning framework and derive a non-asymptotic bound over the gradient of the updates when doing score ascent with the \PPG\ with explicit dependence on the bias and MSE of the estimator. We provide numerical simulations to support our claims, and we show that our algorithm outperforms the current competitors in the two different examples analysed.

\clearpage
\appendix
\onecolumn
\tableofcontents
\section{\PPG}
\label{sec:appendix:ppg}
In this section, we develop the theoretical framework necessary to establish \Cref{theo:bias-mse-rolling}. We recall the notions of  \emph{Feynman--Kac models}, \emph{many-body Feynman--Kac models}, \emph{backward interpretations}, and \emph{conditional dual processes}. Our presentation follows closely \cite{del2016particle} but with a different and hopefully more transparent definition of the many-body extensions. We restate (in \Cref{thm:duality} below) a duality formula of \cite{del2016particle} relating these concepts.
This formula provides a foundation for the \emph{particle Gibbs sampler} described in \Cref{alg:ppg}.
\paragraph{Notations. }
Let $(\zsp{}, \zfd{})$ be a measurable space and $L$ another possibly unnormalised transition kernel on $\ysp{} \times \zfd{}$. Define, with $K$ as above, 
$$
	K L: \xsp{} \times \zfd{} \ni (x, A) \mapsto \int  L(y, A) \, K(x, \rmd y)
$$
and
$$
	K \tensprod L: \xsp{} \times(\yfd{} \tensprod \zfd{}) \ni (x, A) \mapsto \iint \indi{A}(y, z) \, K(x, \rmd y) \, L(y, \rmd z),
$$
whenever these are well defined. This also defines the $\tensprod$ products of a kernel $K$ on $\xsp{} \times \yfd{}$ and a measure $\nu$ on $\xfd{}$ as well as of a kernel $L$ on $\ysp{} \times \xfd{}$ and a measure $\mu$ on $\yfd{}$ as the measures  
\begin{align*}
\nu \tensprod K &: \xfd{} \tensprod \yfd{} \ni A \mapsto \iint \indi{A}(x, y) \, K(x, \rmd y) \, \nu(\rmd x), \\
L \tensprod \mu &: \xfd{} \tensprod \yfd{} \ni A \mapsto \iint \indi{A}(x, y) \, L(y, \rmd x) \, \mu(\rmd y).
\end{align*}

\subsection{Many-body Feynman--Kac models}
\label{sec:mb:FK:models}
In the following, we assume that all random variables are defined on a common probability space $(\Omega, \mathcal{F}, \prob)$.
The distribution flow $\{ \targ{m} \}_{m \in \nset}$ defined in \cref{eq:normalized:F-K:measures} is intractable in general,
but can be approximated  by random samples $\epartmb{m} = \{ \epart{m}{i} \}_{i = 1}^\N$, $m \in \nset$, referred to as  \emph{particles},
where $\N \in \nsetpos$ is a fixed Monte Carlo sample size and each particle $\epart{m}{i}$ is an $\xsp{m}$-valued random variable. Such particle approximation is based on the recursion $\targ{m + 1} = \pd{m}(\targ{m})$, $m \in \nset$, where $\pd{m}$ denotes the mapping
\begin{equation} \label{eq:def:pd:mapping}
\pd{m} : \probmeas(\xfd{m}) \ni \eta \mapsto \frac{\eta \uk{m}}{\eta \pot{m}}
\end{equation}
taking on values in $\probmeas(\xfd{m + 1})$. In order to describe recursively the evolution of the particle population, let $m \in \nset$ and assume that the particles $\epartmb{m}$ form a consistent approximation of $\targ{m}$ in the sense that $\occm(\epartmb{m}) h$, where $\occm(\epartmb{m}) \eqdef N^{-1}\sum_{i = 1}^\N \delta_{\epart{m}{i}}$, with $\delta_x$ denotes the Dirac measure located at $x$, is the occupation measure formed by $\epartmb{m}$, which serves as a proxy for $\targ{m} h$ for all $\targ{m}$-integrable test functions $h$. Under general conditions, $\occm(\epartmb{m}) h$ converges in probability to $\targ{m}$ with $N \rightarrow \infty$; see \cite{delmoral:2004,chopin2020introduction} and references therein.  Then, in order to generate an updated particle sample  approximating $\targ{m + 1}$, new particles $\epartmb{m + 1} = \{ \epart{m + 1}{i} \}_{i = 1}^\N$ are drawn conditionally independently given $\epartmb{m}$ according to
$$
\epart{m + 1}{i} \sim \pd{m}(\occm(\epartmb{m})) = \sum_{\ell = 1}^\N \frac{\pot{m}(\epart{m}{\ell})}{\sum_{\ell' = 1}^\N \pot{m}(\epart{m}{\ell'})} \mk{m}(\epart{m}{\ell}, \cdot), \quad i \in \intvect{1}{\N}.
$$
Since this process of particle updating involves sampling from the mixture distribution $ \pd{m}(\occm(\epartmb{m}))$, it can be naturally decomposed into two substeps: \emph{selection} and \emph{mutation}. The selection step consists of randomly choosing the $\ell$-th mixture stratum with probability $\pot{m}(\epart{m}{\ell}) / \sum_{\ell' = 1}^\N \pot{m}(\epart{m}{\ell'})$ and the mutation step consists of drawing a new particle $\epart{m + 1}{i}$ from the selected stratum $\mk{m}(\epart{m}{\ell}, \cdot)$. In \cite{del2016particle}, the term \emph{many-body Feynman--Kac models} is related to the law of process $\{ \epartmb{m} \}_{m \in \nset}$. For all $m \in \nset$, let $\xspmb{m} \eqdef \xsp{m}^\N$ and $\xfdmb{m} \eqdef \xfd{m}^{\varotimes \N}$; then $\{ \epartmb{m} \}_{m \in \nset}$ is an inhomogeneous Markov chain on $\{ \xspmb{m} \}_{m \in \nset}$ with transition kernels
$$
\mkmb{m} : \xspmb{m} \times \xfdmb{m + 1} \ni (\xmb{m}, A) \mapsto \pd{m}(\occm(\xmb{m}))^{\tensprod \N}(A)
$$
and initial distribution $\initmb = \init^{\tensprod \N}$. Now, denote $\xspmb{0:n} \eqdef \prod_{m = 0}^n \xspmb{m}$ and $\xfdmb{0:n} \eqdef \bigotimes_{m = 0}^n \xfdmb{m}$. In the following, we use a bold symbol to stress that a quantity is related to the many-body process.  The \emph{many-body Feynman--Kac path model} refers to the flows $\{ \untargmb{m} \}_{m \in \nset}$ and $\{ \targmb{m} \}_{m \in \nset}$ of the unnormalised and normalised, respectively, probability distributions on $\{ \xfdmb{0:m} \}_{m \in \nset}$ generated by \eqref{eq:normalized:F-K:measures} and \eqref{eq:unnormalized:F-K:measures} for the Markov kernels $\{ \mkmb{m} \}_{m \in \nset}$, the initial distribution $\initmb$, the potential functions
$$
\potmb{m} : \xspmb{m} \ni \xmb{m} \mapsto \occm(\xmb{m}) \pot{m} = \frac{1}{N} \sum_{i = 1}^\N \pot{m}(x_m^i), \quad m \in \nset,
$$
and the corresponding unnormalised transition kernels
$$
\ukmb{m} : \xspmb{m} \times \xfdmb{m + 1} \ni (\xmb{m}, A) \mapsto \potmb{m}(\xmb{m}) \mkmb{m} (\xmb{m}, A), \quad m \in \nset.
$$

\subsection{Backward interpretation of Feynman--Kac path flows}
Suppose that each kernel $\uk{n}$, $n \in \nset$, defined in \eqref{eq:unnormalised:kernel}, has a transition density $\ud{n}$ with respect to some dominating measure $\refm{n+1} \in \meas(\xfd{n + 1})$. Then for $n \in \nset$ and $\eta \in \probmeas(\xfd{n})$ we may define the \emph{backward kernel}
\begin{equation} \label{def:backward:kernel}
\bk{n, \eta} : \xsp{n + 1} \times \xfd{n} \ni (x_{n + 1}, A) \mapsto \frac{\int \1_A(x_n) \ud{n}(x_n, x_{n + 1}) \, \eta(\rmd x_n)}{\int \ud{n}(x_n', x_{n + 1}) \, \eta(\rmd x_n')}.
\end{equation}
Now, denoting, for $n \in \nsetpos$,
\begin{equation}
\label{eq:bckwd-kernels}
\rk{n} : \xsp{n} \times \xfd{0:n - 1} \ni (x_n, A) \mapsto \idotsint \1_A(x_{0:n - 1}) \prod_{m = 0}^{n - 1} \bk{m, \targ{m}}(x_{m + 1}, \rmd x_m),
\end{equation}

we may state the following---now classical---\emph{backward decomposition} of the Feynman--Kac path measures, a result that  plays a pivotal role in this paper.

\begin{proposition} \label{prop:backward:decomposition}
For every $n \in \nsetpos$ it holds that $\untarg{0:n} = \untarg{n} \tensprod \rk{n}$ and $\targ{0:n} = \targ{n} \tensprod \rk{n}$.
\end{proposition}

Although the decomposition in \Cref{prop:backward:decomposition} is well known (see, \eg, \cite{delmoral:doucet:singh:2010,del2016particle}), we provide a proof in \Cref{sec:proof:backward:decomposition} for completeness. Using the backward decomposition,  a particle approximation of a given Feynman--Kac path measure $\targ{0:n}$ is obtained by first sampling, in an initial forward pass, particle clouds $\{ \epartmb{m} \}_{m = 0}^n$ from $\initmb \tensprod \mkmb{0} \tensprod \cdots \tensprod \mkmb{n - 1}$ and then sampling, in a subsequent backward pass, for instance $N$ conditionally independent paths $\{\eparttd{0:n}{i} \}_{i = 1}^\N$ from $\bdpart{n}(\epartmb{0}, \ldots, \epartmb{n}, \bcdot)$, where
\begin{equation} \label{eq:def:bdpart}
\bdpart{n} : \xspmb{0:n} \times \xfd{0:n} \ni (\xmb{0:n}, A) \mapsto \idotsint  \1_A(x_{0:n}) \left( \prod_{m = 0}^{n - 1} \bk{m,\occm(\xmb{m})}(x_{m + 1}, \rmd x_m) \right) \occm(\xmb{n})(\rmd x_n)
\end{equation}

is a Markov kernel describing the time-reversed dynamics induced by the particle approximations generated in the forward pass. Here and in the following we use blackboard notation to denote kernels related to many-body path spaces. Finally, $\occm(\{\eparttd{0:n}{i} \}_{i = 1}^\N) h$ is returned as an estimator of $\targ{0:n} h$ for any $\targ{0:n}$-integrable test function $h$. This algorithm is in the literature referred to as the \emph{forward--filtering backward--simulation} (FFBSi) \emph{algorithm} and was introduced in \cite{godsill:doucet:west:2004}; see also \cite{cappe:godsill:moulines:2007,douc:garivier:moulines:olsson:2009}.
More precisely, given the forward particles $\{ \epartmb{m} \}_{m = 0}^n$, each path $\eparttd{0:n}{i}$ is generated by first drawing $\eparttd{n}{i}$ uniformly among the particles $\epartmb{n}$ in the last generation and then drawing, recursively,
\begin{equation} \label{eq:backward:sampling:operation}
\eparttd{m}{i} \sim \bk{m,\occm(\epartmb{m})}(\eparttd{m + 1}{i}, \cdot) = \sum_{j = 1}^\N \frac{\ud{m}(\epart{m}{j}, \eparttd{m + 1}{i})}{\sum_{\ell=1}^{N} \ud{m}(\epart{m}{\ell}, \eparttd{m + 1}{i})} \delta_{\epart{m}{j}}(\cdot),
\end{equation}
\emph{i.e.}, given $\eparttd{m + 1}{i}$, $\eparttd{m}{i}$ is picked at random among the $\epartmb{m}$ according to weights proportional to $\{ \ud{m}(\epart{m}{j}, \eparttd{m + 1}{i}) \}_{j = 1}^\N$.
Note that in this basic formulation of the FFBSi algorithm, each backward-sampling operation \eqref{eq:backward:sampling:operation} requires the computation of the normalising constant $\sum_{\ell=1}^{N} \ud{m}(\epart{m}{\ell}, \eparttd{m + 1}{i})$, which implies an overall quadratic complexity of the algorithm. Still, this heavy computational burden can eased by means of an effective accept--reject technique discussed in \Cref{sec:std:paris}.

\subsection{Conditional dual processes and particle Gibbs}
\label{sec:dual:and:Gibbs}
The \emph{dual process} associated with a given Feynman--Kac model (\ref{eq:normalized:F-K:measures}--\ref{eq:unnormalized:F-K:measures}) and a given trajectory $\{ z_n \}_{n \in \nset}$, where $z_n \in \xsp{n}$ for every $n \in \nset$,
is defined as the canonical Markov chain with kernels
\begin{equation}
\label{eq:kernel:cond:dual}
\mkmb{n}[z_{n + 1}] : \xspmb{n} \times \xfdmb{n + 1} \ni 
(\xmb{n}, A) \mapsto \frac{1}{\N} \sum_{i = 0}^{\N - 1} \left(\Phi_n(\occm(\xmb{n}))^{\tensprod i}
\tensprod \delta_{z_{n + 1}} \tensprod \Phi_n(\occm(\xmb{n}))^{\tensprod (\N - i - 1)} \right)(A),
\end{equation}
for $n \in \nset$, and initial distribution
\begin{equation} \label{eq:init:cond:dual}
\initmb[z_0] \eqdef \frac{1}{\N} \sum_{i = 0}^{\N - 1} \left( \init^{\tensprod i} \tensprod \delta_{z_0} \tensprod \init^{\tensprod (\N - i - 1)} \right).
\end{equation}
As clear from (\ref{eq:kernel:cond:dual}--\ref{eq:init:cond:dual}), given $\{ z_n \}_{n \in \nset}$, a realisation $\{ \epartmb{n} \}_{n \in \nset}$ of the dual process is generated as follows. At time zero, the process is initialised by inserting $z_0$ at a randomly selected position in the vector $\epartmb{0}$ while drawing independently the remaining components from $\targ{0}$. Then, given $\epartmb{n}$ at step $n$, $z_{n + 1}$ is inserted at a randomly selected position in $\epartmb{n + 1}$ while drawing independently the remaining components from $\Phi_n(\occm(\epartmb{n}))$.

In order to describe compactly the law of the conditional dual process, we define the Markov kernel
$$
\mbjt{n} : \xsp{0:n} \times \xfdmb{0:n} \ni (\chunk{z}{0}{n}, A) \mapsto \initmb[z_0] \tensprod \mkmb{0}[z_1] \tensprod \cdots \tensprod \mkmb{n - 1}[z_n](A).
$$
The following result elegantly combines the underlying model (\ref{eq:normalized:F-K:measures}--\ref{eq:unnormalized:F-K:measures}), the many-body Feynman--Kac model, the backward decomposition, and the conditional dual process.
\begin{theorem}[\cite{del2016particle}] \label{thm:duality}
For all $n \in \nset$,
\begin{equation}  \label{eq:duality}
\bdpart{n} \tensprod \untargmb{0:n}  = \untarg{0:n} \tensprod \mbjt{n}.
\end{equation}
\end{theorem}
In \cite{del2016particle}, each state $\epartmb{n}$ of the many-body process maps an outcome $\omega$ of the sample space $\Omega$ into an \emph{unordered set} of $N$ elements in $\xsp{n}$. However, we have chosen to let each $\epartmb{n}$ take on values in the standard \emph{product space} $\xsp{n}^\N$ for two reasons: first, the construction of \cite{del2016particle} requires sophisticated measure-theoretic arguments to endow such unordered sets with suitable $\sigma$-fields and appropriate measures; second, we see no need to ignore the index order of the particles as long as the Markovian dynamics (\ref{eq:kernel:cond:dual}--\ref{eq:init:cond:dual}) of the conditional dual process is symmetrised over the particle cloud. Therefore, in \Cref{sec:proof:thm:duality}, we include our own proof of duality \eqref{eq:duality} for completeness. Note that the measure \eqref{eq:duality} on $\xfd{0:n} \tensprod \xfdmb{0:n}$ is unnormalised, but since the kernels $\bdpart{n}$ and $\mbjt{n}$ are both Markovian, normalising the identity with $\untarg{0:n}(\xsp{0:n}) = \untargmb{0:n}(\xspmb{0:n})$ yields immediately
\begin{equation} \label{eq:duality:normalised}
\bdpart{n} \tensprod \targmb{0:n}   = \targ{0:n} \tensprod \mbjt{n}.
\end{equation}
Since the two sides of \eqref{eq:duality:normalised} provide the full conditionals, it is natural to choose a data-augmentation approach and sample the target \eqref{eq:duality:normalised} using a two-stage deterministic-scan Gibbs sampler \cite{andrieu:doucet:holenstein:2010,chopin:singh:2015}. More specifically, assume that we have generated a state
$(\epartmb{0:n}[\ell], \zeta_{0:n}[\ell])$ comprising a dual process with associated path on the basis of $\ell \in \nset$ iterations of the sampler; then the next state $(\epartmb{0:n}[\ell+1], \zeta_{0:n}[\ell+1])$ is generated in a Markovian fashion by sampling first $\epartmb{0:n}[\ell+1] \sim \mbjt{n}(\zeta_{0:n}[\ell], \cdot)$ and then sampling $\zeta_{0:n}[\ell+1] \sim \bdpart{n}(\epartmb{0:n}[\ell+1], \bcdot)$. After arbitrary initialisation (and the discard of possible burn-in iterations), this procedure produces a Markov trajectory $\{ (\epartmb{0:n}[\ell], \chunk{\zeta}{0}{n}[\ell]) \}_{\ell \in \nset}$, and under weak additional technical conditions this Markov chain admits \eqref{eq:duality:normalised} as its unique
invariant distribution. In such a case, the Markov chain is ergodic \cite[Chapter~5]{douc2018markov}, and the marginal distribution of the conditioning path $\chunk{\zeta}{0}{n}[\ell]$ converges to the target distribution $\targ{0:n}$. Therefore, for every $h \in \bmf(\xfd{0:n})$, 
$$
\lim_{L \to \infty} \frac{1}{L} \sum_{\ell=1}^L h(\chunk{\zeta}{0}{n}[\ell]) =
\targ{0:n} h, \quad \prob\mbox{-a.s.}
$$

\subsection{The {\PARIS} algorithm}
\label{sec:std:paris}
In the following, we assume that we are given a sequence $\{ h_n \}_{n \in \nset}$ of \emph{additive state functionals} as in  \eqref{eq:add:functional}.
This problem is particularly relevant in the context of maximum-likelihood-based parameter estimation in general state-space models, \emph{e.g.}, when computing the \emph{score-function}, i.e. the gradient of the log-likelihood function, via the Fisher identity or when computing the intermediate quantity
of the \emph{Expectation Maximization} (EM) \emph{algorithm}, in which case $\targ{0:n}$ and $h_n$ correspond to the joint state posterior and an element of some sufficient statistic, respectively; see \cite{cappe:moulines:2005,douc:garivier:moulines:olsson:2009,delmoral:doucet:singh:2010,poyiadjis:doucet:singh:2011,olsson:westerborn:2017}
and the references therein. Interestingly, as noted in \cite{cappe:2009,delmoral:doucet:singh:2010}, the backward decomposition allows, when applied to additive state functionals, a forward recursion for the expectations $\{ \targ{0:n} \af{n} \}_{n \in \nset}$. More specifically, using the forward decomposition $\af{n + 1}(\chunk{x}{0}{n + 1}) = \af{n}(\chunk{x}{0}{n}) + \afterm{n}(x_n, x_{n + 1})$ and the backward kernel $\rk{n + 1}$ defined in \eqref{eq:bckwd-kernels}, we may write, for $x_{n + 1} \in \xsp{n + 1}$,
\begin{align}
\rk{n + 1} \af{n + 1}(x_{n + 1})  &= \int \bk{n, \targ{n}}(x_{n + 1}, \rmd x_n) \int \left( \af{n}(x_{0:n}) + \afterm{n}(x_n, x_{n + 1}) \right) \rk{n}(x_n, \rmd x_{0:n - 1}) \nonumber \\
&= \bk{n, \targ{n}}(\rk{n} \af{n} + \afterm{n})(x_{n + 1}), \label{eq:backward:forward:recursion}
\end{align}
which by \Cref{prop:backward:decomposition} implies that
\begin{equation}
\label{eq:additive-smoothing-recursion}
\targ{0:n + 1} h_{n + 1} = \targ{n + 1} \bk{n, \targ{n}}(\rk{n} \af{n} + \afterm{n}).
\end{equation}
Since  the marginal flow $\{\targ{n} \}_{n \in \nset}$ can be expressed recursively via the mappings $\{\pd{n} \}_{n \in \nset}$, \eqref{eq:additive-smoothing-recursion} provides, in principle, a basis for online computation of $\{ \targ{0:n} \af{n} \}_{n \in \nset}$. To handle the fact that the marginals are generally intractable we may, following \cite{delmoral:doucet:singh:2010}, plug particle approximations $\occm(\epartmb{n + 1})$ and $\bk{n,\occm(\epartmb{n})}$ (see \eqref{eq:backward:sampling:operation}) of $\targ{n + 1}$ and $\bk{n, \occm(\targ{n})}$, respectively, into the recursion \eqref{eq:additive-smoothing-recursion}. More precisely, we proceed recursively and assume that at time $n$ we  have at hand a sample $\{(\epart{n}{i}, \stat{n}{i})\}_{i=1}^\N$ of particles with associated statistics, where each statistic $\stat{n}{i}$ serves as an approximation of $\rk{n} \af{n}(\epart{n}{i})$; then evolving the particle cloud according to $\epartmb{n + 1} \sim \mkmb{n}(\epartmb{n}, \cdot)$ and updating the statistics using \eqref{eq:backward:forward:recursion}, with $\bk{n, \targ{n}}$ replaced by $\bk{n,\occm(\epartmb{n})}$, yields the particle-wise recursion
\begin{equation} \label{eq:FFBSm:forward:update}
\stat{n + 1}{i} = \sum_{\ell = 1}^\N \frac{\ud{n}(\epart{n}{\ell}, \epart{n + 1}{i})}{\sum_{\ell' = 1}^\N \ud{n}(\epart{n}{\ell'}, \epart{n + 1}{i})} \left( \stat{n}{\ell} + \afterm{n}(\epart{n}{\ell}, \epart{n + 1}{i}) \right), \quad i \in \intvect{1}{\N},
\end{equation}
and, finally, the estimator
\begin{equation}
\label{eq:bckwd-interpretation}
\occm(\statmb{n})(\operatorname{id}) = \frac{1}{N} \sum_{i=1}^\N \stat{n}{i}
\end{equation}
of $\targ{0:n} \af{n}$, where  $\statmb{n} \eqdef (\stat{n}{1}, \ldots, \stat{n}{\N})$, $i \in \intvect{1}{\N}$. The procedure is initialised by simply letting $\stat{0}{i}=0$ for all $i \in \intvect{1}{N}$.
Note that \eqref{eq:bckwd-interpretation} provides a particle interpretation of the backward decomposition in \Cref{prop:backward:decomposition}. This algorithm is a special case of the \emph{forward--filtering backward--smoothing} (FFBSm) \emph{algorithm} (see \cite{andrieu:doucet:2003,godsill:doucet:west:2004,douc:garivier:moulines:olsson:2009,sarkka2013bayesian}) for additive functionals satisfying \eqref{eq:add:functional}. It allows for online processing of the sequence $\{\targ{0:n} \af{n}\}_{n \in \nset}$, but has also the appealing property that only the current particles $\epartmb{n}$ and statistics $\statmb{n}$ need to be stored. However, since each update \eqref{eq:FFBSm:forward:update} requires the summation of $\N$ terms, the scheme has an overall \emph{quadratic} complexity in the number of particles, leading to a computational bottleneck in applications to complex models that require large particle sample sizes $\N$.

In order to detour the computational burden of this forward-only implementation of {FFBSm}, the {\PARIS} algorithm \cite{olsson:westerborn:2017} updates the statistics $\statmb{n}$ by replacing each sum \eqref{eq:FFBSm:forward:update} by a Monte Carlo estimate
\begin{equation}
\label{eq:paris-update}
\stat{n+1}{i} = \frac{1}{\M} \sum_{j=1}^\M \left( \stattd{n}{i, j} + \afterm{n}(\eparttd{n}{i, j}, \epart{n+1}{i}) \right), \quad i \in \intvect{1}{N},
\end{equation}
where $\{(\eparttd{n}{i, j}, \stattd{n}{i, j})\}_{j = 1}^\M$ are drawn randomly among $\{(\epart{n}{i}, \stat{n}{i})\}_{i = 1}^\N$ with replacement, by assigning $(\eparttd{n}{i, j}, \stattd{n}{i, j})$ the value of $(\epart{n}{\ell}, \stat{n}{\ell})$ with probability $\ud{n}(\epart{n}{\ell}, \epart{n + 1}{i}) / \sum_{\ell' = 1}^\N \ud{n}(\epart{n}{\ell'}, \epart{n + 1}{i})$, and the Monte Carlo sample size $\M \in \nsetpos$ is supposed to be much smaller than $\N$ (say, less than $5$). Formally,
$$
\{(\eparttd{n}{i, j}, \stattd{n}{i, j})\}_{j = 1}^\M \sim \left( \sum_{\ell = 1}^\N \frac{\ud{n}(\epart{n}{\ell}, \epart{n + 1}{i})}{\sum_{\ell' = 1}^\N \ud{n}(\epart{n}{\ell'}, \epart{n + 1}{i})} \delta_{(\epart{n}{\ell}, \stat{n}{\ell})} \right)^{\tensprod \M}, \quad i \in \intvect{1}{\N}.
$$
The resulting procedure, summarised in \Cref{alg:paris}, allows for online processing with constant memory requirements, since it only needs to store the current particle cloud and the estimated auxiliary statistics at each iteration. Moreover, in the case where the Markov transition densities of the model can be uniformly bounded, \emph{i.e.} when there exists, for every $n \in \nset$, an upper bound $\mdhigh{n} > 0$ such that for all $(x_n, x_{n + 1}) \in \xsp{n} \times \xsp{n + 1}$, $\md{n}(x_n, x_{n + 1}) \leq \mdhigh{n}$ (a weak assumption satisfied for most models of interest), a sample $(\eparttd{n}{i, j}, \stat{n}{i, j})$ can be generated by drawing, with replacement and until acceptance, candidates $(\eparttd{n}{i, \ast}, \stattd{n}{i, \ast})$ from $\{(\epart{n}{i}, \stat{n}{i})\}_{i = 1}^\N$ according to the normalised particle weights $\{\pot{n}(\epart{n}{\ell}) / \sum_{\ell'} \pot{n}(\epart{n}{\ell'})\}_{\ell = 1}^\N$, obtained as a by-product in the generation of $\epartmb{n + 1}$, and accepting the same with probability $\md{n}(\eparttd{n}{i, \ast}, \epart{n + 1}{i}) /  \mdhigh{n}$. As this sampling procedure bypasses completely the calculation of the normalising constant $\sum_{\ell' = 1}^\N \ud{n}(\epart{n}{\ell'}, \epart{n + 1}{i})$ of the targeted categorical distribution, it yields an overall $\mathcal{O}(\M \N)$ complexity of the algorithm as a whole; see \cite{douc:garivier:moulines:olsson:2009} for details.

Increasing $\M$ improves the accuracy of the algorithm at the cost of additional computational complexity. As shown in \cite{olsson:westerborn:2017}, there is a qualitative difference between the cases $\M=1$ and $\M \geq 2$, and it turns out that the latter is required to keep \PARIS\  numerically stable. More precisely, in the latter case, it can be shown that the \PARIS\  estimator $\occm(\statmb{n})$ satisfies, as $\N$ tends to infinity while $\M$ is held fixed, a central limit theorem (CLT) at the rate $\sqrt{N}$ and with an $n$-normalised asymptotic variance of order $\mathcal{O}(1 - 1/(\M - 1))$. As clear from this bound, using a large $\M$ only yields a waste of computational work, and setting $\M$ to $2$ or $3$ typically works well in practice.

We now introduce the \emph{Parisian particle Gibbs} (\PPG) \emph{algorithm}.
For all $t \in \nsetpos$, let $\ysp{t} \eqdef \xsp{0:t} \times \rset$ and $\yfd{t} \eqdef \xfd{0:t} \tensprod \borel(\rset)$.
Moreover, let $\ysp{0} \eqdef \xsp{0} \times \{ 0 \}$ and $\yfd{0} \eqdef \xfd{0} \tensprod \{ \{0\}, \emptyset \}$.
An element of $\ysp{t}$ will always be denoted by $y_t = (x_{0:t|t}, b_t)$. The Parisian particle Gibbs sampler comprises, as a key ingredient, a \emph{conditional \PARIS\  step}, which updates recursively a set of $\ysp{t}$-valued random variables
$\bpart{t}{i} \eqdef (\epart{0:t|t}{i}, \stat{t}{i})$, $i \in \intvect{1}{\N}$.
Let $(\bpartmb{t})_{t \in \nset}$ denote the corresponding many-body process, each $\bpartmb{t} \eqdef \{(\epart{0:t|t}{i}, \stat{t}{i})\}_{i = 1}^\N$  taking on values in the space $\yspmb{t} \eqdef \ysp{t}^\N$, which we furnish with a $\sigma$-field $\yfdmb{t} \eqdef \yfd{t}^{\tensprod \N}$. The space $\yspmb{0}$ and the corresponding $\sigma$-field $\yfdmb{0}$ are defined accordingly.
For every $t \in \nset$, we write $\epartmb{0:t|t}$ for the collection $\{ \epart{0:t|t}{i} \}_{i = 1}^\N$ of paths in $\bpartmb{t}$, and $\epartmb{t|t}$ for the collection $\{ \epart{t|t}{i} \}_{i = 1}^\N$ of end points of the same.

In the following, we let $t \in \nset$ be a fixed time horizon, and describe in detail how the \PPG\ approximates $\targ{0:t} \af{t}$ iteratively. In short, at each iteration $\ell$, the \PPG\ produces, given an input conditional path $\chunk{\zeta}{0}{t}[\ell]$, a many-body system $\bpartmb{t}[\ell+1]$ by means of a series of conditional \PARIS\  operations; then, an updated path $\chunk{\zeta}{0}{t}[\ell+1]$, serving as input at the next iteration, is generated by picking one of the paths $\epartmb{0:t|t}[\ell+1]$ in $\bpartmb{t}[\ell+1]$ at random. At each iteration, the produced statistics $\statmb{t}$ in $\bpartmb{t}$ provides an approximation of $\targ{0:t} \af{t}$ according to \eqref{eq:bckwd-interpretation}.

More precisely, given the path $\chunk{\zeta}{0}{t}[\ell]$, the conditional \PARIS\  operations are executed as follows. In the initial step, $\epartmb{0|0}[\ell+1]$ are drawn from $\initmb[{\zeta_0[\ell]}]$ defined in \eqref{eq:init:cond:dual} and $\bpart{0}{i}[\ell+1] \gets (\epart{0|0}{i}[\ell+1], 0)$ for all $i \in \intvect{1}{\N}$; then, recursively for $m \in \intvect{0}{t}$, assuming access to $\bpartmb{m}[\ell+1]$, 
\begin{itemize}[noitemsep]
\item[\textbf{(1)}] we generate an updated particle cloud $\epartmb{m + 1}[\ell+1] \sim \mkmb{m}[{\zeta_{m + 1}[\ell]}](\epartmb{m|m}[\ell+1], \cdot)$,
\item[\textbf{(2)}] we pick at random, for each $i \in \intvect{1}{N}$, an ancestor path with associated statistics $(\tilde{\xi}_{0:m}^{i, 1}[\ell+1], \stattd{m}{i, 1}[\ell+1])$ among $\bpartmb{m}[\ell+1]$ by  drawing
$$
(\tilde{\xi}_{0:m}^{i, 1}[\ell+1], \stattd{m}{i, 1}[\ell+1]) \sim
\sum_{s = 1}^\N \frac{\ud{m}(\epart{m|m}{s}[\ell+1], \epart{m + 1}{i}[\ell+1])}{\sum_{s' = 1}^\N \ud{m}(\epart{m|m}{s'}[\ell+1], \epart{m + 1}{i}[\ell+1])} \delta_{\bpart{m}{s}[\ell+1]}, \quad i \in \intvect{1}{N},
$$
\item[\textbf{(3)}] we draw, with replacement, $\M - 1$ ancestor particles and associated statistics $\{ (\eparttd{m}{i, j}[\ell+1], \stattd{m}{i, j}[\ell+1]) \}_{j = 2}^\M$ at random from $\{(\epart{m|m}{s}[\ell+1], \stat{m}{s})[\ell+1]\}_{s = 1}^\N$ according to
$$
\{ (\eparttd{m}{i, j}[\ell+1], \stattd{m}{i, j}[\ell+1]) \}_{j = 2}^\M \sim \left( \sum_{s = 1}^\N \frac{\ud{m}(\epart{m|m}{s}[\ell+1], \epart{m + 1}{i}[\ell+1])}{\sum_{s' = 1}^\N \ud{m}(\epart{m|m}{s'}[\ell+1], \epart{m + 1}{i}[\ell+1])} \delta_{(\epart{m|m}{s}[\ell+1], \stat{m}{s}[\ell+1])} \right)^{\tensprod (\M - 1)},
$$
\item[\textbf{(4)}] we set, for all $i \in \intvect{1}{\N}$, $\epart{0:m + 1|m + 1}{i}[\ell+1] \gets (\tilde{\xi}_{0:m}^{i, 1}[\ell+1], \epart{m + 1}{i}[\ell+1])$ and  $\bpart{m + 1}{i}[\ell+1] \gets (\epart{0:m + 1|m + 1}{i}[\ell+1], \stat{m + 1}{i}[\ell+1])$, where
\[
\stat{m + 1}{i}[\ell+1] \gets \M^{-1} \sum_{j = 1}^\M \left( \stattd{m}{i, j}[\ell+1] + \afterm{m}(\eparttd{m}{i, j}[\ell+1], \epart{m + 1}{i}[\ell+1]) \right).
\]
\end{itemize}
This conditional \PARIS\  procedure is summarised in \Cref{alg:parisian:Gibbs}.

Once the set of trajectories and associated statistics $\bpartmb{t}[\ell+1]$ is formed by means of $n$ recursive conditional {\PARIS}  updates, an updated path $\chunk{\zeta}{0}{t}[\ell+1]$ is drawn from $\occm(\epartmb{0:t|t}[\ell+1])$. A full sweep of the {\PPG} is summarised in \Cref{alg:parisian:particle:Gibbs}.

The following Markov kernels will play an instrumental role in the following. For a given path $\{ z_m \}_{m \in \nset}$, the conditional \PARIS\  update in \Cref{alg:parisian:Gibbs} defines an inhomogeneous Markov chain on the spaces $\{ (\yspmb{m}, \yfdmb{m}) \}_{m \in \nset}$ with kernels
$$
\yspmb{m} \times \yfdmb{m + 1} \ni (\ymb{m}, A) \mapsto \int
\, \mkmb{m}[z_{m + 1}](\xmb{m|m}, \rmd \xmb{m+1}) \, \ck{m}(\ymb{m}, \xmb{m+1}, A), \quad m \in \nset,
$$
where
\begin{align}
\lefteqn{ \ck{m} : \yspmb{m} \times \xspmb{m + 1} \times \yfdmb{m + 1} \ni
(\ymb{m}, \xmb{m+1}, A)} \label{eq:def:ck} \\
&\mapsto \idotsint \indi{A} \left( \Big \{ \Big( (\tilde{x}_{0:m}^{i,1}, x_{m + 1}^i),
\frac{1}{\M} \sum_{j = 1}^\M \left( \tilde{\statl}_m^{i,j} + \afterm{m}(\tilde{x}_m^{i, j}, x_{m + 1}^i) \right) \Big) \Big\}_{i = 1}^\N \right) \nonumber \\
&\times \prod_{i = 1}^\N \left( \sum_{\ell = 1}^\N \frac{\ud{m}(x_{m|m}^\ell, x_{m + 1}^i)}{\sum_{\ell' = 1}^\N \ud{m}(x_{m|m}^{\ell'}, x_{m + 1}^i)}
\delta_{y_m^\ell}(\rmd(\tilde{x}_{0:m}^{i,1}, \tilde{\statl}_m^{i,1})) \right. \nonumber \\
&\left. \times \left( \sum_{\ell = 1}^\N \frac{\ud{m}(x_{m|m}^\ell, x_{m + 1}^i)}{\sum_{\ell' = 1}^\N \ud{m}(x_{m|m}^{\ell'}, x_{m + 1}^i)} \delta_{(x_{m|m}^\ell, \statl_m^{\ell})} \right)^{\tensprod (\M - 1)} (\rmd (\tilde{x}_m^{i,2}, \tilde{\statl}_m^{i, 2}, \ldots, \tilde{x}_{m}^{i, \M}, \tilde{\statl}_m^{i,\M})) \right) \, \nonumber.
\end{align}
In addition, we introduce the joint law
\begin{equation} \label{eq:def:ckjt}
\ckjt{t} : \xspmb{0:t} \times \yfdmb{t} \ni (\xmb{0:t}, A)
\mapsto
\idotsint \indi{A}(\ymb{t}) \, \ck{0}(\koskimat{\N} \xmb{0}, \xmb{1}, \rmd \ymb{1})
\prod_{m = 1}^{t - 1} \ck{m}(\ymb{m}, \xmb{m+1}, \rmd \ymb{m + 1}),
\end{equation}
where we have defined $\koskimat{\N} \eqdef  \Id_{\N} \tensprod (0, 1)^\intercal$.

The kernel $\ckjt{t}$ can be viewed as a \emph{superincumbent sampling kernel} describing the distribution of the output $\bpartmb{t}$
generated by a sequence of \PARIS\  iterates when the many-body process $\{ \epartmb{m} \}_{m = 0}^t$ associated with the underlying SMC algorithm is given. This allows us to describe alternatively the \PPG\ as follows: given $\zeta_{0:t}[\ell]$, draw $\epartmb{0:t}[\ell+1] \sim \mbjt{t}(\zeta_{0:t}[\ell], \cdot)$; then, draw $\bpartmb{t}[\ell+1] \sim \ckjt{t}(\epartmb{0:t}[\ell+1], \cdot)$ and pick a trajectory $\zeta_{0:t}[\ell+1]$ from $\epartmb{0:t|t}[\ell+1]$ at random. The following proposition, which will be instrumental in the coming developments, establishes that the conditional distribution of $\zeta_{0:t}[\ell+1]$ given $\epartmb{0:t}[\ell+1]$ coincides, as expected, with the particle-induced backward dynamics $\bdpart{t}$.
\begin{proposition} \label{prop:backward:marginal}
For all $t \in \nsetpos$, $\N \in \nsetpos$, $\xmb{0:t} \in \xspmb{0:t}$, and $h \in \bmf(\xfd{0:t})$,
$$
\int \ckjt{t}(\xmb{0:t}, \rmd \ymb{t}) \, \occm(\xmb{0:t|t}) h = \bdpart{t}h (\xmb{0:t}).
$$
\end{proposition}
Finally, we define the Markov kernel induced by the \PPG\ as well as the extended probability distribution targeted by the same. For this purpose, we introduce the extended measurable space $(\esp{t}, \efd{t})$ with
$$
\esp{t} \eqdef \yspmb{t} \times \xsp{0:t}, \quad \efd{t} \eqdef \yfdmb{t} \tensprod \xfd{0:t}.
$$
The \PPG\ described in \Cref{alg:parisian:particle:Gibbs} defines a Markov chain on $(\esp{t},  \efd{t})$ with Markov transition kernel
\begin{equation}
\label{eq:ppg_ker_def}
\parisgibbs{t} :  \esp{t} \times \efd{t} \ni (\ymb{t}, z_{0:t}, A)
\mapsto \iiint \1_A(\ymbtd{t}, \tilde{z}_{0:t}) \, \mbjt{t}(z_{0:t}, \rmd \xmbtd{0:t}) \, \ckjt{t}(\xmbtd{0:t}, \rmd \ymbtd{t}) \, \occm(\xmbtd{0:t|t})(\rmd \tilde{z}_{0:t}).
\end{equation}
Note that the values of $\parisgibbs{t}$ defined above do not depend on $\ymb{t}$, but only on $(z_{0:t}, A)$. For any given initial distribution $\upxi \in \probmeas(\xfd{0:t})$, let $\canlaw{\upxi}$ be the distribution of the canonical Markov chain induced by the kernel $\parisgibbs{t}$ and the initial distribution $\upxi$. In the special case where $\upxi = \delta_{z_{0:t}}$ for some given path $z_{0:t} \in \xsp{0:t}$, we use the short-hand notation $\canlaw{\delta_{z_{0:t}}} = \canlaw{z_{0:t}}$.
In addition, denote by
\begin{equation}
\label{eq:gibbs_ker}
\gibbs{t}  :  \xsp{0:t} \times \xfd{0:t} \ni (z_{0:t}, A)
\mapsto \iiint \1_A(\tilde{z}_{0:t}) \, \mbjt{t}(z_{0:t}, \rmd \xmbtd{0:t}) \, \ckjt{t}(\xmbtd{0:t}, \rmd \ymbtd{t}) \, \occm(\xmbtd{0:t|t})(\rmd \tilde{z}_{0:t})
\end{equation}
the path-marginalised version of $\parisgibbs{t}$.
By \Cref{prop:backward:marginal} it holds that $\gibbs{t} = \mbjt{t} \bdpart{t}$, which shows that $\gibbs{t}$ coincides with the Markov transition kernel of the backward-sampling-based particle Gibbs sampler discussed in \Cref{sec:dual:and:Gibbs}.
It is also possible to specify the invariant distribution of $\parisgibbs{t}$.
\begin{proposition}
    \label{prop:inv_measure_parisgibbs}
    For all $t \in \nsetpos$, it holds that
    \begin{equation}
    \targ{0:t}\mbjt{t} \ckjt{t} \parisgibbs{t} = \targ{0:t}\mbjt{t} \ckjt{t} \eqsp.
    \end{equation}
\end{proposition}
\begin{proof}
    Let $f \in \meas(\esp{t}^{\tensprod (k - k_0)})$.
    \begin{align*}
        &\int f(\ymbtd{t}, \tilde{z}_{0:t})\targ{0:t}{}(\rmd z_{0:t}) \mbjt{t}{} \ckjt{t}{} (z_{0:t}, \rmd (\ymb{t}, z'_{0:t})) \parisgibbs{t}{} (z'_{0:t}, \ymb{t}, \rmd (\ymbtd{t}, \tilde{z}_{0:t})) \\
        &= \int f(\ymbtd{t}, \tilde{z}_{0:t})\targ{0:t}{}(\rmd z_{0:t}) \mbjt{t}{} \ckjt{t}{} (z_{0:t}, \rmd (\ymb{t}, z'_{0:t})) \mbjt{t}{} \ckjt{t}{} (z'_{0:t}, \rmd (\ymbtd{t}, \tilde{z}_{0:t})) \\
        &= \int f(\ymbtd{t}, \tilde{z}_{0:t})\targ{0:t}{}(\rmd z_{0:t}) \gibbs{t}{}(z_{0:t}, \rmd z'_{0:t}) \mbjt{t}{} \ckjt{t}{} (z'_{0:t}, \rmd (\ymbtd{t}, \tilde{z}_{0:t}))\\
        &= \int f(\ymbtd{t}, \tilde{z}_{0:t})\targ{0:t}{}(\rmd z'_{0:t}) \mbjt{t}{} \ckjt{t}{} (z'_{0:t}, \rmd (\ymbtd{t}, \tilde{z}_{0:t})) \eqsp.
    \end{align*}
\end{proof}

Finally, in order prepare for the statement of our theoretical results on the \PPG\ we need to introduce the following Feynman--Kac path model \emph{with a frozen path}. More precisely, for a given path $z_{0:t} \in \xpsp{0}{t}$, define, for every $m \in \intvect{0}{t - 1}$, the unnormalised kernel
$$
\uk[z_{m + 1}]{m} : \xsp{m} \times \xfd{m + 1} \ni (x_m, A) \mapsto \left( 1 - \frac{1}{\N} \right) \uk{m}(x_m, A) + \frac{1}{\N} \pot{m}(x_m) \, \delta_{z_{m + 1}}(A)
$$
and the initial distribution $\targ[z_0]{0} : \xfd{0} \ni A \mapsto (1 - 1 / \N) \targ{0}(A) + \delta_{z_0}(A) / \N$.
Given these quantities, define, for $m \in \intvect{0}{t}$,
$\untarg[z_{0:m}]{m} \eqdef \targ[z_0]{0} \uk[z_1]{0} \cdots \uk[z_m]{m - 1}$
along with the normalised counterpart
$\targ[z_{0:m}]{m} \eqdef \untarg[z_{0:m}]{m}/\untarg[z_{0:m}]{m} \indi{\xpsp{0}{m}}$.
Finally, we introduce, for $m \in \intvect{0}{t}$, the kernels
$$
\rk[z_{0:m - 1}]{m} : \xsp{m} \times \xfd{0:m - 1} \ni (x_m, A) \mapsto \idotsint \1_A(x_{0:m - 1}) \prod_{m = 0}^{t - 1} \bk{m, \targ[z_{0:m}]{m}}(x_{m + 1}, \rmd x_m),
$$
as well as the path model $\targ[z_{0:m}]{0:m} \eqdef  \rk[z_{0:m - 1}]{m} \tensprod \targ[z_{0:m}]{m}$.

\subsection{Proof of \Cref{theo:bias-mse-rolling}}
\label{sec:thm:theoretical:results}

We start by establishing bias, MSE and covariance bounds for a fixed iteration of the \PPG\ estimator.

\begin{theorem} \label{thm:bias:bound}
Assume \Cref{assumption:strong_mixing}. Then for every $t \in \nset$ there exist $\cstparisbias{t}$, $\cstparismse{t}$, and $\cstpariscov{t}$ in $\rset_+^\ast$ such that for every $\M \in \nsetpos$,
$\upxi \in \probmeas(\xpfd{0}{t})$, $\ell \in \nsetpos$, $s \in \nsetpos$, and $\N \in \nsetpos$ such that $\N > \N_t$,
\begin{align}
\label{eq:bias:bound}
\left|\E_\upxi \left[ \occm(\statmb{t}[\ell])(\operatorname{id}) \right] - \targ{0:t} \af{t} \right| &\leq
\cstparisbias{t} \left(\sum_{m=0}^{t-1} \| \afterm{m} \|_\infty \right) \N^{-1} \mixrate{t}{\N}^\ell, \\
\label{eq:mse:bound}
\E_\upxi \left[ \left( \occm(\statmb{t}[\ell])(\operatorname{id}) - \targ{0:t} \af{t} \right)^2 \right]&\leq
\cstparismse{t} \left(\sum_{m=0}^{t - 1} \| \afterm{m} \|_\infty \right)^2 \N^{-1}, \\
\label{eq:cov:bound}
\left|\E_\upxi \left[ \left( \occm(\statmb{t}[\ell])(\operatorname{id}) - \targ{0:t} \af{t} \right)  \left( \occm(\statmb{t}[\ell+s])(\operatorname{id})
- \targ{0:t} \af{t} \right) \right]\right|
&\leq \cstpariscov{t}  \left(\sum_{m=0}^{t - 1} \| \afterm{m} \|_\infty \right)^2 \N^{-3/2} \mixrate{t}{\N}^{s}.
\end{align}
\end{theorem}
\noindent
The constants $\cstparisbias{t}$, $\cstparismse{t}$, and $\cstpariscov{t}$ are explicitly given in the proof.
Since the focus of this paper is on the dependence on $\N$ and the index $\ell$,
we have made no attempt to optimise the dependence of these constants on $t$ in our proofs; still, we believe that it is possible to prove,
under the stated assumptions, that this dependence is linear.
The proof of the bound in \Cref{thm:bias:bound} is based on four key ingredients. The first is the following unbiasedness property of the {\PARIS} under the many-body Feynman--Kac path model.
\begin{theorem} \label{thm:unbiasedness}
For every $t \in \nset$, $\N \in \nsetpos$, and $\ell \in \nsetpos$,
$$
\E_{\targ{0:t}} \left[ \occm(\statmb{t}[\ell])(\operatorname{id}) \right] =
\int \targ{0:t} \mbjt{t} \ckjt{t}(\rmd \statlmb{t}) \, \occm(\statlmb{t})(\operatorname{id}) = \int \targmb{0:t}\ckjt{t}(\rmd \statlmb{t}) \, \occm(\statlmb{t})(\operatorname{id})  = \targ{0:t} h_t.
$$
\end{theorem}
The proof of \Cref{thm:unbiasedness} is postponed to  \Cref{sec:proof:unbiasedness}. The second ingredient of the proof of \Cref{thm:bias:bound} is the uniform geometric ergodicity of the particle Gibbs with backward sampling established in \cite{del2018sharp}.
\begin{theorem} \label{thm:geometric:ergodicity:particle:Gibbs}
Assume \Cref{assumption:strong_mixing}. Then, for every $t \in \nset$, $(\mu, \nu) \in \probmeas(\xfd{0:t})^2$, $\ell \in \nsetpos$,
and $\N \in \nsetpos$ such that $N > 1 + 5 \rho_t^2 t / 2$,
$\| \mu \gibbs{t}^\ell - \nu \gibbs{t}^\ell \|_{\mathrm{TV}} \leq \mixrate{t}{\N}^\ell$, where $\mixrate{t}{\N}$ is defined in \eqref{eq:def:kappa}.
\end{theorem}

As a third ingredient, we require the following uniform exponential concentration inequality of the conditional {\PARIS} with respect to the frozen-path Feynman--Kac model defined in the previous section.
\begin{theorem} \label{cor:exp:conc:cond:paris}
For every $t \in \nset$ there exist $\cstcondparisc_t > 0$ and $\cstcondparisd_t > 0$ such that for every $\M \in \nsetpos$, $z_{0:t} \in \xpsp{0}{t}$, $\N \in \nsetpos$, and $\varepsilon > 0$,
$$
\int \mbjt{t} \ckjt{t}(z_{0:t}, \rmd \statlmb{t}) \indin{\left| \occm( \statlmb{t})(\operatorname{id})  - \targ[z_{0:t}]{0:t}\af{t} \right|\geq \varepsilon } 
\leq \cstcondparisc_t \exp \left( -   \frac{\cstcondparisd_t \N \varepsilon^2}{2 (\sum_{m = 0}^{t - 1} \| \afterm{m} \|_\infty)^2}  \right).
$$
\end{theorem}
\Cref{cor:exp:conc:cond:paris}, whose proof is postponed to  \Cref{sec:proof:exp:concentration}, implies, in turn, the following conditional variance bound.
\begin{proposition} \label{cor:Lp:cond:paris}
For every $t \in \nset$, $M \in \nsets$, $z_{0:t} \in \xpsp{0}{t}$, and $\N \in \nsetpos$,
$$
\int \mbjt{t} \ckjt{t}(z_{0:t}, \rmd \statlmb{t}) \left| \occm( \statlmb{t})(\operatorname{id})  - \targ[z_{0:t}]{0:t}\af{t} \right|^2 \leq \frac{\cstcondparisc_t}{\cstcondparisd_t} \left( \sum_{m = 0}^{t - 1} \| \afterm{m} \|_\infty \right)^2 \N^{-1}.
$$
\end{proposition}
Using \Cref{cor:Lp:cond:paris}, we deduce, in turn, the following bias bound, whose proof is postponed to \Cref{subsec:prop:bias:cond:paris}.
\begin{proposition}
\label{prop:bias:cond:paris}
For every $t \in \nset$ there exists $\cstcondparisbias{t} > 0$  such that for every $\M \in \nsetpos$, $z_{0:t} \in \xpsp{0}{t}$, and $\N \in \nsetpos$,
\begin{align*}
  \left|\int \mbjt{t} \ckjt{t}(z_{0:t}, \rmd \statlmb{t}) \, \occm(\statlmb{t})(\operatorname{id}) - \targ[z_{0:t}]{0:t} \af{t} \right| \leq
\cstcondparisbias{t} \N^{-1} \left( \sum_{m = 0}^{t - 1} \| \afterm{m} \|_\infty \right) .
\end{align*}
\end{proposition}
A fourth and last ingredient in the proof of \Cref{thm:bias:bound} is the following bound on the discrepancy between additive expectations under the original and frozen-path Feynman--Kac models. This bound is established using novel results in \cite{gloaguen:lecorff:olsson:2022}. More precisely, since for every $m \in \nset$, $(x, z) \in \xsp{m}^2$, $\N \in \nsetpos$, and $h \in \bmf(\xfd{m + 1})$, using \Cref{assumption:strong_mixing},
$$
\left| \uk[z]{m}h(x) - \uk{m}h(x) \right| \leq \frac{1}{\N} \| \pot{m} \|_\infty \|h \|_\infty \leq \frac{1}{\N} \gsupbound{m} \| h \|_\infty,
$$
applying \cite[Theorem~4.3]{gloaguen:lecorff:olsson:2022} yields the following.
\begin{proposition} \label{prop:error:conditional:model}
Assume \Cref{assumption:strong_mixing}. Then there exists $\constpdist > 0$ such that for every $t \in \nset$, $\N \in \nset$, and $z_{0:t} \in \xsp{0:t}$,
$$
\left| \targ[z_{0:t}]{0:t} \af{t} - \targ{0:t} \af{t} \right| \leq
\constpdist \N^{-1} \sum_{m = 0}^{t - 1} \| \afterm{m} \|_\infty. 
$$
\end{proposition}
Note that assuming, in addition, that $\sup_{t \in \nset} \| \afterm{t} \|_\infty < \infty$ yields an $\mathcal{O}(n / \N)$ bound in \Cref{prop:error:conditional:model}.

Finally, by combining these ingredients we are now ready to present a proof of \Cref{thm:bias:bound}.

\begin{proof}[Proof of \Cref{thm:bias:bound}]
Write, using the tower property,
$$
\E_{\upxi} \left[ \occm(\statmb{t}\left[\ell\right])(\operatorname{id}) \right] = \E_{\upxi} \left[\E_{\zeta_{0:t}\left[\ell\right]} \left[\occm(\statmb{t}\left[0\right])(\operatorname{id})\right] \right] = \int \upxi \gibbs{t}^{\ell}\mbjt{t} \ckjt{t}(\rmd \statlmb{t}) \, \occm(\statlmb{t})(\operatorname{id}).
$$
Thus, by the unbiasedness property in \Cref{thm:unbiasedness},
\begin{align}
\left|\E_{\upxi} \left[ \occm(\statmb{t}\left[\ell\right])(\operatorname{id}) \right]- \targ{0:t} \af{t} \right|
&=  \left| \int \upxi \gibbs{t}^\ell \mbjt{t} \ckjt{t}(\rmd \statlmb{t}) \, \occm(\statlmb{t})(\operatorname{id}) - \int \targ{0:t} \mbjt{t} \ckjt{t}(\rmd \statlmb{t}) \, \occm(\statlmb{t})(\operatorname{id}) \right| \nonumber \\
&\leq \big\| \upxi \gibbs{t}^\ell - \targ{0:t} \big\|_{\mathrm{TV}}  \operatorname{osc} \left( \int \mbjt{t} \ckjt{t}(\cdot, \rmd \statlmb{t}) \, \occm(\statlmb{t})(\operatorname{id}) \right), \nonumber
\end{align}
where, by \Cref{thm:geometric:ergodicity:particle:Gibbs}, $\| \upxi \gibbs{t}^\ell - \targ{0:t}\|_{\mathrm{TV}} \leq \mixrate{t}{\N}^\ell$. Moreover, to derive an upper bound on the oscillation, we consider the decomposition
\begin{equation*} \label{eq:osc:norm:bound}
\operatorname{osc} \left( \int \mbjt{t} \ckjt{t}(\cdot, \rmd \statlmb{t}) \, \occm(\statlmb{t})(\operatorname{id}) \right) 
\leq 2 \left( \left\| \int \mbjt{t} \ckjt{t}(\cdot, \rmd \statlmb{t}) \, \occm(\statlmb{t})(\operatorname{id}) - \targ[\cdot]{0:t} \af{t} \right\|_\infty  + \left \| \targ[\cdot]{0:t} \af{t}  - \targ{0:t} \af{t} \right \|_\infty \right),
\end{equation*}
where the two terms on the right-hand side can be bounded using \Cref{prop:error:conditional:model} and \Cref{prop:bias:cond:paris}, respectively. This completes the proof of \eqref{eq:bias:bound}. We now consider the proof of \eqref{eq:mse:bound}. Writing 
\begin{equation*}
\E_\upxi \left[ \left( \occm(\statmb{t}[\ell])(\operatorname{id})  - \targ{0:t} \af{t} \right)^2 \right]
= \int \upxi \gibbs{t}^{\ell}(\rmd \chunk{z}{0}{t}) \mbjt{t} \ckjt{t}(z_{0:t}, \rmd \statlmb{t}) \left( \occm( \statlmb{t})(\operatorname{id})  -  \targ{0:t} \af{t} \right)^2,
\end{equation*}
we may establish \eqref{eq:mse:bound} using \Cref{cor:Lp:cond:paris} and \Cref{prop:error:conditional:model}. We finally consider \eqref{eq:cov:bound}.
Using the Markov property we obtain 
\begin{multline*}
\E_\upxi \left[ \left( \occm(\statmb{t}[\ell])(\operatorname{id})  - \targ{0:t} \af{t} \right)  \left( \occm(\statmb{t}[\ell+s])(\operatorname{id})  - \targ{0:t} \af{t} \right) \right] 
\\ =
\E_\upxi \left[ \left( \occm(\statmb{t}[\ell])(\operatorname{id})  - \targ{0:t} \af{t} \right) \left(\E_{\chunk{\zeta}{0}{t}[\ell]}[\occm(\statmb{t}[s])(\operatorname{id})] - \targ{0:t} \af{t} \right)\right], 
\end{multline*}
from which \eqref{eq:cov:bound} follows by \eqref{eq:bias:bound} and \eqref{eq:mse:bound}.
\end{proof}


We are finally equipped to prove \Cref{theo:bias-mse-rolling}.
\begin{proof}[Proof of \Cref{theo:bias-mse-rolling}]
\label{sec:proof:theo:bias-mse-rolling}
We first consider the bias, which can be bounded according to
\begin{align*}
    \left| \E_{\upxi}[\rollingestim[\ki_0][\ki][N][f]] - \targ{0:t} \af{t} \right| &\leq  (\ki - \ki_0)^{-1}  \sum_{\ell=\ki_0 + 1}^{\ki} \left| \E_{\upxi} \occm(\statmb{t}[\ell])(\operatorname{id})  - \targ{0:t} \af{t} \right| \\
    &\leq (\ki - \ki_0)^{-1} N^{-1} \cstparisbias{t}   \left(\sum_{m=0}^{t-1} \| \afterm{m} \|_\infty \right)
    \sum_{\ell=\ki_0 + 1}^{\ki} \mixrate{t}{\N}^{\ell},
\end{align*}
from which the bound \eqref{eq:theo:bias-mse-rolling:bias} follows immediately.

We turn to the MSE. Using the decomposition
\begin{align*}
    \E_{\upxi}[(\rollingestim[\ki_0][\ki][N][f] - \targ{0:t} \af{t})^2] \leq (\ki-\ki_0)^{-2} \left\{ \sum_{\ell=\ki_0 + 1}^{\ki}   \E_{\upxi}[ (\occm(\statmb{t}[\ell])(\operatorname{id}) - \targ{0:t} \af{t} )^2] \right.\\
    + \left. 2 \sum_{\ell=\ki_0 + 1}^{\ki} \sum_{j= \ell + 1}^{\ki} \E_{\upxi}[(\occm(\statmb{t}[\ell])(\operatorname{id}) - \targ{0:t} \af{t})(\occm(\statmb{t}[j])(\operatorname{id}) - \targ{0:t} \af{t})] \right\},
\end{align*}
the MSE bound in \Cref{thm:bias:bound} implies that
\begin{equation*}
\sum_{\ell=\ki_0 + 1}^{\ki} \E_{\upxi}[(\occm(\statmb{t}[\ell])(\operatorname{id}) - \targ{0:t} \af{t})^2] \leq
\cstparismse{t} \left(\sum_{m=0}^{t - 1} \| \afterm{m} \|_\infty \right)^2 \N^{-1} (\ki - \ki_0).
\end{equation*}
Moreover, using the covariance bound in \Cref{thm:bias:bound}, we deduce that
\begin{equation*}
    \sum_{\ell=\ki_0 + 1}^{\ki} \sum_{j= \ell + 1}^{\ki} \E_{\upxi}[(\occm(\statmb{t}[\ell])(\operatorname{id}) - \targ{0:t} \af{t})(\occm(\statmb{t}[j])(\operatorname{id}) - \targ{0:t} \af{t})]
    \leq \cstpariscov{t}  \left(\sum_{m=0}^{t - 1} \| \afterm{m} \|_\infty \right)^2 \N^{-3/2}  \left( \sum_{\ell=\ki_0 + 1}^{\ki} \sum_{j= \ell + 1}^{\ki} \mixrate{t}{\N}^{(j - \ell)} \right)  .
\end{equation*}
Thus, the proof is concluded by noting that  $\sum_{\ell=\ki_0 + 1}^{\ki} \sum_{j= \ell + 1}^{\ki} \mixrate{t}{\N}^{(j - \ell)}  \leq (\ki - \ki_0) / (1- \mixrate{t}{\N})$.
\end{proof}

\subsection{Proofs of intermediate results}
\subsubsection{Proof of \Cref{prop:backward:decomposition}} \label{sec:proof:backward:decomposition}

Using the identity
$$
\targ{0} \uk{0} \cdots \uk{t - 1} \indi{\xsp{t}} = \prod_{m = 0}^{t - 1} \targ{m} \uk{m} \indi{\xsp{m + 1}}
$$
and the fact that each kernel $\uk{m}$ has a transition density, write, for $h \in \bmf(\xfd{0:t})$,
\begin{align}
\targ{0:t} h &= \idotsint h(x_{0:t}) \, \targ{0}(\rmd x_0) \prod_{m = 0}^{t - 1} \left(\frac{\targ{m}[\ud{m}(\bcdot, x_{m + 1})] \, \refm{m+1}(\rmd x_{m + 1})}{\targ{m} \uk{m} \indi{\xsp{m + 1}}} \right) \left( \frac{\ud{m}(x_m, x_{m + 1})}{\targ{m}[\ud{m}(\bcdot, x_{m + 1})]} \right) \nonumber \\
&= \idotsint h(x_{0:t}) \, \targ{t}(\rmd x_t) \prod_{m = 0}^{t - 1} \frac{\targ{m}(\rmd x_m) \, \ud{m}(x_m, x_{m + 1})}{\targ{m}[\ud{m}(\bcdot, x_{m + 1})]} \\
&= \left( \bk{0,\targ{0}}  \tensprod \cdots \tensprod \bk{n-1,\targ{t - 1}} \tensprod  \targ{t} \right) h, \nonumber
\end{align}
which was to be established.


\subsubsection{Proof of \Cref{thm:duality}}
\label{sec:proof:thm:duality}

\begin{lemma}
\label{lem:transport:id}
For all $t \in \nset$, $\xmb{t} \in \xspmb{t}$, and $h \in \bmf(\xfdmb{t + 1} \varotimes \xfd{t + 1})$,
\begin{equation} \label{eq:transport:id}
\iint h(\xmb{t + 1}, z_{t + 1}) \, \ukmb{t}(\xmb{t}, \rmd \xmb{t + 1}) \, \occm(\xmb{t + 1})(\rmd z_{t + 1})
= \iint h(\xmb{t + 1}, z_{t + 1}) \, \occm(\xmb{t}) \uk{t}(\rmd z_{t + 1}) \, \mkmb{t}[z_{t + 1}](\xmb{t}, \rmd \xmb{t + 1}).
\end{equation}
In addition, for all $h \in \bmf(\xfdmb{0} \varotimes \xfd{0})$,
\begin{equation} \label{eq:transport:id:init}
\iint h(\xmb{0}, z_0) \, \initmb(\rmd \xmb{0}) \, \occm(\xmb{0})(\rmd z_0) = \iint h(\xmb{0}, z_0) \, \initmb[z_0](\rmd \xmb{0}) \, \init(\rmd z_0).
\end{equation}
\end{lemma}

\begin{proof}
Since
$
\occm(\xmb{t}) \, \uk{t}(\rmd z_{t + 1}) = \potmb{t}(\xmb{t}) \, \pd{t}(\occm(\xmb{t}))(\rmd z_{t + 1}),
$
we may rewrite the right-hand side of \eqref{eq:transport:id} according to
\[
\begin{split}
\lefteqn{\iint h(\xmb{t + 1}, z_{t + 1}) \, \occm(\xmb{t}) \uk{t}(\rmd z_{t + 1}) \, \mkmb{t}[z_{t + 1}](\xmb{t}, \rmd \xmb{t + 1})} \hspace{15mm} \\
&= \potmb{t}(\xmb{t}) \frac{1}{\N} \sum_{i = 0}^{\N - 1} \iint h(\xmb{t + 1}, z_{t + 1}) \, \pd{t}(\occm(\xmb{t}))(\rmd z_{t + 1}) \\
&\hspace{15mm} \times \left( \Phi_t(\occm(\xmb{t}))^{\tensprod i} \tensprod \delta_{z_{t + 1}} \tensprod \Phi_t(\occm(\xmb{t}))^{\tensprod (\N - i - 1)} \right)(\rmd \xmb{t + 1}) \\
&= \potmb{t}(\xmb{t}) \frac{1}{\N} \sum_{i = 1}^\N \idotsint h((x_{t + 1}^{1}, \ldots, x_{t + 1}^{i - 1}, z_{t + 1}, x_{t + 1}^{i + 1}, \ldots, x_{t + 1}^{\N}), z_{t + 1}) \\
&\hspace{15mm} \times \pd{t}(\occm(\xmb{t}))(\rmd z_{t + 1}) \prod_{\ell \neq i} \pd{t}(\occm(\xmb{t}))(\rmd x_{t + 1}^{\ell}) \\
&= \potmb{t}(\xmb{t}) \frac{1}{\N} \sum_{i = 1}^\N \int h(\xmb{t + 1}, x_{t + 1}^{i}) \, \mkmb{t}(\xmb{t}, \rmd \xmb{t + 1}).
\end{split}
\]

On the other hand, note that the left-hand side of \eqref{eq:transport:id} can be expressed as
$$
\iint h(\xmb{t + 1}, z_{t + 1}) \, \ukmb{t}(\xmb{t}, \rmd \xmb{t + 1}) \, \occm(\xmb{t + 1})(\rmd z_{t + 1}) 
= \potmb{t}(\xmb{t}) \frac{1}{\N} \sum_{i = 1}^\N \int h(\xmb{t + 1}, x_{t + 1}^{i}) \, \mkmb{t}(\xmb{t}, \rmd \xmb{t + 1}),
$$
which establishes the identity.
The identity \eqref{eq:transport:id:init} is established along similar lines.
\end{proof}

We establish \Cref{thm:duality} by induction; thus, assume that the claim holds true for $n$ and show that for all $h \in \bmf(\xfdmb{0:t + 1} \varotimes \xfd{0:t + 1})$,
\begin{multline} \label{eq:induction:step}
\iint h(\xmb{0:t + 1}, z_{0:t + 1}) \, \untargmb{0:t + 1}(\rmd \xmb{0:t + 1}) \, \bdpart{t + 1}(\xmb{0:t + 1}, \rmd z_{0:t + 1}) \\
= \iint h(\xmb{0:t + 1}, z_{0:t + 1}) \, \untarg{0:t + 1}(\rmd \chunk{z}{0}{t + 1}) \, \mbjt{t + 1}(z_{0:t + 1}, \rmd \xmb{0:t + 1}).
\end{multline}
To prove this, we process, using definition \eqref{eq:def:bdpart}, the left-hand side of \eqref{eq:induction:step} according to
\begin{equation} \label{eq:duality:lhs}
\begin{split}
\lefteqn{\iint h(\xmb{0:t + 1}, z_{0:t + 1}) \, \untargmb{0:t + 1}(\rmd \xmb{0:t + 1}) \, \bdpart{t + 1}(\xmb{0:t + 1}, \rmd z_{0:t + 1})} \\
&= \iint \untargmb{0:t}(\rmd \xmb{0:t}) \, \bdpart{t}(\xmb{0:t}, \rmd {\chunk{z}{0}{t}}) \\
&\hspace{20mm} \times \iint \bar{h}(\xmb{0:t + 1}, z_{0:t + 1}) \, \ukmb{t}(\xmb{t}, \rmd \xmb{t + 1}) \, \occm(\xmb{t + 1})(\rmd z_{t + 1}),
\end{split}
\end{equation}
where we have defined the function
$$
\bar{h} (\xmb{0:t + 1}, z_{0:t + 1}) \eqdef \frac{\ud{t}(z_t, z_{t + 1}) h(\xmb{0:t + 1}, z_{0:t + 1})}{\occm(\xmb{t})[\ud{t}(\cdot, z_{t + 1})]}.
$$
Now, applying \Cref{lem:transport:id} to the inner integral and using that
$$
\occm(\xmb{t}) \uk{t}(\rmd z_{t + 1}) = \occm(\xmb{t})[\ud{t}(\cdot, z_{t + 1})] \, \refm{t + 1}(\rmd z_{t + 1})
$$
yields, for every $\xmb{0:t}$ and ${\chunk{z}{0}{t}}$,
\[
\begin{split}
\lefteqn{\iint \bar{h}(\xmb{0:t + 1}, z_{0:t + 1}) \, \ukmb{t}(\xmb{t}, \rmd \xmb{t + 1}) \, \occm(\xmb{t + 1})(\rmd z_{t + 1})} \\
&= \iint \bar{h}(\xmb{0:t + 1}, z_{0:t + 1}) \, \occm(\xmb{t}) \uk{t}(\rmd z_{t + 1}) \, \mkmb{t}[z_{t + 1}](\xmb{t}, \rmd \xmb{t + 1}) \\
&= \iint h(\xmb{0:t + 1}, z_{0:t + 1}) \, \uk{t}(z_t, \rmd z_{t + 1}) \, \mkmb{t}[z_{t + 1}](\xmb{t}, \rmd \xmb{t + 1}).
\end{split}
\]
Inserting the previous identity into \eqref{eq:duality:lhs} and using the induction hypothesis provides
\[
\begin{split}
\lefteqn{\iint h(\xmb{0:t + 1}, z_{0:t + 1}) \, \untargmb{0:t + 1}(\rmd \xmb{0:t + 1}) \, \bdpart{t + 1}(\xmb{0:t + 1}, \rmd z_{0:t + 1})} \\
&= \iint \untarg{0:t}(\rmd {\chunk{z}{0}{t}}) \, \mbjt{t}({\chunk{z}{0}{t}}, \rmd \xmb{0:t})   \\
& \hspace{20mm} \times \iint h(\xmb{0:t + 1}, z_{0:t + 1}) \, \uk{t}(z_t, \rmd z_{t + 1}) \,  \mkmb{t}[z_{t + 1}](\xmb{t}, \rmd \xmb{t + 1})  \\
&= \iint h(\xmb{0:t + 1}, z_{0:t + 1}) \, \untarg{0:t + 1}(\rmd z_{0:t + 1}) \, \mbjt{t + 1}(z_{0:t + 1}, \rmd \xmb{0:t + 1}),
\end{split}
\]
which establishes \eqref{eq:induction:step}.

\subsubsection{Proof of \Cref{thm:unbiasedness}}
\label{sec:proof:unbiasedness}
First, define, for $m \in \nset$,
\begin{equation} \label{eq:def:P}
\pk{m} : \yspmb{m} \times \yfdmb{m + 1} \ni (\ymb{m}, A) \mapsto \int
\, \mkmb{m}(\xmb{m|m}, \rmd \xmb{m + 1}) \, \ck{m}(\ymb{m}, \xmb{m + 1}, A).
\end{equation}
For any given initial distribution $\pinit \in \probmeas(\yfdmb{0})$, let $\canlaw[\pkl]{\pinit}$ be the distribution of the canonical Markov chain induced by the Markov kernels $\{ \pk{m} \}_{m \in \nset}$ and the initial distribution $\pinit$. By abuse of notation we write, for $\initmb \in \probmeas(\xfdmb{0})$, $\canlaw[\pkl]{\initmb}$ instead of $\canlaw[\pkl]{\pinit[\initmb]}$, where we have defined the extension $\pinit[\initmb](A) = \int \1_A(\koskimat{\N} \xmb{0}) \, \initmb(\rmd \xmb{0})$, $A \in \yfdmb{0}$. We preface the proof of \Cref{thm:unbiasedness} by some technical lemmas and a proposition.
\begin{lemma} \label{lem:key:identity:untarg:affine}
For all $t \in \nset$ and $(f_{t + 1}, \tilde{f}_{t + 1}) \in \bmf(\xfd{t + 1})^2$,
$$
 \untarg{t + 1}(f_{t + 1} \rk{t + 1} \af{t + 1} + \tilde{f}_{t + 1}) = \untarg{t}\{ \uk{t} f_{t + 1} \rk{t} \af{t} + \uk{t}(\afterm{t} f_{t + 1} + \tilde{f}_{t + 1}) \}.
$$
\end{lemma}
\begin{proof}
Pick arbitrarily $\varphi \in \bmf(\xfd{t:t+1})$ and write, using definition \eqref{eq:bckwd-kernels} and the fact that $\uk{t}$ has a transition density, 
\begin{align}
    \lefteqn{\iint \varphi(x_{t:t+1}) \, \untarg{t}(\rmd x_t) \, \uk{t}(x_t, \rmd x_{t + 1})} \hspace{10mm} \nonumber \\
    &= \iint \varphi(x_{t:t+1}) \untarg{t}[\ud{t}(\cdot, x_{t + 1})] \, \refm{t + 1}(\rmd x_{t + 1}) \, \frac{\untarg{t}(\rmd x_t) \ud{t}(x_t, x_{t + 1})}{\untarg{t}[\ud{t}(\cdot, x_{t + 1})]} \nonumber \\
    &= \iint \varphi(x_{t:t+1}) \, \untarg{t+1}(\rmd x_{t+1}) \, \bk{n, \targ{t}}(x_{t+1}, \rmd x_{t}). \label{eq:key_relation_lemma2} 
\end{align}

    Now, by \eqref{eq:backward:forward:recursion} it holds that 
    \begin{equation*}
        \rk{t+1} \af{t+1}(x_{t+1}) = \int \bk{n, \targ{t}} (x_{t+1}, \rmd x_{t}) \left( \afterm{t}(x_{t:t+1})
        + \int \af{t}(x_{0:t}) \, \rk{t}(x_{t}, \rmd x_{0:t-1}) \right);
    \end{equation*}
    therefore, by applying \eqref{eq:key_relation_lemma2} with
    \begin{equation*}
        \varphi(x_{t:t+1}) \eqdef f_{t+1}(x_{t+1}) \left( \afterm{t}(x_{t:t+1}) + \int \af{t}(x_{0:t}) \, \rk{t}(x_{t}, \rmd x_{0:t-1})\right)
    \end{equation*}
    we obtain that
    \begin{align*}
    \untarg{t + 1}(f_{t + 1} \rk{t + 1} \af{t + 1}) &= \iint \varphi(x_{t:t+1}) \, \untarg{t + 1}(\rmd x_{t + 1}) \, \bk{n, \targ{t}} (x_{t+1}, \rmd x_{t}) \\
    &= \iint \varphi(x_{t:t+1}) \, \untarg{t}(\rmd x_t) \, \uk{t}(x_t, \rmd x_{t + 1}) \\
    &= \untarg{t}(\uk{t} f_{t + 1} \rk{t} \af{t} + \uk{t}\afterm{t} f_{t + 1}).
    \end{align*}
    Now the proof is concluded by noting that since $\untarg{t+1} = \untarg{t} \uk{t}$, $\untarg{t + 1}\tilde{f}_{t + 1} = \untarg{t}\uk{t}\tilde{f}_{t + 1}$.
\end{proof}

\begin{lemma}
\label{lem:unbiasedness-1}
For every $t \in \nsetpos$, $h_t \in \bmf(\yfd{t})$, and $\initmb \in \probmeas(\xfdmb{0})$ it holds that
\[
\cPE[\initmb][\pkl]{h_t(\bpartmb{t})}{\epartmb{0|0},\dots,\epartmb{t|t}}=
\ckjt{t} h_t(\epartmb{0|0},\dots,\epartmb{t|t}), \quad \canlaw[\pkl]{\initmb}\mbox{-a.s.}
\]
\end{lemma}

\begin{proof}
Pick arbitrarily $v_t \in \bmf(\xfd{0:t})$. We show that
\begin{equation}
\label{eq:proof-objective}
\E_{\initmb}^{\pkl}[ v_t(\epartmb{0|0},\dots,\epartmb{t|t}) h_t(\bpartmb{t}) ]=
\E_{\initmb}^{\pkl}[ v_t(\epartmb{0|0},\dots,\epartmb{t|t}) \ckjt{t} h_t(\epartmb{0|0},\dots,\epartmb{t|t})], 
\end{equation}
from which the claim follows. Using the definition \eqref{eq:def:P}, the left-hand side
of the previous identity may be rewritten as
\begin{align*}
\lefteqn{\idotsint \pinit[\initmb](\rmd \ymb{0}) \prod_{m=0}^{t-1} \pk{m}(\ymb{m}, \rmd \ymb{m+1}) \, h_t(\ymb{t})
v_t(\xmb{0|0},\dots,\xmb{t|t})} \hspace{10mm} \\
&=\idotsint \initmb( \rmd \xmb{0|0}) \prod_{m=0}^{t-1} \mkmb{m}(\xmb{m|m}, \rmd \xmb{m+1}) \, \ck{0}(\koskimat{\N} \xmb{0|0}, \xmb{1}, \rmd \ymb{1})  \\
&\hspace{10mm} \times \prod_{m=0}^{t-1}  \ck{m}(\ymb{m}, \xmb{m+1}, \rmd \ymb{m+1}) \, h_t(\ymb{t}) v_t(\xmb{0|0},\dots,\xmb{t|t}) \\
&=\idotsint \initmb( \rmd \xmb{0}) \prod_{m=0}^{t-1} \mkmb{m}(\xmb{m}, \rmd \xmb{m+1}) \, \ck{0}(\koskimat{\N} \xmb{0}, \xmb{1}, \rmd \ymb{1})  \\
&\hspace{10mm} \times \prod_{m=0}^{t-1}  \ck{m}(\ymb{m}, \xmb{m+1}, \rmd \ymb{m+1}) \, h_t(\ymb{t}) v_t(\xmb{0},\dots,\xmb{t}).
\end{align*}
Thus, we may conclude the proof by using the definition \eqref{eq:def:ckjt} of $\ckjt{t}$  together with Fubini's theorem.
\end{proof}

\begin{lemma}
\label{lem:unbiasedness-2}
For every $t \in \nsetpos$ and $h_t \in \bmf(\yfd{t})$,
\[
\E_{\initmb} \left[\left(\prod_{m=0}^{t-1} \potmb{m}(\epartmb{m|m})\right) h_t(\bpartmb{t})\right]
= \int  \untargmb{0:t} \ckjt{t}(\rmd  \ymb{t}) \, h_t(\ymb{t}).
\]
\end{lemma}
\begin{proof}
The claim of the lemma is a direct implication of \Cref{lem:unbiasedness-1}; indeed, by applying the tower property and the latter we obtain
\begin{align*}
\lefteqn{\E_{\initmb}^{\pkl} \left[\left(\prod_{m=0}^{t-1} \potmb{m}(\epartmb{m|m})\right) h_t(\bpartmb{t})\right]} \hspace{15mm} \\
&= \E_{\initmb}^{\pkl} \left[\left(\prod_{m=0}^{t-1} \potmb{m}(\epartmb{m|m})\right) \ckjt{t} h_t(\epartmb{0|0},\dots,\epartmb{t|t}) \right]\\
&= \idotsint \initmb(\rmd \xmb{0})  \prod_{m=0}^{t-1} \potmb{m}(\xmb{m}) \,  \mkmb{m}(\xmb{m},\rmd \xmb{m+1}) \, \ckjt{t} h_t(\xmb{0:t}) \\
&= \int  \untargmb{0:t} \ckjt{t}(\rmd  \ymb{t}) \, h_t(\ymb{t}).
\end{align*}
\end{proof}

\begin{proposition} \label{prop:unbiasedness:affine}
For all $t \in \nsetpos$, $(\N, \M) \in (\nsetpos)^2$, and $(f_t, \tilde{f}_t) \in \bmf(\xfd{t})^2$,
$$
\int \untargmb{0:t} \ckjt{t}(\rmd \ymb{t}) \,
\left( \frac{1}{\N} \sum_{i = 1}^\N \{\statl_t^i f_t(x_{t|t}^i) + \tilde{f}_t(x_{t|t}^i)\} \right)
= \untarg{t}(f_t \rk{t} \af{t} + \tilde{f}_t).
$$
\end{proposition}

\begin{proof}
Applying \Cref{lem:unbiasedness-2} yields
\begin{equation}
\int \untargmb{0:t} \ckjt{t}(\rmd \ymb{t}) \,
\left( \frac{1}{\N} \sum_{i = 1}^\N \{\statl_t^i f_t(x_{t|t}^i) + \tilde{f}_t(x_{t|t}^i)\} \right) 
= \E_{\initmb}^{\pkl}\left[ \left( \prod_{m = 0}^{t - 1} \potmb{m}(\epartmb{m|m}) \right) \frac{1}{\N} \sum_{i = 1}^\N \{\stat{t}{i} f_t(\epart{t|t}{i}) + \tilde{f}_t(\epart{t|t}{i})\} \right] \, .
\end{equation}
In the following we will use repeatedly the following filtrations. Let
$\partfiltbar{t} \eqdef \sigma( \{ \bpartmb{m} \}_{m = 0}^t)$
be the $\sigma$-field generated by the output of the {\PARIS} (\Cref{alg:paris}) during the first $t$ iterations. In addition, let $\partfilt{t} \eqdef \partfiltbar{t - 1} \vee \sigma(\epartmb{t|t})$.

We proceed by induction. Thus, assume that the statement of the proposition holds true for a given $t \in \nsetpos$ and consider, for arbitrarily chosen $(f_{t + 1}, \tilde{f}_{t + 1}) \in \bmf(\xfd{t + 1})^2$,
\begin{multline*}
\CPE[\initmb][\pkl]{\left( \prod_{m = 0}^t \potmb{m}(\epartmb{m|m}) \right) \frac{1}{\N}
\sum_{i = 1}^\N \{\stat{t + 1}{i} f_{t + 1}(\epart{t + 1|t + 1}{i}) + \tilde{f}_{t + 1}(\epart{t + 1|t + 1}{i})\}}{\partfiltbar{t}} \\
= \left( \prod_{m = 0}^t \potmb{m}(\epartmb{m|m}) \right) \cPE[\initmb][\pkl]{\stat{t + 1}{1} f_{t + 1}(\epart{t + 1|t + 1}{1}) + \tilde{f}_{t + 1}(\epart{t + 1|t + 1}{1})}{\partfiltbar{t}} \,,
\end{multline*}
where we used that the variables $\{\stat{t + 1}{i} f_{t + 1}(\epart{t + 1|t + 1}{i}) + \tilde{f}_{t + 1}(\epart{t + 1|t + 1}{i})\}_{i = 1}^\N$ are conditionally i.i.d. given $\partfiltbar{t}$. Note that, by symmetry,
\begin{align}
\CPE[\initmb][\pkl]{\stat{t + 1}{1}}{\partfilt{t + 1}} &= \int \ck{t}(\bpartmb{t}, \epartmb{t + 1|t + 1}, \rmd \ymb{t + 1}) \, \statl_{t + 1}^1 \nonumber \\
&= \idotsint \left( \prod_{j = 1}^\M \sum_{\ell = 1}^\N \frac{\ud{t}(\epart{t|t}{\ell}, \epart{t + 1|t + 1}{1})}{\sum_{\ell' = 1}^\N \ud{t}(\epart{t|t}{\ell'}, \epart{t + 1|t + 1}{1})} \delta_{(\epart{t|t}{\ell}, \stat{t}{\ell})}  (\rmd \tilde{x}_t^{1,j}, \rmd \tilde{\statl}_t^{1,j}) \right) \nonumber \\
& \hspace{50mm} \times \frac{1}{\M} \sum_{j = 1}^\M \left( \tilde{\statl}_t^{1,j} + \afterm{t}(\tilde{x}_t^{1,j}, \epart{t + 1|t + 1}{1}) \right) \nonumber \\
&= \sum_{\ell = 1}^\N \frac{\ud{t}(\epart{t|t}{\ell}, \epart{t + 1|t + 1}{1})}{\sum_{\ell' = 1}^\N \ud{t}(\epart{t|t}{\ell'}, \epart{t + 1|t + 1}{1})} \left( \stat{t}{\ell} + \afterm{t}(\epart{t|t}{\ell}, \epart{t + 1|t + 1}{1}) \right). \label{eq:cond:exp:beta}
\end{align}
Thus, using the tower property,
\begin{multline*}
\CPE[\initmb][\pkl]{\stat{t + 1}{1} f_{t + 1}(\epart{t + 1|t + 1}{1})}{\partfiltbar{t}} \\
= \int \pd{t}(\occm(\epartmb{t|t}))(\rmd x_{t + 1}) \, f_{t + 1}(x_{t + 1}) \sum_{\ell = 1}^\N \frac{\ud{t}(\epart{t|t}{\ell}, x_{t + 1})}{\sum_{\ell' = 1}^\N \ud{t}(\epart{t|t}{\ell'}, x_{t + 1})} \left( \stat{t}{\ell} + \afterm{t}(\epart{t|t}{\ell}, x_{t + 1}) \right),
\end{multline*}
and consequently, using definition \eqref{eq:def:pd:mapping},
\begin{align}
&\left( \prod_{m = 0}^t \potmb{m}(\epartmb{m|m}) \right) \CPE[\initmb][\pkl]{\stat{t + 1}{1} f_{t + 1}(\epart{t + 1|t + 1}{1})}{\partfiltbar{t}} \nonumber \\
&=\left( \prod_{m = 0}^{t - 1} \potmb{m}(\epartmb{m|m}) \right) \int \frac{1}{\N} \sum_{i = 1}^\N \ud{t}(\epart{t|t}{i}, x_{t + 1}) \nonumber \\
&\hspace{6mm} \times f_{t + 1}(x_{t + 1}) \sum_{\ell = 1}^\N \frac{\ud{t}(\epart{t|t}{\ell}, x_{t + 1})}{\sum_{\ell' = 1}^\N \ud{t}(\epart{t|t}{\ell'}, x_{t + 1})} \left( \stat{t}{\ell} + \afterm{t}(\epart{t|t}{\ell}, x_{t + 1}) \right) \, \refm{t+1}(\rmd x_{t + 1}) \nonumber \\
&= \left( \prod_{m = 0}^{t - 1} \potmb{m}(\epartmb{m|m}) \right) \frac{1}{\N} \sum_{\ell = 1}^\N \left( \stat{t}{\ell} \uk{t} f_{t + 1} (\epart{t|t}{\ell}) + \uk{t}(\afterm{t} f_{t + 1})(\epart{t|t}{\ell}) \right). \nonumber
\end{align}
Thus, applying the induction hypothesis,
\begin{align}
\lefteqn{\E_{\initmb}^{\pkl} \left[ \left( \prod_{m = 0}^t \potmb{m}(\epartmb{m|m}) \right) \frac{1}{\N} \sum_{i = 1}^\N \stat{t + 1}{i} f_{t + 1}(\epart{t + 1|t + 1}{i}) \right]} \nonumber \\
&= \E_{\initmb}^{\pkl} \left[ \left( \prod_{m = 0}^{t - 1} \potmb{m}(\epartmb{m|m}) \right) \frac{1}{\N} \sum_{\ell = 1}^\N \left( \stat{t}{\ell} \uk{t} f_{t + 1} (\epart{t|t}{\ell}) +\uk{t}(\afterm{t} f_{t + 1})(\epart{t|t}{\ell}) \right) \right] \nonumber \\
&= \untarg{t} \left( \uk{t} f_{t + 1} \rk{t} \af{t} + \uk{t}(\afterm{t} f_{t + 1}) \right). \label{eq:unbiasedness:first:term}
\end{align}
In the same manner, it can be shown that
\begin{equation} \label{eq:unbiasedness:second:term}
\E_{\initmb}^{\pkl} \left[ \left( \prod_{m = 0}^t \potmb{m}(\epartmb{m|m}) \right) \frac{1}{\N} \sum_{i = 1}^\N  \tilde{f}_{t + 1}(\epart{t + 1|t + 1}{i}) \right] = \untarg{t} \uk{t} \tilde{f}_{t + 1}.
\end{equation}
Now, by (\ref{eq:unbiasedness:first:term}--\ref{eq:unbiasedness:second:term}) and \Cref{lem:key:identity:untarg:affine},
\begin{align}
\lefteqn{\hspace{-30mm}\E_{\initmb}^{\pkl} \left[ \left( \prod_{m = 0}^t \potmb{m}(\epartmb{m|m}) \right) \frac{1}{\N} \sum_{i = 1}^\N \{\stat{t + 1}{i} f_{t + 1}(\epart{t + 1|t + 1}{i}) + \tilde{f}_{t + 1}(\epart{t + 1|t + 1}{i})\} \right]} \nonumber \\
&\hspace{-30mm} = \untarg{t} \left( \uk{t} f_{t + 1} \rk{t} \af{t} + \uk{t}(\afterm{t} f_{t + 1} + \uk{t} \tilde{f}_{t + 1}) \right) \nonumber \\
&\hspace{-30mm} = \untarg{t + 1}(f_{t + 1} \rk{t + 1} \af{t + 1} + \tilde{f}_{t + 1}), \nonumber
\end{align}
which shows that the claim of the proposition holds at time $n + 1$.

It remains to check the base case $n = 0$, which holds trivially true as $\statmb{0} = \boldsymbol{0}$, $\rk{0} \af{0} = 0$ by convention, and the initial particles $\epartmb{0|0}$ are drawn from $\initmb$. This completes the proof.
\end{proof}

\begin{proof}[Proof of  \Cref{thm:unbiasedness}]
The identity $\int \targmb{0:t}(\rmd \xmb{0:t}) \, \ckjt{t}(\xmb{0:t}, \rmd \statlmb{t}) \, \occm(\statlmb{t})(\operatorname{id}) =  \targ{0:t} h_t$ follows immediately by letting $f_t \equiv 1$ and $\tilde{f}_t \equiv 0$ in \Cref{prop:unbiasedness:affine} and using that $\untargmb{0:t}(\xspmb{0:t}) = \untarg{0:t}(\xsp{0:t})$. Moreover, applying  \Cref{thm:duality} yields
\begin{align}
\int \targ{0:t} \mbjt{t} \ckjt{t}(\rmd \statlmb{t}) \, \occm(\statlmb{t})(\operatorname{id}) &= \iint \targ{0:t}(\rmd {\chunk{z}{0}{t}}) \, \mbjt{t}({\chunk{z}{0}{t}}, \rmd \xmb{0:t}) \int \ckjt{t}(\xmb{0:t}, \rmd \statlmb{t}) \, \occm(\statlmb{t})(\operatorname{id}) \nonumber \\
&= \iint \targmb{0:t}(\rmd \xmb{0:t}) \, \bdpart{t}(\xmb{0:t}, \rmd {\chunk{z}{0}{t}}) \int \ckjt{t}(\xmb{0:t}, \rmd \statlmb{t}) \, \occm(\statlmb{t})(\operatorname{id}) \nonumber \\
&= \int \targmb{0:t} \ckjt{t}(\rmd \statlmb{t}) \, \occm(\statlmb{t})(\operatorname{id}). \nonumber
\end{align}
Finally, the first identity holds true since $\gibbs{t}$ leaves $\targ{0:t}$ invariant.
\end{proof}


\subsubsection{Proof of \Cref{prop:backward:marginal}}
\label{sec:proof:prop:backward:marginal}
First, note that, by definitions \eqref{eq:def:ck} and \eqref{eq:def:ckjt},
\begin{align*} 
H_t(\xmb{0:t}) &\eqdef \int \ckjt{t}(\xmb{0:t}, \rmd \ymb{t}) \, \occm(\xmb[0:n|n]) h \\
&= \idotsint \left( \frac{1}{\N} \sum_{j_t = 1}^\N h(x_{0:t-1|t}^{j_t}, x_{t}^{j_t}) \right)
\\
&\quad \times
\prod_{m = 0}^{t - 1} \prod_{i_{m + 1} = 1}^\N \int \sum_{j_m = 1}^\N \frac{\ud{m}(x_m^{j_m}, x_{m + 1}^{i_{m + 1}})}{\sum_{j'_m = 1}^\N \ud{m}(x_m^{j'_m}, x_{m + 1}^{i_{m + 1}})}
\delta_{x_{0:m|m}^{j_m}}(\rmd x_{0:m | m+1}^{i_{m + 1}}),
\end{align*}
where $x_{0:-1|0}^i = \emptyset$ for all $i \in \intvect{1}{\N}$ by convention.
We will show that for every $k \in \intvect{0}{t}$, $H_{k, t} \equiv H_t$, where
\begin{equation*} \label{eq:induction:hyp}
H_{k, n}(\xmb{0:t}) \eqdef \frac{1}{\N} \sum_{j_t = 1}^\N \cdots \sum_{j_k = 1}^\N \prod_{\ell = k}^{t - 1} \frac{\ud{\ell}(x_\ell^{j_\ell}, x_{\ell + 1}^{j_{\ell + 1}})}{\sum_{j_\ell' = 1}^\N \ud{\ell}(x_\ell^{j'_\ell}, x_{\ell + 1}^{j_{\ell + 1}})}
a_{k,n}(\xmb{0}, \ldots, \xmb{k - 1}, x_k^{j_k}, \ldots, x_t^{j_t})
\end{equation*}
with
\begin{multline*}
    a_{k,n}(\xmb{0}, \ldots, \xmb{k - 1}, x_k^{j_k}, \ldots, x_t^{j_t}) \\
    = \int \prod_{m = 0}^{k - 1} \prod_{i_{m + 1} = 1}^\N
\sum_{j_m = 1}^\N \frac{\ud{m}(x_m^{j_m}, x_{m + 1}^{i_{m + 1}})}{\sum_{j'_m = 1}^\N \ud{m}(x_m^{j'_m}, x_{m + 1}^{i_{m + 1}})}
\delta_{x_{0:m|m}^{j_m}}(\rmd x_{0:m|m+1}^{i_{m + 1}}) h(x_{0:k-1|k}^{j_k}, x_{k}^{j_k}, \ldots, x_t^{j_t}).
\end{multline*}
Since, by convention, $\prod_{\ell = n}^{t - 1} \ldots = 1$, $H_{n, n}(\xmb{0:t}) = \N^{-1} \sum_{j_t = 1}^\N a_{n,n}(\xmb{0}, \ldots, \xmb[n - 1], x_t^{j_t})$,
and we note that
$H_t \equiv H_{n, n}$.
We now show that $H_{k, n} \equiv H_{k - 1, n}$ for every $k \in \intvect{1}{t}$; for this purpose, note that
\begin{multline*}
    a_{k,n}(\xmb{0}, \ldots, \xmb{k - 1}, x_k^{j_k}, \ldots, x_t^{j_t}) \\ = \int \prod_{m = 0}^{k - 2} \prod_{i_{m + 1} = 1}^\N \sum_{j_m = 1}^\N \frac{\ud{m}(x_m^{j_m}, x_{m + 1}^{i_{m + 1}})}{\sum_{j'_m = 1}^\N \ud{m}(x_m^{j'_m}, x_{m + 1}^{i_{m + 1}})}
\delta_{x_{0:m|m}^{j_m}}(\rmd x_{0:m|m+1}^{i_{m + 1}}) \\
    \times \int \prod_{i_{k} = 1}^\N
    \sum_{j_{k-1} = 1}^\N \frac{\ud{k-1}(x_{k-1}^{j_{k-1}}, x_{k}^{i_{k}})}{\sum_{j'_{k-1} = 1}^\N \ud{k-1}(x_{k-1}^{j'_{k-1}}, x_{k}^{i_{k}})}
    \delta_{x_{0:k-1|k-1}^{j_{k-1}}}(\rmd x_{0:{k-1}|k}^{i_{k}}) \, h(x_{0:k-1|k}^{j_k}, x_{k}^{j_k}, \ldots, x_t^{j_t}),
\end{multline*}
and since $x_{0:k-1|k-1}^{j_{k-1}} = (x_{0:k-2|k-1}^{j_{k-1}}, x_{k-1}^{j_{k-1}})$, it holds that
\begin{multline*}
\int \prod_{i_{k} = 1}^\N
\sum_{j_{k-1} = 1}^\N \frac{\ud{k-1}(x_{k-1}^{j_{k-1}}, x_{k}^{i_{k}})}{\sum_{j'_{k-1} = 1}^\N \ud{k-1}(x_{k-1}^{j'_{k-1}}, x_{k}^{i_{k}})}
\delta_{x_{0:k-1|k-1}^{j_{k-1}}}(\rmd x_{0:{k-1}|k}^{i_{k}}) \, h(x_{0:k-1|k}^{j_k}, x_{k}^{j_k}, \ldots, x_t^{j_t}) \\
    =\sum_{j_{k-1} = 1}^\N \frac{\ud{k-1}(x_{k-1}^{j_{k-1}}, x_{k}^{j_{k}})}{\sum_{j'_{k-1} = 1}^\N \ud{k-1}(x_{k-1}^{j'_{k-1}}, x_{k}^{j_{k}})}  h(x_{0:k-2|k-1}^{j_{k-1}}, x_{k-1}^{j_{k-1}}, x_{k}^{j_k}, \ldots, x_t^{j_t}).
\end{multline*}
Therefore, we obtain
\begin{multline*}
a_{k,n}(\xmb{0}, \ldots, \xmb{k - 1}, x_k^{j_k}, \ldots, x_t^{j_t}) \\
= \int \prod_{m = 0}^{k - 2} \prod_{i_{m + 1} = 1}^\N
\sum_{j_m = 1}^\N \frac{\ud{m}(x_m^{j_m}, x_{m + 1}^{i_{m + 1}})}{\sum_{j'_m = 1}^\N \ud{m}(x_m^{j'_m}, x_{m + 1}^{i_{m + 1}})}
\delta_{x_{0:m|m}^{j_m}}(\rmd x_{0:m|m + 1}^{i_{m + 1}}) \\
\times \sum_{j_{k - 1} = 1}^\N \frac{\ud{k - 1}(x_{k - 1}^{j_{k - 1}}, x_k^{j_k})}{\sum_{j'_{k - 1} = 1}^\N \ud{k - 1}(x_{k - 1}^{j'_{k - 1}}, x_k^{j_k})} h(x_{0:k - 2|k - 1}^{j_{k - 1}}, x_{k - 1}^{j_{k - 1}}, x_k^{j_k},
\ldots, x_t^{j_t}).
\end{multline*}
Now, changing the order of summation with respect to $j_{k - 1}$ and integration on the right hand side of the previous display yields
\begin{multline*}
a_{k,n}(\xmb{0}, \ldots, \xmb{k - 1}, x_k^{j_k}, \ldots, x_t^{j_t})
\\ = \sum_{j_{k - 1} = 1}^\N \frac{\ud{k - 1}(x_{k - 1}^{j_{k - 1}}, x_k^{j_k})}{\sum_{j'_{k - 1} = 1}^\N \ud{k - 1}(x_{k - 1}^{j'_{k - 1}}, x_k^{j_k})} a_{k - 1,n}(\xmb{0}, \ldots, \xmb{k - 2}, x_{k - 1}^{j_{k - 1}}, \ldots, x_t^{j_t}).
\end{multline*}
Thus,
\begin{align*}
\lefteqn{H_{k, n}(\xmb{0:t})} \\
&= \frac{1}{\N} \sum_{j_t = 1}^\N \cdots \sum_{j_k = 1}^\N \prod_{\ell = k}^{t - 1} \frac{\ud{\ell}(x_\ell^{j_\ell}, x_{\ell + 1}^{j_{\ell + 1}})}{\sum_{j_\ell' = 1}^\N \ud{\ell}(x_\ell^{j'_\ell}, x_{\ell + 1}^{j_{\ell + 1}})} \\
& \quad \times \sum_{j_{k - 1} = 1}^\N \frac{\ud{k - 1}(x_{k - 1}^{j_{k - 1}}, x_k^{j_k})}{\sum_{j'_{k - 1} = 1}^\N \ud{k - 1}(x_{k - 1}^{j'_{k - 1}}, x_k^{j_k})} a_{k - 1,n}(\xmb{0}, \ldots, \xmb{k - 2}, x_{k - 1}^{j_{k - 1}}, \ldots, x_t^{j_t}) \\
&= \frac{1}{\N} \sum_{j_t = 1}^\N \cdots \sum_{j_{k - 1} = 1}^\N \prod_{\ell = k - 1}^{t - 1} \frac{\ud{\ell}(x_\ell^{j_\ell}, x_{\ell + 1}^{j_{\ell + 1}})}{\sum_{j_\ell' = 1}^\N \ud{\ell}(x_\ell^{j'_\ell}, x_{\ell + 1}^{j_{\ell + 1}})} a_{k - 1,n}(\xmb{0}, \ldots, \xmb{k - 2}, x_{k - 1}^{j_{k - 1}}, \ldots, x_t^{j_t}) \\
&= H_{k-1, n}(\xmb{0:t}),
\end{align*}
which establishes the recursion. Therefore, $H_t \equiv H_{0, n}$ and we may now conclude the proof by noting that $\bdpart{t}h \equiv H_{0, n}$.


\subsubsection{Proof of \Cref{cor:exp:conc:cond:paris}}
\label{sec:proof:exp:concentration}

In order to establish \Cref{cor:exp:conc:cond:paris} we will prove the following more general result, of which \Cref{cor:exp:conc:cond:paris} is a direct consequence.

\begin{proposition}
\label{thm:exp:conc:cond:paris}
For every $t \in \nset$ and $M \in \nsetpos$ there exist $\cstcondparisc_t > 0$ and $\cstcondparisd_t > 0$ such that for every $\N \in \nsetpos$, $z_{0:t} \in \xpsp{0}{t}$, $(f_t, \tilde{f}_t) \in \bmf(\xfd{t})^2$, and $\varepsilon > 0$,
\begin{multline*}
\int \mbjt{t} \ckjt{t}(z_{0:t}, \rmd \statlmb{t}) \indin{\left|\frac{1}{\N} \sum_{i = 1}^\N \{b_{t}^i f_t(x_{t|t}^{i}) + \tilde{f}_t(x_{t|t}^{i}) \} - \targ[z_{0:t}]{t}(f_t \rk[z_{0:t - 1}]{t} \af{t} + \tilde{f}_t) \right|\geq \varepsilon} \\
\leq \cstcondparisc_t \exp \left( -   \frac{\cstcondparisd_t \N \varepsilon^2}{2 \upkappa_t^2} \right),
\end{multline*}
where
\begin{equation}
\label{eq:definition:upkappa}
\upkappa_t \eqdef \| f_t \|_\infty \sum_{m = 0}^{t - 1} \| \afterm{m} \|_\infty + \| \tilde{f}_t \|_\infty.
\end{equation}
\end{proposition}

To prove \Cref{thm:exp:conc:cond:paris} we need the following technical lemma.

\begin{lemma} \label{lem:key:identity:untarg:affine:ii}
For every $t \in \nset$, $(f_{t + 1}, \tilde{f}_{t + 1}) \in \bmf(\xfd{t + 1})^2$, $z_{0:t + 1} \in \xpsp{0}{t + 1}$, and $\N \in \nsetpos$,
\begin{multline*}
\untarg[z_{0:t + 1}]{t + 1}(f_{t + 1} \rk[{\chunk{z}{0}{t}}]{t + 1} \af{t + 1} + \tilde{f}_{t + 1}) \\
= \left( 1- \frac{1}{\N} \right) \untarg[{\chunk{z}{0}{t}}]{t}\{ \uk{t} f_{t + 1} \rk[z_{0:t - 1}]{t} \af{t} + \uk{t}(\afterm{t} f_{t + 1} + \tilde{f}_{t + 1}) \} \\
+ \frac{1}{\N} \untarg[{\chunk{z}{0}{t}}]{t} \pot{t} \left( f_{t + 1}(z_{t + 1}) \rk[{\chunk{z}{0}{t}}]{t + 1} \af{t + 1}(z_{t + 1}) + \tilde{f}_{t + 1}(z_{t + 1}) \right).
\end{multline*}
\end{lemma}

\begin{proof}
    Since \Cref{lem:key:identity:untarg:affine} holds also for the Feynman--Kac model with a frozen path, we obtain
$$
\untarg[{\chunk{z}{0}{t+1}}]{t + 1}(f_{t + 1} \rk[{\chunk{z}{0}{t}}]{t + 1} \af{t + 1} + \tilde{f}_{t + 1}) \\
        = \untarg[{\chunk{z}{0}{t}}]{t}\{ \uk[z_{t+1}]{t} f_{t + 1} \rk[{\chunk{z}{0}{t}}]{t} \af{t} + \uk[z_{t+1}]{t}(\afterm{t} f_{t + 1} + \tilde{f}_{t + 1}) \}.
$$
Thus, the proof is concluded by noting that for every $x_t \in \xsp{t}$ and $h \in \bmf(\xfd{t:t+1})$,
    $$
    \uk[z_{t+1}]{t}h(x_t) = \left(1 - \frac{1}{\N} \right) \uk{t} h(x_t) + \frac{1}{\N} g(x_t) h(x_t, z_{t + 1}).
    $$
\end{proof}

Finally, before proceeding to the proof of \Cref{thm:exp:conc:cond:paris}, we introduce the law of the {\PARIS} evolving conditionally on a frozen path $\zpath = \{z_m\}_{m \in \nset}$.
Define, for $m \in \nset$ and $z_{m + 1} \in \xsp{m + 1}$,
\begin{equation*} 
\pk{m}[z_{m + 1}] : \yspmb{m} \times \yfdmb{m + 1} \ni (\ymb{m}, A) \mapsto \int
\, \mkmb{m}[z_{m + 1}](\xmb{m|m}, \rmd \xmb{m + 1}) \, \ck{m}(\ymb{m}, \xmb{m + 1}, A).
\end{equation*}
For any given initial distribution $\pinit \in \probmeas(\yfdmb{0})$, let $\canlaw[\pkl, \zpath]{\pinit}$ be the distribution of the canonical Markov chain induced by the Markov kernels $\{ \pk{m}[z_{m + 1}] \}_{m \in \nset}$ and the initial distribution $\pinit$. By abuse of notation we write
$\canlaw[\pkl, \zpath]{\initmb}$ instead of $\canlaw[\pkl, \zpath]{\pinit[\initmb[z_0]]}$, where the extension $\pinit[\initmb]$ is defined in \Cref{sec:proof:unbiasedness}.

\begin{proof}[Proof of \Cref{thm:exp:conc:cond:paris}]
We proceed by forward induction over $t$. Let the $\sigma$-fields $\partfiltbar{t}$ and $\partfilt{t}$ be defined as in the proof of \Cref{thm:unbiasedness}, but for the conditional {\PARIS} dual process. Then, under the law $\canlaw[\pkl, \zpath]{\initmb}$, reusing \eqref{eq:cond:exp:beta},
\begin{align*}
&\canlawexp[\pkl, \zpath]{\initmb} \left[ \stat{t}{1} f_t(\epart{t}{1}) + \tilde{f}_t(\epart{t}{1}) \mid \partfiltbar{t - 1} \right] \nonumber \\
&= \canlawexp[\pkl, \zpath]{\initmb} \left[ \canlawexp[\pkl, \zpath]{\initmb} \left[ \stat{t}{1} \mid \partfilt{t} \right] f_t(\epart{t}{1}) + \tilde{f}_t(\epart{t}{1}) \mid \partfiltbar{t - 1} \right] \nonumber \\
&= \canlawexp[\pkl, \zpath]{\initmb} \left[ f_t(\epart{t}{1}) \sum_{\ell = 1}^\N \frac{\ud{t - 1}(\epart{t - 1}{\ell}, \epart{t}{1})}{\sum_{\ell' = 1}^\N \ud{t - 1}(\epart{t - 1}{\ell'}, \epart{t}{1})} \left( \stat{t - 1}{\ell} + \afterm{t - 1}(\epart{t - 1}{\ell}, \epart{t}{1}) \right)
\vphantom{\sum_{\ell' = 1}^\N} + \tilde{f}_t(\epart{t}{1}) \mid \partfiltbar{t - 1} \right].
\end{align*}
Using \eqref{eq:kernel:cond:dual}, we get
\begin{multline} \label{eq:cond:exp:split}
\canlawexp[\pkl, \zpath]{\initmb} \left[ \stat{t}{1} f_t(\epart{t}{1}) + \tilde{f}_t(\epart{t}{1}) \mid \partfiltbar{t - 1} \right] \\
= \left( 1 - \frac{1}{\N} \right) \frac{\sum_{\ell = 1}^\N \{\stat{t - 1}{\ell} \uk{t - 1} f_t (\epart{t - 1}{\ell}) + \uk{t - 1}(\afterm{t - 1} f_t + \tilde{f}_t)(\epart{t - 1}{\ell})\}}{\sum_{\ell' = 1}^\N \pot{t - 1}(\epart{t - 1}{\ell'})} \\
+ \frac{1}{\N} \left( f_t(z_t) \sum_{\ell = 1}^\N \frac{\ud{t - 1}(\epart{t - 1}{\ell}, z_t)}{\sum_{\ell' = 1}^\N \ud{t - 1}(\epart{t - 1}{\ell'}, z_t)} \left( \stat{t - 1}{\ell} + \afterm{t}(\epart{t - 1}{\ell}, z_t) \right) + \tilde{f}_t(z_t) \right).
\end{multline}
In order to apply the induction hypothesis to each term on the right-hand side of the previous identity, note that
$$
\rk[z_{0:t - 1}]{t} \af{t}(z_t)
= \frac{\targ[z_{0:t - 1}]{t - 1}[ \ud{t - 1}(\cdot, z_t) \{ \rk[z_{0:t - 2}]{t - 1} \af{t - 1}(\cdot) + \afterm{t - 1}(\cdot, z_t) \} ] }{\targ[z_{0:t - 1}]{t - 1} [ \ud{t - 1}(\cdot, z_t) ]}.
$$
Therefore, using \Cref{lem:key:identity:untarg:affine:ii} and noting that
$\untarg[{\chunk{z}{0}{t}}]{t}\indi{\xsp{t}}/ \untarg[{\chunk{z}{0}{t}}]{t-1} \indi{\xsp{t-1}} = \targ[{\chunk{z}{0}{t-1}}]{t-1}\pot{t-1}$ yields
\begin{multline}
    \label{eq:cond:exp:recursion}
    \targ[{\chunk{z}{0}{t}}]{t}(f_{t} \rk[\chunk{z}{0}{t-1}]{t} \af{t} + \tilde{f}_{t}) = \frac{1}{\N} \left( f_{t}(z_{t}) \rk[\chunk{z}{0}{t-1}]{t} \af{t}(z_{t}) + \tilde{f}_{t}(z_{t}) \right) \\
    + \left( 1- \frac{1}{\N} \right) \frac{\targ[z_{0:t - 1}]{t - 1} \{ \uk{t - 1} f_t \rk[z_{0:t - 2}]{t-1} \af{t} + \uk{t - 1}(\afterm{t - 1} f_t + \tilde{f}_t) \}}{\targ[z_{0:t - 1}]{t - 1} \pot{t - 1}}.
\end{multline}
By combining \eqref{eq:cond:exp:split} with \eqref{eq:cond:exp:recursion}, we decompose the error according to
\begin{align} \label{eq:error_decomposition_paris}
\lefteqn{\frac{1}{\N} \sum_{i = 1}^\N \{\stat{t}{i} f_t(\epart{t|t}{i}) + \tilde{f}_t(\epart{t|t}{i}) \} - \targ[{\chunk{z}{0}{t}}]{t}(f_t \rk[z_{0:t - 1}]{t} \af{t} + \tilde{f}_t)} \hspace{10mm} \nonumber \\
&= \frac{1}{\N} \sum_{i = 1}^\N \{\stat{t}{i} f_t(\epart{t|t}{i}) + \tilde{f}_t(\epart{t|t}{i}) \} -
\canlawexp[\pkl, \zpath]{\initmb} \left[ \stat{t}{1} f_t(\epart{t}{1}) + \tilde{f}_t(\epart{t}{1})
\mid \partfiltbar{t - 1} \right] \nonumber \\
&\hspace{10mm} + \canlawexp[\pkl, \zpath]{\initmb} \left[
\stat{t}{1} f_t(\epart{t}{1}) + \tilde{f}_t(\epart{t}{1}) \mid \partfiltbar{t - 1} \right] - \targ[{\chunk{z}{0}{t}}]{t}(f_t \rk[z_{0:t - 1}]{t} \af{t} + \tilde{f}_t) \nonumber \\
&= \operatorname{I}^{(1)}_{\N} + \left( 1 - \frac{1}{\N} \right) \operatorname{I}^{(2)}_{\N} + \frac{1}{\N} \operatorname{I}^{(3)}_{\N},
\end{align}
where
\begin{align}
\operatorname{I}^{(1)}_{\N} &\eqdef \frac{1}{\N} \sum_{i = 1}^\N \{\stat{t}{i} f_t(\epart{t}{i}) + \tilde{f}_t(\epart{t}{i}) \}
-  \canlawexp[\pkl, \zpath]{\initmb} \left[ \stat{t}{1} f_t(\epart{t}{1}) + \tilde{f}_t(\epart{t}{1}) \mid \partfiltbar{t - 1} \right], \nonumber \\
\operatorname{I}^{(2)}_{\N} &\eqdef \frac{\sum_{\ell = 1}^\N \{\stat{t - 1}{\ell} \uk{t - 1} f_t (\epart{t - 1}{\ell}) + \uk{t - 1}(\afterm{t - 1} f_t + \tilde{f}_t)(\epart{t - 1}{\ell})\}}{\sum_{\ell' = 1}^\N \pot{t - 1}(\epart{t - 1}{\ell'})} \nonumber \\
&\hspace{10mm} - \frac{\targ[z_{0:t - 1}]{t - 1} \{ \uk{t - 1} f_t \rk[z_{0:t - 1}]{t} \af{t} + \uk{t - 1}(\afterm{t - 1} f_t + \tilde{f}_t) \}}{\targ[z_{0:t - 1}]{t - 1} \pot{t - 1}}, \label{eq:def:b_N}
\end{align}
and
\begin{multline} \label{eq:def:c_N}
\operatorname{I}^{(3)}_{\N} \eqdef f_t(z_t) \sum_{\ell = 1}^\N \frac{\ud{t - 1}(\epart{t - 1}{\ell}, z_t)}{\sum_{\ell' = 1}^\N \ud{t - 1}(\epart{t - 1}{\ell'}, z_t)} \left( \stat{t - 1}{\ell} + \afterm{t - 1}(\epart{t - 1}{\ell}, z_t) \right) \\
- f_t(z_t) \frac{\targ[z_{0:t - 1}]{t - 1}[ \ud{t - 1}(\cdot, z_t) \{ \rk[z_{0:t - 2}]{t - 1} \af{t - 1}(\cdot) + \afterm{t - 1}(\cdot, z_t) \} ] }{\targ[z_{0:t - 1}]{t - 1} [ \ud{t - 1}(\cdot, z_t) ]}.
\end{multline}

The proof is now completed by treating the terms $\operatorname{I}^{(1)}_{\N}$, $\operatorname{I}^{(2)}_{\N}$, and $\operatorname{I}^{(3)}_{\N}$ separately, using Hoeffding's inequality and its generalisation in \cite[Lemma~4] {douc:garivier:moulines:olsson:2009}. Choose $\varepsilon > 0$; then, by Hoeffding's inequality,
\begin{equation} \label{eq:a_N:bound}
\canlaw[\pkl, \zpath]{\initmb} \left( | \operatorname{I}^{(1)}_{\N} |\geq \varepsilon \right) \leq 2\exp \left( - \frac{1}{2} \frac{\varepsilon^2}{\upkappa_t^2}  \N \right).
\end{equation}
To treat $\operatorname{I}^{(2)}_{\N}$, we apply the induction hypothesis to the numerator and denominator, each normalised by $1 / \N$, yielding, since $\| \uk{t - 1} h \|_\infty \leq \pothigh{t - 1} \| h \|_\infty$ for all $h \in \bmf(\xfd{t - 1} \tensprod \xfd{t})$,
\begin{multline*}
\canlaw[\pkl, \zpath]{\initmb}  \left( \left| \frac{1}{\N} \sum_{\ell = 1}^\N \{\stat{t - 1}{\ell} \uk{t - 1} f_t (\epart{t - 1}{\ell}) + \uk{t - 1}(\afterm{t - 1} f_t + \tilde{f}_t)(\epart{t - 1}{\ell}) \} \right. \right. \\
\left. \left. - \targ[z_{0:t - 1}]{t - 1} \{ \uk{t - 1} f_t \rk[z_{0:t - 1}]{t} \af{t} + \uk{t - 1}(\afterm{t - 1} f_t + \tilde{f}_t) \} \vphantom{\sum_{\ell = 1}^\N} \right| \geq \varepsilon \right) \\
\leq \cstcondparisc_{t - 1} \exp \left( - \cstcondparisd_{t - 1} \frac{\varepsilon^2}{\pothigh{t - 1}^2 \upkappa_t^2}  \N \right)
\end{multline*}
and
$$
\canlaw[\pkl, \zpath]{\initmb} \left( \left| \frac{1}{\N} \sum_{\ell = 1}^\N \pot{t - 1}(\epart{t - 1}{\ell}) - \targ[z_{0:t - 1}]{t - 1} \pot{t - 1} \right|\geq \varepsilon \right)
\leq \cstcondparisc_{t - 1} \exp \left( - \cstcondparisd_{t - 1} \frac{\varepsilon^2}{\pothigh{t - 1}^2} \N \right).
$$
Combining the previous two bounds with the generalised Hoeffding inequality in \cite[Lemma~4]{douc:garivier:moulines:olsson:2009} yields, using also the bounds
$$
\frac{\sum_{\ell = 1}^\N \{\stat{t - 1}{\ell} \uk{t - 1} f_t (\epart{t - 1}{\ell}) + \uk{t - 1}(\afterm{t - 1} f_t + \tilde{f}_t)(\epart{t - 1}{\ell})\}}{\sum_{\ell' = 1}^\N \pot{t - 1}(\epart{t - 1}{\ell'})} \\
\leq \upkappa_t
$$
and $\targ[z_{0:t - 1}]{t - 1} \pot{t - 1} \geq \potlow{t - 1}$,
the inequality
\begin{equation} \label{eq:b_N:bound}
\canlaw[\pkl, \zpath]{\initmb} \left( | \operatorname{I}^{(2)}_{\N} |\geq \varepsilon \right)
\leq \cstcondparisc_{t - 1} \exp \left( - \cstcondparisd_{t - 1}  \frac{\potlow{t - 1}^2 \varepsilon^2}{ \pothigh{t - 1}^2 \upkappa^2_t}  \N \right).
\end{equation}
The last term $\operatorname{I}^{(3)}_{\N}$ is treated along similar lines; indeed, by the induction hypothesis, since $\|\ud{t - 1} \|_\infty \leq \pothigh{t - 1} \mdhigh{t - 1}$,
\begin{multline*}
\canlaw[\pkl, \zpath]{\initmb} \left( \left| \frac{1}{\N} \sum_{\ell = 1}^\N \ud{t - 1}(\epart{t - 1}{\ell}, z_t) \left( \stat{t - 1}{\ell} + \afterm{t - 1}(\epart{t - 1}{\ell}, z_t) \right) \right. \right. \\
\left. \left. \vphantom{\sum_{\ell = 1}^\N} - \targ[z_{0:t - 1}]{t - 1}[ \ud{t - 1}(\cdot, z_t) \{ \rk[z_{0:t - 1}]{t - 1} \af{t - 1}(\cdot) + \afterm{t - 1}(\cdot, z_t) \} ] \right| \geq \varepsilon \right) \\
\leq \cstcondparisc_{t - 1} \exp \left( - \cstcondparisd_{t - 1} \left( \frac{\varepsilon}{\pothigh{t - 1} \mdhigh{t - 1} \sum_{m = 0}^{t - 1} \| \afterm{m} \|_\infty} \right)^2 \N \right)
\end{multline*}
and
\begin{equation*}
\canlaw[\pkl, \zpath]{\initmb} \left( \left| \frac{1}{\N} \sum_{\ell = 1}^\N \ud{t - 1}(\epart{t - 1}{\ell}, z_t) - \targ[z_{0:t - 1}]{t - 1} [ \ud{t - 1}(\cdot, z_t) ] \right| \geq \varepsilon \right)
\leq \cstcondparisc_{t - 1} \exp \left( - \cstcondparisd_{t - 1} \left( \frac{\varepsilon}{\pothigh{t - 1} \mdhigh{t - 1}} \right)^2 \N \right).
\end{equation*}
Thus, since
$$
\sum_{\ell = 1}^\N \frac{\ud{t - 1}(\epart{t - 1}{\ell}, z_t)}{\sum_{\ell' = 1}^\N \ud{t - 1}(\epart{t - 1}{\ell'}, z_t)} \left( \stat{t - 1}{\ell} + \afterm{t - 1}(\epart{t - 1}{\ell}, z_t) \right) \leq \sum_{m = 0}^{t - 1} \| \afterm{m} \|_\infty
$$
and $\targ[z_{0:t - 1}]{t - 1} [ \ud{t - 1}(\cdot, z_t) ] \geq \potlow{t - 1}$, the generalised Hoeffding inequality provides 
\begin{equation} \label{eq:c_N:bound}
\canlaw[\pkl, \zpath]{\initmb} \left( | \operatorname{I}^{(3)}_{\N} |\geq \varepsilon \right)
\leq \cstcondparisc_{t - 1} \exp \left( - \cstcondparisd_{t - 1} \left( \frac{\potlow{t - 1} \varepsilon}{2 \pothigh{t - 1} \mdhigh{t - 1} \|f_t \|_\infty \sum_{m = 0}^{t - 1} \| \afterm{m} \|_\infty} \right)^2 \N \right).
\end{equation}
Finally, combining the bounds (\ref{eq:a_N:bound}--\ref{eq:c_N:bound}) completes the proof. 
\end{proof}


\subsubsection{Proof of \Cref{cor:Lp:cond:paris}}
\label{subsec:proof:cor:Lp:cond:paris}

The statement of \Cref{cor:Lp:cond:paris} is implied by the following more general result, which we will prove below.

\begin{proposition} \label{cor:Lp:cond:paris:affine}
For every $t \in \nset$, $M \in \nsets$, $\N \in \nsetpos$, $z_{0:t} \in \xpsp{0}{t}$, $(f_t, \tilde{f}_t) \in \bmf(\xfd{t})^2$, and $p \geq 2$, it holds that
\begin{equation*}
\int \mbjt{t} \ckjt{t}(z_{0:t}, \rmd \statlmb{t}) \left| \frac{1}{\N} \sum_{i = 1}^\N \{b_{t}^i f_t(x_{t|t}^{i}) + \tilde{f}_t(x_{t|t}^{i}) \} - \targ[z_{0:t}]{t}(f_t \rk[z_{0:t - 1}]{t} \af{t} + \tilde{f}_t) \right|^p 
\leq \cstcondparisc_t (p / \cstcondparisd_t)^{p/2} 
\N^{- p/2} \upkappa_t^p,
\end{equation*}
where $\cstcondparisc_t > 0$, $\cstcondparisd_t > 0$ and $\upkappa_t$ are defined in \Cref{thm:exp:conc:cond:paris} and \eqref{eq:definition:upkappa}, respectively.
\end{proposition}

Before proving \Cref{cor:Lp:cond:paris:affine}, we establish the following result.
\begin{lemma}
\label{lem:Lp}
 Let $X$ be an $\rset^d$-valued random variable, defined on some probability space $(\Omega, \mathcal{F}, \prob)$, satisfying $\mathbb{P}(|X| \geq t) \leq  c \exp(-t^2 / (2 \sigma^{2}))$ for every $t \geq 0$ and some $c > 0$ and $\sigma > 0$. Then for every $p \geq 2$ it holds that $\mathbb{E}[|X|^p] \leq c p^{p / 2} \sigma^{p}$.
 \end{lemma}
\begin{proof}
Using Fubini's theorem and the change of variable formula,
$$
\E \left[ |X|^p \right]=\int_{0}^{\infty} p t^{p-1} \mathbb{P}(|X| \geq t) \, \rmd t = c p 2^{p / 2 - 1} \sigma^{p} \Gamma(p / 2),
$$
where $\Gamma$ is the Gamma function. It remains to apply the bound $\Gamma(p / 2) \leq(p / 2)^{p / 2-1}$ (see \cite{anderson:qiu:1997}), which holds for $p \geq 2$ by [2, Theorem 1.5].
\end{proof}
\begin{proof}[Proof of \Cref{cor:Lp:cond:paris:affine}]
By combining \Cref{thm:exp:conc:cond:paris} and  \Cref{lem:Lp} we obtain
\begin{multline*}
\N \, \int \mbjt{t} \ckjt{t}(z_{0:t}, \rmd \statlmb{t}) \left| \frac{1}{\N} \sum\nolimits_{i = 1}^\N \{b_{t}^i f_t(x_{t|t}^{i}) + \tilde{f}_t(x_{t|t}^{i}) \} - \targ[z_{0:t}]{t}(f_t \rk[z_{0:t - 1}]{t} \af{t} + \tilde{f}_t) \right|^2 \\ 
\leq \cstcondparisc_t (p / \cstcondparisd_t)^{p/2} 
\N^{- p/2}
\left( \| f_t \|_\infty \sum_{m = 0}^{t - 1} \| \afterm{m} \|_\infty + \| \tilde{f}_t \|_\infty \right)^p,
\end{multline*}
which was to be established.
\end{proof}

\subsubsection{Proof of \Cref{prop:bias:cond:paris}}
\label{subsec:prop:bias:cond:paris}

Like previously, we establish \Cref{prop:bias:cond:paris} via a more general result, namely the following.

\begin{proposition} \label{prop:bias:cond:paris:affine}
For every $t \in \nset$, the exists $\cstcondparisbias{t}< \infty$ such that for every $M \in \nsets$, $\N \in \nsetpos$, $z_{0:t} \in \xpsp{0}{t}$, and $(f_t, \tilde{f}_t) \in \bmf(\xfd{t})^2$,
\begin{equation*}
 \left| \int \mbjt{t} \ckjt{t}(z_{0:t}, \rmd \statlmb{t}) \frac{1}{\N} \sum_{i = 1}^\N \{b_{t}^i f_t(x_{t|t}^{i}) + \tilde{f}_t(x_{t|t}^{i}) \} - \targ[z_{0:t}]{t}(f_t \rk[z_{0:t - 1}]{t} \af{t} + \tilde{f}_t) \right| 
\leq \cstcondparisbias{t} \upkappa_t  \N^{- 1},
\end{equation*}
where $\upkappa_t$ is defined in \eqref{eq:definition:upkappa}.
\end{proposition}

We preface the proof of \Cref{prop:bias:cond:paris:affine} by a technical lemma providing a bound on the bias of ratios of random variables.
\begin{lemma}
\label{lemma:bias-estimator-general}
Let $\upalpha$ and $\upbeta$ be (possibly dependent) random variables defined on some probability space $(\Omega, \mathcal{F}, \prob)$ and such that $\E[\upalpha^2] < \infty$ and $\E[\upbeta^2] < \infty$. Moreover, assume that there exist $c > 0$ and $d > 0$ such that $|\upalpha / \upbeta| \leq c$, $\prob$-a.s., $|a/b| \leq c$, $\E[ (\upalpha - a)^2] \leq c^2 d^2$, and $\E[ (\upbeta - b)^2] \leq d^2$. Then
\begin{equation}
\left| \E[ \upalpha / \upbeta] - a / b  \right| \leq  2 c(d/b)^2 + c |\E[\upbeta - b]|/|b| + |\E[\upalpha - a]|/|b|.
\end{equation}
\end{lemma}

\begin{proof}
Using the identity
$$
\E[\upalpha / \upbeta] - a/b = \E[ (\upalpha/ \upbeta)(b - \upbeta)^2] / b^2 + \E[(\upalpha - a)(b - \upbeta)] / b^2  + a\E[b - \upbeta] / b^2 + \E[\upalpha - a]/b,
$$
the claim is established by applying the Cauchy--Schwarz inequality and the assumptions of the lemma according to
\begin{align}
\lefteqn{\left|\E[\upalpha / \upbeta] - a/b\right|} \hspace{5mm}  \nonumber \\
&\leq c \E[ (\upbeta - b)^2] / b^2 + \{\E[(\upalpha - a)^2] \E[(\upbeta - b)^2]\}^{1/2} / b^2 + |a||\E[\upbeta - b]|/b^2 + |\E[\upalpha - a]|/b^2  \nonumber \\
&\leq 2 c (d / b)^2 + c|\E[\upbeta - b]|/|b| + |\E[\upalpha - a]|/|b|. \nonumber
\end{align}
\end{proof}

\begin{proof}[Proof of \Cref{prop:bias:cond:paris}]
We proceed by induction and assume that the claim holds true for $n - 1$. Reusing the error decomposition \eqref{eq:error_decomposition_paris}, it is enough to bound the expectations of the terms $\operatorname{I}^{(2)}_\N$ and $\operatorname{I}^{(3)}_\N$ given in \eqref{eq:def:b_N} and \eqref{eq:def:c_N}, respectively (since $\canlawexp[\pkl, \zpath]{\initmb}[\operatorname{I}^{(1)}_\N] = 0$). This will be done using the induction hypothesis, \Cref{lemma:bias-estimator-general}, and \Cref{cor:Lp:cond:paris:affine}. More precisely, to bound the expectation of $\operatorname{I}^{(2)}_\N$, we use \Cref{lemma:bias-estimator-general} with $\upalpha \leftarrow \upalpha_t$, $\upbeta \leftarrow \upbeta_t$, $a \leftarrow a_t$, and $b \leftarrow b_t$, where
\begin{align*}
&\upalpha_t \eqdef \frac{1}{\N} \sum_{\ell = 1}^\N \{\stat{t - 1}{\ell} \uk{t - 1} f_t (\epart{t - 1}{\ell}) + \uk{t - 1}(\afterm{t - 1} f_t + \tilde{f}_t)(\epart{t - 1}{\ell})\}, \quad
\upbeta_t \eqdef \frac{1}{\N} \sum_{\ell = 1}^\N \pot{t - 1}(\epart{t - 1}{\ell}), \\
& a_t \eqdef \targ[z_{0:t - 1}]{t - 1} \{ \uk{t - 1} f_t \rk[z_{0:t - 1}]{t} \af{t} + \uk{t - 1}(\afterm{t - 1} f_t + \tilde{f}_t) \}, \quad
b_t \eqdef \targ[z_{0:t - 1}]{t - 1} \pot{t - 1}.
\end{align*}
For this purpose, note that $|\upalpha_t / \upbeta_t | \leq \upkappa_t$ and $|a_t/b_t| \leq \upkappa_t$, where $\upkappa_t$ is defined in \eqref{eq:definition:upkappa}. On the other hand, using \Cref{cor:Lp:cond:paris:affine}  (applied with $p = 2$), we obtain
\begin{equation*}
\canlawexp[\pkl, \zpath]{\initmb}[(\upalpha_t - a_t)^2] \leq d_t^2 \upkappa_t^2  \quad \text{and} \quad \canlawexp[\pkl, \zpath]{\initmb}[(\upbeta_t - b_t )^2] \leq d_t^2,
\end{equation*}
where $d_t^2 \eqdef \cstcondparisc_t \gsupbound{t-1}^2 / (\cstcondparisd_t \N)$.
Using the induction assumption, we get
\begin{equation*}
|\canlawexp[\pkl, \zpath]{\initmb}[\upalpha_t] - a_t| \leq \cstcondparisbias{t-1} \N^{-1} \gsupbound{t-1} \upkappa_t
\quad \text{and} \quad
|\canlawexp[\pkl, \zpath]{\initmb}[\upbeta_t] - b_t| \leq \cstcondparisbias{t-1} \N^{-1} \gsupbound{t-1}.
\end{equation*}
Hence, the conditions of \Cref{lemma:bias-estimator-general} are satisfied and we deduce that
\[
|\canlawexp[\pkl, \zpath]{\initmb}[\operatorname{I}^{(2)}_\N] | = |\canlawexp[\pkl, \zpath]{\initmb}[\upalpha_t/\upbeta_t] - a_t/b_t| \leq 2 \upkappa_t \frac{\cstcondparisc_t}{\cstcondparisd_t \N} \frac{\gsupbound{t-1}^2}{\ginfbound{t-1}^2} +
2 \cstcondparisbias{t-1} \upkappa_t  \frac{\gsupbound{t-1}}{\ginfbound{t-1} \N} .
\]
The bound on $|\canlawexp[\pkl, \zpath]{\initmb}[\operatorname{I}^{(2)}_\N]|$ is obtained along the same lines.
\end{proof}

\section{Learning with \PPG\ }
\label{sec:appendix:learning_with_ppg}
This section is divided into three subsections.
 \Cref{sec:appendix:learning:general_non_as_bound}  establishes, following closely \cite{pmlr-v99-karimi19a}, a non-asymptotic bound for stochastic approximation schemes under general assumptions.
 \Cref{sec:appendix:learning:bound_grad} shows how assumptions \cref{assumption:lyapunov_smoothness} and \cref{assumption:strong_mixing} imply the assumptions provided in \Cref{sec:appendix:learning:general_non_as_bound} and therefore allow to establish  \Cref{theo:bound_grad}.
 Finally, \Cref{sec:appendix:learning:model_conditions} provides sufficient  assumptions on the model  ensuring that \Cref{assumption:lyapunov_smoothness} holds.

\subsection{Non-asymptotic bound}
\label{sec:appendix:learning:general_non_as_bound}

We follow closely \cite{pmlr-v99-karimi19a}. Consider the recursion 
$$
\paramopt{n+1} = \paramopt{n} - \lr{n+1} \gradest{}_{\paramopt{n}}(X_{n+1}), \quad n \in \nset, 
$$ 
where $\paramopt{n} \in \paramspace \subset \rset^d$ for some $d \in \nsetpos$ and $\{X_n\}_{n \in \nset}$ is a \emph{state-dependent} Markov chain on some measurable space $(\mathsf{X}, \mathcal{X})$ in the sense that $X_{n+1} \sim \mathbb{P}_{\paramopt{n}}(X_n, \cdot)$ with $\mathbb{P}_{\paramopt{}}$ being some Markov kernel on $(\mathsf{X}, \mathcal{X})$. Let $h(\theta) = \int \gradest{}_\theta(x) \, \pi_\theta (\rmd x)$, where $\pi_\theta$ is the invariant measure of $\mathbb{P}_{\paramopt{}}$
and $e_{n+1} \eqdef \gradest{}_{\paramopt{n}}(X_{n+1}) - h(\paramopt{n})$. As all norms are equivalent in finite dimensional vector spaces, we use $\|\cdot\|$ to denote a generic norm.
We denote by $\{\partfilt{n}\}_{n \in \nset}$ the natural filtration of the Markov chain $\{X_n\}_{n \in \nset}$.
\begin{assumption}
    \label{assumption:nonas:mean_field_unbiased}
    There exists a Borel measurable function $V : \paramspace \to \rset$ such that for every $\paramopt{} \in \paramspace$, $\nabla V(\paramopt{}) = h(\paramopt{})$.
\end{assumption}

\begin{assumption}
    \label{assumption:nonas:lyapunov_l_lipschitz}
    There exists $\lyapunovlip \in \rsetpos$ such that for every $(\paramopt{}, \paramopt{}') \in \paramspace^2$,
    \[
        \|\nabla V(\paramopt{}) - \nabla V(\paramopt{}')\| \leq \lyapunovlip \|\paramopt{} - \paramopt{}'\|.
    \]
\end{assumption}

\begin{assumption}
    \label{assumption:nonas:poisson_eq}
    There exists a Borel measurable function $\widehat{H}: \paramspace \times \xsp{} \to \paramspace$ such that for every $\paramopt{} \in \paramspace$ and
    $x \in \xsp{}$,
    \begin{equation*}
        \gradestP{}(x) - \mathbb{P}_{\theta}\gradestP{}(x) = \gradest{}_\theta(x) - h(\paramopt{}) \eqsp.
    \end{equation*}
\end{assumption}

\begin{assumption}
    \label{assumption:nonas:lip_noisegradestp}
    There exists $\kergradestplip \in \rsetpos$ such that for every $(\paramopt{0}, \paramopt{1}) \in \paramspace^2$,
    \begin{equation*}
        \sup_{x \in \xsp{}}\| \mathbb{P}_{\theta_0} \gradestP{0}(x) - \mathbb{P}_{\theta_0} \gradestP{1}(x) \| \leq \kergradestplip \|\paramopt{0} - \paramopt{1}\| \eqsp.
    \end{equation*}
\end{assumption}

\begin{assumption}
    \label{assumption:nonas:boundness_noisegradestp}
    There exists $\kergradestpbound \in \rsetpos$ such that
    \begin{equation*}
        \sup_{\paramopt{} \in \paramspace}\| \mathbb{P}_{\theta} \gradestP{} \| \leq \kergradestpbound \eqsp.
    \end{equation*}
\end{assumption}

\begin{assumption}
    \label{assumption:nonas:bound_mse}
    There exists $\constmsealg \in \rsetpos$ such that for every $x \in \xsp{}$ and $\paramopt{} \in \paramspace$,
    \begin{equation*}
         \int \|\gradest{}_{\paramopt{}}(x') - h(\paramopt{})\|^2 \, \mathbb{P}_\theta (x, \rmd x') \leq \constmsealg^2 \eqsp.
    \end{equation*}
    
\end{assumption}

\begin{assumption}
    \label{assumption:nonas:stochastic_update_bounded}
    There exists $\boundgradestp \in \rsetpos$ such that for every $x \in \xsp{}$, 
    \begin{equation*}
    \sup_{\paramopt{} \in \paramspace} \int \|\gradestP{}\| \, \mathbb{P}_\theta(x, \rmd x')   \leq \boundgradestp \eqsp.
    \end{equation*}
\end{assumption}

\begin{theorem}
    \label{thm:non_ass}
    Assume that \cref{assumption:nonas:mean_field_unbiased}--\cref{assumption:nonas:stochastic_update_bounded} hold. In addition, assume that there exist $a> 0$ and $a'>0$ such that for all $n \in \nset$,
    $$
    \lr{n+1} \leq \lr{n}\leq a\lr{n+1}\eqsp,\quad \lr{n}-\lr{n+1} \leq a'\lr{n}^2\eqsp,\quad \lr{1} \leq (\lyapunovlip+C_h)^{-1}/2\eqsp. 
    $$
    Moreover, for any $n \in \nsetpos$, let $\varpi$ be a $\intvect{0}{n}$-valued random variable, independent of $\{\mathcal{F}_\ell\}_{\ell\geq 0}$ and such that  $\mathbb{P}(\varpi=k) = \lr{k+1}/\sum_{\ell=0}^n\lr{\ell+1}$ for $k \in \intvect{0}{n}$. Then,
    \begin{equation*}
      \PE{\|h(\paramopt{\varpi})\|^2}{} \leq 2\frac{V_{0,n} + C_{0,n} + (\boundmse^2 \lyapunovlip + C_{\gamma})\sum_{k=0}^{n} \lr{k+1}^2}{\sum_{k=0}^{n} \lr{k+1}} \eqsp,
    \end{equation*}
    where $V_{0, n} \eqdef \PE{V(\paramopt{}) - V(\paramopt{n})}{}$ and
    \begin{align}
        C_{0,n} &\eqdef \lr{1} h(\paramopt{0}) \boundgradestp + \kergradestpbound (\lr{1} - \lr{n+1} + 1)\eqsp, \label{eq:non_ass:c_0}\\
        C_{\gamma} &\eqdef \boundmse \kergradestplip + (1 + \boundmse)\lyapunovlip\kergradestpbound\eqsp, \label{eq:non_ass:c_gamma}\\
        C_{h} &\eqdef \kergradestplip \left((a+1)/2 + a \boundmse \right) + (\lyapunovlip + a' + 1) \kergradestpbound \eqsp.\label{eq:non_ass:c_h}
    \end{align}
\end{theorem}

\begin{proof}
    We follow closely the proof of \cite[Theorem 2]{pmlr-v99-karimi19a} and adapt it to our setting. First, note that by \cref{assumption:nonas:mean_field_unbiased}, assumptions {\bf A1} and {\bf A2} of \cite[Theorem 2]{pmlr-v99-karimi19a} hold with $c_0=d_0 = 0$ and $c_1=d_1=1$. In addition, the claim in \cite[Lemma 1]{pmlr-v99-karimi19a} holds true since by \Cref{assumption:nonas:lyapunov_l_lipschitz}, {\bf A3} holds. Moreover, \cite[Equation 17]{pmlr-v99-karimi19a} can also be established under \cref{assumption:nonas:bound_mse}, as we may rewrite it as
    \begin{equation*}
        \sum_{\ell=0}^n \lr{\ell+1}^2 \PE{\|e_{\ell+1}\|^2}{} = \sum_{\ell=0}^{n} \lr{\ell+1}^2 \PE{\CPE{\|e_{\ell+1}\|^2}{\partfilt{\ell}}}{} \leq \constmsealg^2 \sum_{\ell=0}^{n} \lr{\ell+1}^2 \eqsp .
    \end{equation*}
    Following the proof of \cite[Lemma 2]{pmlr-v99-karimi19a}, consider the decomposition
    $$
    \PE{-\sum_{\ell=0}^{n} \lr{\ell + 1} \left< \nabla V(\paramopt{\ell}), e_{\ell+1}\right> }{} = \PE{A_1 + A_2 + A_3 + A_4 + A_5}{},
    $$
    where
    \begin{align*}
        A_1 & \eqdef -\sum_{\ell=1}^{n}\lr{\ell + 1}\left< \nabla V(\paramopt{\ell}), \gradestP{\ell}(X_{\ell+1}) - \noisekerhom{\ell}\gradestP{\ell}(X_\ell)\right>,\\
        A_2 & \eqdef -\sum_{\ell=1}^{n}\lr{\ell + 1}\left< \nabla V(\paramopt{\ell}), \noisekerhom{\ell}\gradestP{\ell}(X_\ell) - \noisekerhom{\ell-1}\gradestP{\ell-1}(X_\ell)\right>,\\
        A_3 & \eqdef -\sum_{\ell=1}^{n}\lr{\ell + 1}\left< \nabla V(\paramopt{\ell})-\nabla V(\paramopt{\ell-1}), \noisekerhom{\ell-1}\gradestP{\ell-1}(X_\ell)\right>,\\
        A_4 & \eqdef -\sum_{\ell=1}^{n}\left(\lr{\ell + 1}-\lr{\ell}\right)\left< \nabla V(\paramopt{\ell-1}), \noisekerhom{\ell-1}\gradestP{\ell-1}(X_\ell)\right>,\\
        A_5 & \eqdef -\lr{1}\left< \nabla V(\paramopt{0}), \gradestP{0}(X_1)\right> + \lr{n+1}\left< \nabla V(\paramopt{n}), \noisekerhom{n}\gradestP{n}(X_{n+1})\right> .
    \end{align*}
    As $\gradestP{\ell}(X_{\ell+1}) - \noisekerhom{\ell}\gradestP{\ell}(X_\ell)$ is a martingale difference, it holds that $\PE{A_1}{} = 0$.
    The upper bounds
    on the expectations of $A_2$, $A_3$ and $A_4$ are obtained similarly as in \cite{pmlr-v99-karimi19a}. Using \cref{assumption:nonas:lip_noisegradestp},
    $$
    A_2\leq\kergradestplip \left(\boundmse\sum_{k=1}^n\lr{k}^2 + \frac{1}{2}\left(1+2a\boundmse+a\right)\sum_{k=0}^n\lr{k+1}^2\|h(\paramopt{k})\|^2\right)\eqsp.
    $$
    By \cref{assumption:nonas:lyapunov_l_lipschitz,assumption:nonas:boundness_noisegradestp},
    $$
    A_3\leq \lyapunovlip  \kergradestpbound \left((1+\boundmse) \sum_{k=1}^n \lr{k}^2 + \sum_{k=1}^n \lr{k}^2\|h(\paramopt{k})\|^2)\right)\eqsp.
    $$
    On the other hand,
    $$
    A_4\leq \kergradestpbound \left(\lr{1}-\lr{n+1} + a'\sum_{k=1}^n \lr{k}^2\|h(\paramopt{k-1})\|^2\right)\eqsp.
    $$
    We now focus on $A_5$. As in the proof of \cite[Lemma 2]{pmlr-v99-karimi19a}, 
  the expectation of the first term can be straightforwardly bounded by $ \lr{1}\|h(\paramopt{0})\| \boundgradestp$ using the Cauchy--Schwarz inequality and \cref{assumption:nonas:stochastic_update_bounded}. The second term can, using \cref{assumption:nonas:boundness_noisegradestp} and $\lr{n+1}\|h(\paramopt{n})\| \leq 1 + \lr{n+1}^2\|h(\paramopt{n})\|^2$, be bounded in the same way according to
    \begin{multline*}
        \lr{n+1}\left< \nabla V(\paramopt{n}), \noisekerhom{n}\gradestP{n}(X_{n+1})\right> \leq  \kergradestpbound \lr{n+1}\|h(\paramopt{n})\| \leq \kergradestpbound \left(1 + \lr{n+1}^2 \|h(\paramopt{n})\|^2 \right) \\ 
        \leq  \kergradestpbound \left(1 +  \sum_{\ell=0}^{n} \lr{\ell+1}^2 \|h(\paramopt{\ell})\|^2 \right) \eqsp.
    \end{multline*}
    The rest of the proof follows that of \cite[Theorem 2]{pmlr-v99-karimi19a}.
\end{proof}

\subsection{Application to \Cref{theo:bound_grad}}
\label{sec:appendix:learning:bound_grad}
The goal of this section is to establish that the assumptions of \Cref{theo:bound_grad} ensure all the assumptions in \cref{sec:appendix:learning:general_non_as_bound}, which in turn allows \Cref{thm:non_ass} to be applied. First, we start by explicitly defining the kernel $\mathbb{P}_\theta$ and the function $h$ in terms of the kernels presented in \cref{sec:appendix:ppg}. We write $\mathbb{P}_{\theta, t}$ instead of $\mathbb{P}_\theta$ to explicit the dependence of the kernel on the \emph{fixed} number of observations $t$.
\subsubsection{Verification of the assumptions of \Cref{thm:non_ass}}
\label{sec:appendix:learning:bound_grad:assumptions_check}

For $(k_0, k) \in (\nsetpos)^2$ such that $k_0 < k$, define
\begin{equation}
\label{eq:def:bbP}
    \noiseker{t}{}: \esp{t}^{k - k_0} \times \efd{t}^{\tensprod (k - k_0)} \ni
    (\ymb{t}[k_0:k], z_{0:t}[k_0:k], A) 
    \mapsto
    \parisgibbs{\paramopt{}, t}^{k_0} \tensprod \parisgibbs{\paramopt{}, t}^{\tensprod(k - k_0)}(z_{0:t}[k], A),
\end{equation}
where $\parisgibbs{\paramopt{}, t}$ is the \PPG\ kernel defined in \eqref{eq:ppg_ker_def}. Note that $\noiseker{t}{}$ depends only on the last frozen path, namely $z_{0:t}[k]$. Note also that, since $\parisgibbs{\paramopt{}, t}$ depends only on the paths, there is no dependence between $\ymb{t,\ell}[k_0:k]$ and $\ymb{t,\ell + 1}[k_0:k]$.
The score ascent algorithm (\Cref{alg:scoreasc:full}) can be formulated as follows.
\begin{enumerate}
    \item Sample $(z_{0:t, \ell}[k_0:k], \ymb{t, \ell}[k_0:k]) \sim \noiseker{t}{\ell} \big((z_{0:t, \ell-1}[k_0:k], \ymb{t, \ell-1}[k_0:k]), \cdot \big)$.
    \item Update the parameter according to $\eta_{\ell + 1} = \eta_{\ell} + \gamma_{\ell+1} \gradest(z_{0:t, \ell}[k_0:k], \ymb{t, \ell}[k_0:k])$, where
    \[
        \gradest(z_{0:t, \ell}[k_0:k], \ymb{t, \ell}[k_0:k]) = \frac{1}{k-k_0+1} \sum_{i = k_0}^k \mu(\statmb{t, \ell}[i])(\mathrm{id}) = \rollingestim[\ki_0-1][\ki][N][\af{t}],
    \]
\end{enumerate}
where $\rollingestim[\ki_0-1][\ki][N][\af{t}]$ is defined in 
\eqref{eq:rolling-estimator}. 
 We denote by $\invar{t}{}$ the invariant distribution of $\noiseker{t}{}$, which, by \Cref{prop:inv_measure_parisgibbs}, is given by $\invar{t}{} = (\targ{0:t}{} \tensprod \mbjt{t}{} \ckjt{t}{})^{\tensprod (k - k_0)}$.

We also require the strong mixing assumption to hold uniformly in $\paramopt{}$.

\begin{assumption}[Strong mixing uniformly in $\paramopt{}$] 
    \label{assumption:strong_mixing_uniform}
  For every $s \in \nset$ there exist $\potlow{s}$, $\pothigh{s}$, $\mdlow{s}$, and $\mdhigh{s}$ in $\rset_+^\ast$ such that for all $\paramopt{} \in \paramspace$, 
   \begin{enumerate}[label=(\roman*),nosep]
   \item  $\potlow{s} \leq \pot{s, \paramopt{}}(x_s) \leq \pothigh{s}$ for every $x_s \in \xsp{s}$, 
   \item  $\mdlow{s} \leq \md{s, \paramopt{}}(x_s, x_{s + 1}) \leq \mdhigh{s}$ for every $(x_s, x_{s + 1}) \in \xsp{s:s+1}$.
   \end{enumerate}
\end{assumption}
Note that the assumption above implies that $\mixrate{t}{\N}$ is also uniform in $\paramopt{}$.

\paragraph{Proof that \cref{assumption:nonas:mean_field_unbiased} holds. }
\begin{proposition}
    \label{thm:unbiased_mean_field}
    For all $\paramopt{} \in \paramspace$, $h(\paramopt{}) = \nabla V(\paramopt{})$, where $V(\theta) = \log  \untarg{0:t,\paramopt{}}{} (\xsp{0:t})$ is the log-likelihood function.  
\end{proposition} 

\begin{proof}
    By \Cref{thm:unbiasedness},
    \begin{align*}
        h(\paramopt{}) &= \int \gradest(\ymbtd{t}[k_0:k], \tilde{x}_{0:t}[k_0:k]) \, \invar{t}{}(\rmd( \ymbtd{t}[k_0:k], \tilde{x}_{0:t}[k_0:k])) \\
        & = \frac{1}{k-k_0+1} \sum_{i = k_0}^k \int \left[\targ{0:t,\paramopt{}}{} \tensprod \mbjt{t, \paramopt{}} \ckjt{t, \paramopt{}}\right](\rmd (\ymbtd{t}[i], \tilde{x}_{0:t}[i])) \mu(\statmbtd{t, \ell}[i])(\mathrm{id}) \\
        & = \targ{0:t,\paramopt{}}{} \left( \score{}{0:t} \right) = \nabla V(\paramopt{}).
    \end{align*}
\end{proof}

\paragraph{Proof that \cref{assumption:nonas:lyapunov_l_lipschitz} holds. }
\cref{assumption:nonas:lyapunov_l_lipschitz} is trivially implied by \cref{assumption:lyapunov_smoothness}(i).

\paragraph{Proof that \cref{assumption:nonas:poisson_eq,assumption:nonas:boundness_noisegradestp} hold.}
Let $\gradestP{}$ be given by 
\begin{equation}
    \label{eq:poissoneq}
\gradestP{} : \esp{t}^{k - k_0} \ni (\ymb{t}[k_0:k], z_{0:t}[k_0:k])  \mapsto \sum_{r = 0}^\infty \lbrace \noiseker{t}{}^r \gradest(\ymb{t}[k_0:k], z_{0:t}[k_0:k]) - h(\paramopt{}) \rbrace .
\end{equation}
Then the following holds true.
\begin{lemma}
    \label{lem:boundness_p_gradestp}
    Assume \cref{assumption:strong_mixing_uniform}. Then for all $\paramopt{} \in \paramspace$ and $t \in \nsetpos$,
    \begin{equation*}
       \|\noiseker{t}{}\gradestP{}\|_{\infty} \leq \constbiasalg (1 - \kappa_{N,t}^{k})^{-1}  \eqsp.
    \end{equation*}
\end{lemma}
\begin{proof}
By \Cref{theo:bias-mse-rolling}, we have for any $r > 0$
\[ 
    \big| \noiseker{t}{}^r \gradest(\ymb{t}[k_0:k], z_{0:t}[k_0:k]) - h(\paramopt{}) \big| \leq \constbiasalg \kappa_{N,t}^{(r-1)k}
\]
and thus
$$
\|\noiseker{t}{}\gradestP{} \|_\infty  
\leq \sum_{r = 1}^{\infty} \left\| \noiseker{t}{}^r H - h(\paramopt{}) \right\|_\infty  
\leq \constbiasalg \sum_{r = 0}^{\infty} \kappa_{N,t} ^{rk} \leq \constbiasalg (1 - \kappa_{N,t}^{k})^{-1} \eqsp,
$$
where $\kappa_{N,t} \in (0, 1)$.
\end{proof}
\Cref{lem:boundness_p_gradestp} proves \cref{assumption:nonas:poisson_eq,assumption:nonas:boundness_noisegradestp} with $\kergradestpbound\eqdef \constbiasalg  (1 - \kappa_{N,t}^{k})^{-1}$. 

\paragraph{Proof that \cref{assumption:nonas:lip_noisegradestp} holds. }

\begin{theorem}
    \label{theo:lip_cont_phhat}
    Assume \cref{assumption:strong_mixing_uniform} and \cref{assumption:lyapunov_smoothness}. Then for every $t \in \nset$, $\paramopt{} \in \Theta$ and $\N \in \nsetpos$ such that $N > 1 + 5 \rho_t^2 t / 2$, 
    \begin{equation*}
    \left\| \noiseker{t}{1} \gradestP{1} - \noiseker{t}{2} \gradestP{2} \right \|_\infty \leq \kergradestplip \|\paramopt{1} - \paramopt{2}\|\eqsp,
    \end{equation*}
    where
    \begin{multline}
        \label{eq:lip:PHhat}
         \kergradestplip \eqdef \|L^{P}_2\|_{\infty} \left[1 + \mixrate{t}{\N}^{\ki}(1 - \mixrate{t}{\N}^{\ki})\right] + \lyapunovlip +\\  \constbiasalg  (1 - \mixrate{t}{\N})^{-1}  (1 - \mixrate{t}{\N}^{k})^{-1} \left[\|L^P_{1}\|_\infty (1 - \mixrate{t}{\N}^{k})^{-1} + \targlip \mixrate{t}{\N}^{\ki}\right] \eqsp .
\end{multline}
\end{theorem}
\begin{proof}
    We establish the claim by adapting the proof of \cite[Lemma 7]{pmlr-v99-karimi19a}. First, recall that the kernel $\gibbs{\paramopt{}, t}$ defined in  \eqref{eq:gibbs_ker} is the path marginalized version of $\parisgibbs{\paramopt{}, t}$ given in \eqref{eq:ppg_ker_def}.
    Note that for every $x \in \esp{t}^{k - k_0}$, $$
        \noiseker{t}{1} \gradestP{1}(x) = \sum_{n=0}^\infty \delta_x \noiseker{t}{1} \left\{ \noiseker{t}{1}^n \gradest - h(\paramopt{1})\right\}   = \sum_{n = 0}^\infty \delta_x \gibbs{\paramopt{1},t}^{kn} \left\{ \noiseker{t}{1} \gradest - \targ{0:t, \paramopt{1}} \noiseker{t}{1} \gradest \right\}\eqsp,
    $$
    where we have used
    (i) the fact that the backward statistics output by $\noiseker{t}{}$ are independent of the input backward statistics and (ii)
    the penultimate line in the computation of $h(\paramopt{})$ above. We follow the proof of \cite[Lemma 4.2]{Fort_2011} and consider the following decomposition: for $n \in \nsetpos$,
    \begin{align}
         \lefteqn{\delta_x \gibbs{\paramopt{1},t}^{kn} \left( \noiseker{t}{1} \gradest - \targ{0:t, \paramopt{1}} \noiseker{t}{1} \gradest \right)  - \delta_x \gibbs{\paramopt{2},t}^{kn} \left( \noiseker{t}{2} \gradest - \targ{0:t, \paramopt{2}} \noiseker{t}{2} \gradest \right)} \hspace{20mm} \label{eq:lipP}\\
        & = \sum_{j = 0}^{n-1} \left(\delta_x \gibbs{\paramopt{1},t}^{kj} - \targ{0:t, \paramopt{1}} \right) \left(\gibbs{\paramopt{1},t}^{kj} - \gibbs{\paramopt{2},t}^{kj}\right) \left( \gibbs{\paramopt{2},t}^{k(n-j-1)} \noiseker{t}{1} \gradest - \targ{0:t, \paramopt{2}} \noiseker{t}{1} \gradest \right)\nonumber \\
        &\quad  - \left( \delta_x \gibbs{\paramopt{2},t}^{kn}  \noiseker{t}{2} \gradest - \targ{0:t, \paramopt{2}} \noiseker{t}{2} \gradest\right) + \left( \delta_x \gibbs{\paramopt{2},t}^{kn}  \noiseker{t}{1} \gradest - \targ{0:t, \paramopt{2}} \noiseker{t}{1} \gradest\right)\nonumber \\
        &\quad - \targ{0:t, \paramopt{1}} \left( \gibbs{\paramopt{2},t}^{kn}  \noiseker{t}{1} \gradest - \targ{0:t, \paramopt{2}} \noiseker{t}{1} \gradest \right)
        \nonumber
        \eqsp.
    \end{align}
    Applying \Cref{thm:geometric:ergodicity:particle:Gibbs} with $\mu=\delta_x$ and  $\nu=\targ{0:t, \paramopt{}}$ and using the fact that $\targ{0:t, \paramopt{}} \gibbs{\paramopt{}, t}^\ell = \targ{0:t, \paramopt{}}$ for all $\ell \in \nset$, we obtain that for all $\ell \in \nset$ and all $\paramopt{} \in \paramspace$,  $\tv{\delta_x \gibbs{\paramopt{}, t}^\ell}{\targ{0:t, \paramopt{}}} \leq \mixrate{t}{N}^\ell$.
    Note that by \cref{assumption:lyapunov_smoothness}(iii), $\gibbs{\paramopt{}, t}$ is Lipschitz; therefore, for all $r \in \nsetpos$,  by \Cref{lem:lip:ergodic}, $\gibbs{\paramopt{}, t}^r$ is Lipschitz with constant $\|L^P_{1}\|_{\infty} (1 - \mixrate{t}{\N})^{-1}$. Combining all this together, we obtain
    \begin{align*}
        \lefteqn{\left|  \left(\delta_x \gibbs{\paramopt{1},t}^{kj} - \targ{0:t, \paramopt{1}} \right)
        \left(\gibbs{\paramopt{1},t}^{kj} - \gibbs{\paramopt{2},t}^{kj}\right)
        \left(\gibbs{\paramopt{2},t}^{k(n-j-1)} \noiseker{t}{1} \gradest - \targ{0:t, \paramopt{2}} \noiseker{t}{1} \gradest \right)
        \right|} \hspace{10mm}
        \\& = \left| \left(\delta_x \gibbs{\paramopt{1},t}^{kj} - \targ{0:t, \paramopt{1}} \right)
        \left(\gibbs{\paramopt{1},t}^{kj} - \gibbs{\paramopt{2},t}^{kj}\right)
        \left\{\gibbs{\paramopt{2},t}^{k(n-j-1)}
        \left[\noiseker{t}{1} \gradest - h(\paramopt{1})\right] -
        \targ{0:t, \paramopt{2}} \left[\noiseker{t}{1} \gradest - h(\paramopt{1})\right] \right\} \right|
        \\ &\leq \|L^P_{1}\|_{\infty} (1 - \mixrate{t}{\N})^{-1} \mixrate{t}{N}^{k j} \mixrate{t}{N}^{k(n-j-1)}\| \noiseker{t}{1} \gradest - h(\paramopt{1}) \|_\infty \| \paramopt{1} - \paramopt{2} \|\\
        &\leq  \constbiasalg \|L^P_{1}\|_{\infty} (1 - \mixrate{t}{\N})^{-1}  \mixrate{t}{N}^{k(n-1)} \| \paramopt{1} - \paramopt{2} \|
        \eqsp,
    \end{align*}
    where the last inequality is due to \Cref{theo:bias-mse-rolling}.
    Therefore, the first term of the right side of \eqref{eq:lipP} is upper bounded by $ \constbiasalg \|L^P_{1}\|_{\infty} (1 - \mixrate{t}{\N})^{-1}  n \kappa_{N, t}^{k(n-1)} \| \paramopt{1} - \paramopt{2} \|$. The second term of \eqref{eq:lipP} can be written
    \begin{multline*}
        - \left( \delta_x \gibbs{\paramopt{2},t}^{kn}  \noiseker{t}{2} \gradest - \targ{0:t, \paramopt{2}} \noiseker{t}{2} \gradest\right) + \left( \delta_x \gibbs{\paramopt{2},t}^{kn}  \noiseker{t}{1} \gradest - \targ{0:t, \paramopt{2}} \noiseker{t}{1} \gradest\right)\\ = \left( \delta_x \gibbs{\paramopt{2},t}^{kn}   - \targ{0:t, \paramopt{2}} \right) \left(\noiseker{t}{1} \gradest -\noiseker{t}{2} \gradest\right)\eqsp,
    \end{multline*}    
    and using again the ergodicity of $\gibbs{\paramopt{},t}{}$ and the fact that $\paramopt{} \mapsto \noiseker{t}{}\gradest$  is uniformly Lipschitz by \cref{assumption:lyapunov_smoothness}(iv), we may conclude that it is upper bounded by $\|L^P_2\|_{\infty}\kappa_{N, t}^{kn} \| \paramopt{1} - \paramopt{2} \|$.
    Finally, for the last term, using the facts that $K^k _{\paramopt{},t}$ is $\targ{0:t,\paramopt{}}{}$-invariant and geometrically ergodic
    and that $\paramopt{} \mapsto \targp{0:t}{}$ is Lipschitz by \cref{assumption:lyapunov_smoothness}(iv) yields
    \begin{align*}
       \lefteqn{\left| \targ{0:t, \paramopt{1}} \left( \gibbs{\paramopt{2},t}^{kn}  \noiseker{t}{1} \gradest - \targ{0:t, \paramopt{2}} \noiseker{t}{1} \gradest \right) \right|} \hspace{10mm}
        \\ 
        &= \left| \left( \targ{0:t, \paramopt{1}} - \targ{0:t, \paramopt{2}}\right) \left\{ \gibbs{\paramopt{2},t}^{kn}  \left[\noiseker{t}{1} \gradest - h(\paramopt{1}) \right] - \targ{0:t, \paramopt{2}}  \left[\noiseker{t}{1} \gradest - h(\paramopt{1}) \right] \right\} \right|
        \\ 
        &\leq \targlip \kappa_{N, t}^{kn} \| \noiseker{t}{1} \gradest - h(\paramopt{1}) \|_\infty \| \paramopt{1} - \paramopt{2} \|
        \\ &\leq  \targlip \constbiasalg (1 - \kappa_{N, t})^{-1} \kappa_{N, t}^{kn}\| \paramopt{1} - \paramopt{2} \|\eqsp.
    \end{align*}
    Therefore, we have that
    \begin{multline*}
         \delta_x \gibbs{\paramopt{1},t}^{kn} \left( \noiseker{t}{1} \gradest - \targ{0:t, \paramopt{1}} \noiseker{t}{1} \gradest \right) - \delta_x \gibbs{\paramopt{2},t}^{kn} \left( \noiseker{t}{2} \gradest - \targ{0:t, \paramopt{2}} \noiseker{t}{2} \gradest \right)
         \\
         \leq \left\{ \constbiasalg \|L^P_{1}\|_{\infty} (1 - \mixrate{t}{\N})^{-1}   n \kappa_{N, t}^{k(n-1)}
        + \left[\|L^P_2\|_{\infty} + \targlip \constbiasalg (1 - \kappa_{N, t})^{-1}\right] \kappa_{N, t}^{kn} \right\} \| \paramopt{1} - \paramopt{2} \| \eqsp.
    \end{multline*}
    Therefore, we obtain
    \begin{align*}
        \lefteqn{\left| \noiseker{t}{1} \gradestP{1}(x) -  \noiseker{t}{2} \gradestP{2}(x) \right|} \\
        & \leq \left| \delta_x \noiseker{t}{1} \gradest - \delta_x \noiseker{t}{2} \gradest \right| + \left| \targ{0:t, \paramopt{1}} \noiseker{t}{1}\gradest - \targ{0:t, \paramopt{2}} \noiseker{t}{2} \gradest \right| \\
        & \hspace{1cm}+ \left |\sum_{n = 1}^\infty \delta_x \gibbs{\paramopt{1},t}^{kn} \left( \noiseker{t}{1} \gradest - \targ{0:t, \paramopt{1}} \noiseker{t}{1} \gradest \right) - \delta_x \gibbs{\paramopt{2},t}^{kn} \left( \noiseker{t}{2} \gradest - \targ{0:t, \paramopt{2}} \noiseker{t}{2} \gradest \right) \right | \\
        &\leq \left| \delta_x \noiseker{t}{1} \gradest - \delta_x \noiseker{t}{2} \gradest \right| + \left| \targ{0:t, \paramopt{1}} \noiseker{t}{1}\gradest - \targ{0:t, \paramopt{2}} \noiseker{t}{2} \gradest \right| \\
        & + \bigg\{ \constbiasalg \|L^P_{1}\|_{\infty} (1 - \mixrate{t}{\N})^{-1}  (1 - \mixrate{t}{\N}^{k})^{-2} \\
        &+ \left[\|L^P_2\|_{\infty}  + \targlip \constbiasalg (1 - \kappa_{N, t})^{-1}\right] \mixrate{t}{\N}^{k} (1 - \mixrate{t}{\N}^{k})^{-1}\bigg\} \|\paramopt{1} - \paramopt{2} \| \eqsp .
    \end{align*}
    To conclude, note that by \cref{assumption:lyapunov_smoothness}(iv), $\left\| \delta_x \noiseker{t}{1} \gradest - \delta_x \noiseker{t}{2} \gradest \right\| \leq \|L^{P}_2\|_{\infty} \|\paramopt{1} - \paramopt{2} \|$. Furthermore, note that by \Cref{thm:unbiasedness} we obtain that for all $\paramopt{} \in \paramspace$, $\targ{0:t, \paramopt{}} \noiseker{t}{}\gradest = \targ{0:t, \paramopt{}} \score{}{0:t} = \nabla V(\paramopt{})$.
    Therefore, by \cref{assumption:lyapunov_smoothness}(i) we obtain that $\left\| \targ{0:t, \paramopt{1}} \noiseker{t}{1}\gradest - \targ{0:t, \paramopt{2}} \noiseker{t}{2} \gradest \right\| \leq \lyapunovlip \|\paramopt{1} - \paramopt{2} \|$, concluding the proof.
\end{proof}

\paragraph{Proof that \cref{assumption:nonas:bound_mse} holds. }

\Cref{assumption:nonas:bound_mse} is simply a bound on the MSE of the roll-out \PPG\ estimator, given by \Cref{theo:bias-mse-rolling}. 

\paragraph{Proof that \cref{assumption:nonas:stochastic_update_bounded} holds. }
\begin{proposition}
    \label{prop:gradestp_bound}
    For all $\paramopt{} \in \paramspace$ and all $\ell \in \intvect{1}{t-1}$  
    \begin{equation*}
         \CPE{\|\gradestP{}\|}{\partfilt{\ell}} \leq 2 \|\score{}{0:t}\|_{\infty} +  \constbiasalg (1 - \mixrate{t}{\N}^{k})^{-1} \eqsp .
    \end{equation*}
\end{proposition}
\begin{proof}
    Note that for all $x \in \esp{t}^{k - k_0}$ and all $\paramopt{} \in \paramspace$, 
    \begin{equation}
        \gradestP{}(x) = \gradest(x) - h(\paramopt{}) + \noiseker{t}{} \gradestP{}(x) \eqsp.
    \end{equation}
    \Cref{lem:boundness_p_gradestp} shows that $\|\noiseker{t}{} \gradestP{}\|_{\infty} \leq \constbiasalg (1 - \mixrate{t}{\N}^{k})^{-1}$. Note that $h(\paramopt{}) \leq \|\score{}{0:t}\|_{\infty}$
    We write
    \begin{align*}
        \CPE{\|\gradest\|}{\partfilt{\ell}} \leq \frac{1}{(k-k_0+1)N} \sum_{i = k_0}^k \sum_{j=1}^{N} \CPE{\|\stat{t, \ell}{j}[i]\|}{\partfilt{\ell}} \eqsp .
    \end{align*}
    By \Cref{prop:lip_cont_ppg}, $\CPE{\|\stat{t, \ell}{j}[i]\|}{\partfilt{\ell}} \leq \|\score{}{0:t}\|_{\infty}$, concluding the proof.
\end{proof}

\Cref{assumption:nonas:stochastic_update_bounded} follows directly by \Cref{prop:gradestp_bound} and by considering $\sup_{\paramopt{} \in \paramspace} \|\score{}{0:t}\|_{\infty}$.

\subsubsection{Proof of \Cref{theo:bound_grad}}
We have shown in \Cref{sec:appendix:learning:bound_grad:assumptions_check} that under \Cref{assumption:lyapunov_smoothness,assumption:strong_mixing_uniform}, it is possible to apply \Cref{thm:non_ass}. To conclude the proof of \Cref{theo:bound_grad} we just have to rearrange the constants.
We start by rewriting the constant in \Cref{theo:lip_cont_phhat}
\begin{equation*}
    \kergradestplip = C_1 + \constbiasalg (1 - \mixrate{t}{\N})^{-1}(1 - \mixrate{t}{\N}^{\ki})^{-1} C_2,
\end{equation*}
with 
\begin{align*}
    C_1 &= \left\|L_2^P\right\|_{\infty}\left[1 + \mixrate{t}{\N}^{k}(1 - \mixrate{t}{\N}^{k})^{-1}\right] + \lyapunovlip \\
    C_2 &= \left\|L_1^P\right\|_\infty (1-\mixrate{t}{\N}^{k})^{-1} + \targlip\mixrate{t}{\N}^{k} \eqsp .
\end{align*}
By \eqref{eq:non_ass:c_gamma} and \Cref{lem:boundness_p_gradestp},
\begin{align*}
    C_{\gamma} &= \boundmse \kergradestplip + (1 + \boundmse)\lyapunovlip\kergradestpbound \\
    &= \boundmse \left[ C_1 + \constbiasalg (1 - \mixrate{t}{\N})^{-1}(1 - \mixrate{t}{\N}^{\ki})^{-1} C_2\right] + (1 + \boundmse)\lyapunovlip \constbiasalg (1 - \kappa_{N,t}^{k})^{-1} \\
    &= \boundmse C_1 + \boundmse \constbiasalg (1 - \kappa_{N,t}^{k})^{-1}\left[\lyapunovlip  + (1 - \mixrate{t}{\N})^{-1} C_2\right] + \constbiasalg\lyapunovlip (1 - \kappa_{N,t}^{k})^{-1} \eqsp .
\end{align*}
Therefore,
\begin{align*}
    C_{0, \gamma} &\eqdef \boundmse^2 \lyapunovlip + C_{\gamma} \\
    &= \boundmse^2 \lyapunovlip + \boundmse C_1 + \boundmse \constbiasalg (1 - \kappa_{N,t}^{k})^{-1}\left[\lyapunovlip  + (1 - \mixrate{t}{\N})^{-1} C_2\right] + \constbiasalg\lyapunovlip (1 - \kappa_{N,t}^{k})^{-1} \eqsp .
\end{align*}
In the same way, we can rewrite \eqref{eq:non_ass:c_h} as
\begin{align*}
        C_{h} & = \kergradestplip \left[(a+1)/2 + a \boundmse \right] + (\lyapunovlip + a' + 1) \kergradestpbound \\
        &= \left[ C_1 + \constbiasalg (1 - \mixrate{t}{\N})^{-1}(1 - \mixrate{t}{\N}^{\ki})^{-1} C_2 \right]\left[(a+1)/2 + a \boundmse \right] + (\lyapunovlip + a' + 1) \constbiasalg (1 - \mixrate{t}{\N}^{\ki})^{-1} \eqsp.
\end{align*}
The constant $C_0$ from \Cref{theo:bound_grad} is $\boundgradestp = 2\sup_{\paramopt{} \in \paramspace} \|\score{}{0:t}\|_{\infty} + \constbiasalg (1 - \kappa_{N, t}^k)^{-1}$ which completes the proof.

\subsection{ Conditions on the model to verify \Cref{assumption:lyapunov_smoothness}}
\label{sec:appendix:learning:model_conditions}

In our specific application to score ascent, we work with the following assumptions.

\begin{assumption}[Lipschitz]
    \label{assumption:problem_lipschitz}
    \begin{itemize}[label=(\roman*),nosep]
        \item[(i)] For all $t \in \nset$, there exists $L^s _t \in \mathsf{M}(\xsp{t:t+1})$ such that for all $(x_{t}, x_{t+1}) \in \xsp{t:t+1}$, the function $\paramopt{} \mapsto \score{}{t}(x_t, x_{t+1})$ is $L^s _t(x_t, x_{t+1})$-Lipschitz and $\xsp{t:t+1} \ni (x_t, x_{t+1}) \mapsto \score{}{t}(x_t, x_{t+1})$ is bounded by $\| s_t(\paramopt{}) \|_\infty$ for all $\paramopt{} \in \paramspace$. Furthermore, $\|\scorelip{k}\|_{\infty} < \infty$.
        \item[(ii)] For all $t \in \nset$, there exists $\udlip{t} \in \mathsf{\xsp{t:t+1}}$ such that $\|\udlip{t}\|_{\infty} < \infty$ and that for all $(x_t, x_{t+1}) \in \xsp{t:t+1}$, $\paramopt{} \mapsto \ud{t, \paramopt{}}(x_t, x_{t+1})$ is $\udlip{t}(x_t, x_{t+1})$-Lipschitz.
    \end{itemize}
\end{assumption}


\begin{lemma}[\Cref{assumption:nonas:lyapunov_l_lipschitz}(i) holds]
\label{lem:V:smooth}
Assume \cref{assumption:strong_mixing_uniform} and \cref{assumption:lyapunov_smoothness}. There exists a constant $\lyapunovlip$ such that the Lyapunov function $V$ satisfies, for all $(\paramopt{1}$, $\paramopt{2}) \in \paramspace^2$,
$$
\| \nabla V(\paramopt{1}) - \nabla V(\paramopt{2}) \|\leq \lyapunovlip\|\paramopt{1} - \paramopt{2}\|.
$$
\end{lemma}

\begin{proof}
For all $\paramopt{1}$, $\paramopt{2}$,
\begin{align*}
    \| \nabla V(\paramopt{1}) - \nabla V(\paramopt{2}) \| 
    &= \|\targp{0:t}{1}(\score{1}{0:t}) - \targp{0:t}{2}(\score{2}{0:t}) \| \\
    &\leq \|\targp{0:t}{1}(\score{1}{0:t}) - \targp{0:t}{1}(\score{2}{0:t}) \| + \|\targp{0:t}{1}(\score{2}{0:t}) - \targp{0:t}{2}(\score{2}{0:t})\|\,.
\end{align*}
By \eqref{assumption:strong_mixing} and by \cite[Theorem 4.10]{gloaguen:lecorff:olsson:2022} there exists a constant $c$ such that
$$
\|\targp{0:t}{1}(\score{2}{0:t}) - \targp{0:t}{2}(\score{2}{0:t})\|
\leq c t  \|\paramopt{1} - \paramopt{2} \|\operatorname{sup}_{\paramopt{}}\operatorname{sup}_{k} \|s_{k}(\paramopt{}) \|_{\infty}\,,
$$
Using \cref{assumption:strong_mixing} and \cref{assumption:lyapunov_smoothness}[i], we can write:
\begin{align*}
    \|\targp{0:t}{1}(\score{1}{0:t}) - \targp{0:t}{1}(\score{2}{0:t}) \| &\leq \sum_{u=0}^{t-1} \targp{0:t}{1}\left[\|\score{1}{u}(x_{u:u + 1}) - \score{2}{u}(x_{u:u + 1}) \right\|],\\
    &\leq \sum_{u=0}^{t-1} \targp{0:t}{1} \left[\scorelip{u}(x_{u:u + 1})\right] \|\paramopt{1} - \paramopt{2}\|,\\
    &\leq \frac{\sigma_+}{\sigma_-}\operatorname{sup}_{u \in \intvect{0}{t-1}} \left[\scorelip{u}\right]\|\paramopt{1} - \paramopt{2}\|t.
\end{align*}
\end{proof}

\begin{theorem}[Lipschitz continuity of Particle Gibbs with Backward Sampling]
    \label{theo:lip_cont_ppg}
    Assume \cref{assumption:problem_lipschitz}.
    For every $t \in \nset$, $\paramopt{} \in \Theta$ and $\N \in \nsetpos$
    \begin{equation*}
        \sup_{x_{0:t} \in \xsp{0:t}}\tv{\gibbs{\paramopt{1}, t}(x_{0:t}, .)}{\gibbs{\paramopt{2}, t}(x_{0:t}, .)} \leq \gibbslip{t,N}\|\paramopt{1} - \paramopt{2}\| \,,
    \end{equation*}
    where 
    \begin{equation} 
        \label{eq:lip:gibbs}
        \gibbslip{t,N} \eqdef  \sum_{\ell=0}^{t-1} \pothigh{\ell}^{-1}\left[\mdhigh{\ell}^{-1} + (N-1) \right] \|\udlip{\ell}\|_{\infty} \eqsp. 
    \end{equation} 
\end{theorem}
\begin{proof}
    We know that $\gibbs{\paramopt{}, t} = \mbjt{m, \paramopt{}} \bdpart{t, \paramopt{}}$.
    Therefore, by \Cref{lem:mk:lip_comp,lem:lip:mbjt,lem:lip:bdpart}, we have that $\gibbs{\paramopt{}, t}$ is Lipschitz with constant equals $\mbjtlip{t} + \sup_{\paramopt{}} \mbjt{t, \paramopt{}} \bdpartlip{t}$.
\end{proof}

\begin{corollary}[\Cref{assumption:lyapunov_smoothness}(iii) holds.]
    \label{cor:lip:gibbs:iterate}
    Assume \cref{assumption:problem_lipschitz}. For every $t \in \nset$, $\paramopt{} \in \Theta$, $r \in \nsetpos$ and $\N \in \nsetpos$ such that $N > 1 + 5 \rho_t^2 t / 2$
    \begin{equation*}
        \sup_{x_{0:t} \in \xsp{0:t}}\tv{\gibbs{\paramopt{1}, t}^r(x_{0:t}, .)}{\gibbs{\paramopt{2}, t}^r(x_{0:t}, .)} \leq 
        \noisekerlip{t, N} \|\paramopt{1} - \paramopt{2}\| \,
    \end{equation*}
where 
\begin{equation} 
    \label{eq:lip:noisekerlip}
\noisekerlip{t,N} \eqdef (1 - \kappa_{t,N})^{-1}\| \gibbslip{t,N} \|_\infty 
\end{equation}
where $\gibbslip{t, N}$ is defined in \eqref{eq:lip:gibbs}.
\end{corollary}
\begin{proof}
    Under \ref{assumption:strong_mixing_uniform}, the Particle Gibbs with backward sampling is geometrically ergodic with contraction rate $\kappa_{t,N}$ and thus $\gibbslip{t,N}$ is bounded and the result follows from \Cref{lem:lip:ergodic}
\end{proof}
\begin{corollary}[\Cref{assumption:lyapunov_smoothness}(i)]
\label{lem:targ_lip}
Assume \cref{assumption:strong_mixing_uniform} and  \cref{assumption:problem_lipschitz}. For all $t \in \nsetpos$, $(\paramopt{0}$, $\paramopt{1}) \in \paramspace^2$,
$$
\tv{\targ{0:t, \paramopt{0}}}{\targ{0:t, \paramopt{1}}}\leq \targlip\|\paramopt{0} - \paramopt{1}\|,
$$
where
\begin{equation}
    \label{eq:lip:targ}
\targlip \eqdef \noisekerlip{t, N^*} \eqsp,
\end{equation}
and $\noisekerlip{t,N}$ is defined in \eqref{eq:lip:noisekerlip} and $N^* = \lceil 1 + 5 \rho^2 _t / 2\rceil.$
\end{corollary}
\begin{proof}
    Consider the following decomposition, valid for all $k \in \nset^{*}$ and $N \geq 1 + 5 \rho^2 _t / 2$, and all $x_{0:t} \in \xsp{0:t}$,
    \begin{align*} 
        \tv{\targp{0:t}{1}}{\targp{0:t}{2}} & \leq \tv{\targp{0:t}{1}}{\gibbs{\paramopt{1}, t}^k(x_{0:t},\cdot)} + \tv{\targp{0:t}{2}}{\gibbs{\paramopt{2}, t}^k(x_{0:t},\cdot)} + \tv{\gibbs{\paramopt{1}, t}^k(x_{0:t},\cdot)}{\gibbs{\paramopt{2}, t}^k(x_{0:t},\cdot)} \\
        & \leq \tv{\targp{0:t}{1}}{\gibbs{\paramopt{1}, t}^k(x_{0:t},\cdot)} + \tv{\targp{0:t}{2}}{\gibbs{\paramopt{2}, t}^k(x_{0:t},\cdot)} + L^P _{t,N} \| \paramopt{1} - \paramopt{2} \| \eqsp,
    \end{align*}
    where we applied \Cref{cor:lip:gibbs:iterate}. Since the Lipschitz constant of $\gibbs{\paramopt{}, t}$ is independent of $k$, and $\gibbs{\paramopt{}, t}$ is geometrically ergodic for all $\paramopt{}$, we obtain by taking the limit when $k$ goes to infinity with $N$ fixed,
    \[ 
        \tv{\targp{0:t}{1}}{\targp{0:t}{2}} \leq \frac{\| \gibbslip{t,N} \|_\infty}{1 - \kappa_{t,N}}\| \paramopt{1} - \paramopt{2} \| \eqsp,
    \]
    for all $N \geq 1 + 5 \rho^2 _t / 2$, where the dependence in $N$ is hidden in $L^P _{t,N}$. The result follows by choosing $N = \lceil 1 + 5 \rho^2 _t / 2 \rceil$.
\end{proof}
\begin{remark}
  As noted by \cite{lindholm2018}, the Lipschitz constant appearing in \Cref{cor:lip:gibbs:iterate} possesses an unexpected dependence on $N-1$. 
One would expect it not to be true, in that we know that $\parisgibbs{\paramopt{}, t}$ converges geometrically fast and uniformly to $\targ{0:t}{}$ and this is faster as $N$ gets bigger. Therefore, for large $N$ the Lipschitz constant is expected to converge to that of $\targ{0:t}$ whose Lipschitz constant is independent of $N$.
\end{remark}
\begin{proposition}[Lipschitz continuity of $\paramopt{} \mapsto \parisgibbs{\paramopt{}, t}\occm(\statmb{t})(\operatorname{id})$]
    \label{prop:lip_parisgibbs_mean}
     Assume \cref{assumption:problem_lipschitz}. For every $t \in \nset$, $\paramopt{} \in \Theta$ and $\N \in \nsetpos$,
    \label{prop:parisgibbs:}
    \[
     \left\|  \parisgibbs{\paramopt{1}, t}\occm(\statmb{t})(\operatorname{id}) -  \parisgibbs{\paramopt{2}, t}\occm(\statmb{t})(\operatorname{id}) \right\|_\infty \leq
       \parisgibbslip{t} \| \paramopt{1} - \paramopt{2} \| \eqsp,
      \]
    where 
    \begin{equation}
        \label{eq:lip:parisgibbs}
        \parisgibbslip{t} \eqdef (N - 1) \sum_{\ell=0}^{t-1}\pothigh{\ell}\|\udlip{\ell}\|_{\infty} + \sum_{j=1}^{m} \| \bklip{j} \| _\infty \left[\sum_{\ell=0}^{m-1}\supscore{\ell}\right] + \sum_{j=1}^{m}\| \scorelip{j}\|_\infty    \eqsp.
    \end{equation}
\end{proposition}
\begin{proof}
    Consider $e = (x_{0:t}, \ymb{0:t}) \in \esp{t}$ and $f_{\paramopt{}}(e) \eqdef \int \ckjt{m, \paramopt{}}(x_{0:t}, \rmd \ymbtd{t}) \occm(\statlmb{t})(id)$.
    Then $\parisgibbs{\paramopt{}, t}\occm(\statlmb{t})(id) = \mbjt{m, \paramopt{}} f_{\paramopt{}}(x_{0:t})$ is a composition of a Markov kernel and a Lipschitz function, therefore Lipschitz.
\end{proof}

\begin{corollary}[\Cref{assumption:lyapunov_smoothness}(iv) holds.]
    \label{prop:lip_ph}
    Assume \cref{assumption:problem_lipschitz}. For every $t \in \nset$, $\paramopt{} \in \Theta$ and $\N \in \nsetpos$
    \label{prop:noisekergradestlip}
    \[
    \sup_{x_{0:t} \in \xsp{0:t}} \left\| \noiseker{t}{1}\gradest - \noiseker{t}{2} \gradest \right\| \leq
       L^{P}_{2}\| \paramopt{1} - \paramopt{2} \| \eqsp,
      \]
    where \begin{equation}
    \label{eq:lip:PH}
    L^{P}_{2}= \noisekerlip{t, N} + \parisgibbslip{t} \eqsp,
    \end{equation}
    with $\noisekerlip{}$ and $\parisgibbslip{t}$ are defined in \eqref{eq:lip:parisgibbs} and \eqref{eq:lip:noisekerlip}.
\end{corollary}
\begin{proof}
    Let $\tilde{f}: \esp{}^{\ki - \ki_0} \ni (x_{0:t}[\ki_0:\ki], \xmb{0:t|t}[\ki_0:\ki], \statlmb{t}[\ki_0:\ki]) \mapsto (\ki - \ki_0)^{-1} \sum_{\ell=\ki_0 + 1}^{\ki} \occm(\statlmb{t}[\ell])(\operatorname{id})$.
    As $ \parisgibbs{\paramopt{}, t}$ depends only on the path, with a slight abuse of notation, we can define $f_{\paramopt{}}(x_{0:t}) \eqdef  \parisgibbs{\paramopt{}, t}^{\tensprod \ki - \ki_0}(\tilde{f})(x_{0:t})$.
    By \cref{prop:lip_parisgibbs_mean}, we have that $f_{\paramopt{}}$ is Lipschitz with $L^f = \parisgibbslip{t}$.
    Note that $\noiseker{t}{}\gradest(x_{0:t}, \ymb{t}) = \gibbs{\paramopt{}, t}^{k_0}f_{\paramopt{}}(x_{0:t})$, therefore, by \cref{lem:mk:lip_comp} Lipschitz with constant $\noisekerlip{} + \parisgibbslip{t}$.
\end{proof}

\section{Lipschitz properties}
\subsection{Lipschitz continuity of $\noiseker{}{}$}

In this section we prove the following items:
\begin{itemize}
    \item $\mbjt{m, \paramopt{}}(z_{0:m}, \cdot)$ is Lipschitz, see \Cref{sec:Clip}
    \item $\bdpart{m,\paramopt{}}(\xmb{0:m}, \cdot)$ is Lipschitz, see \Cref{sec:Blip}
    \item $\int\ckjt{m, \paramopt{}}(\xmb{0:m}, \rmd \statlmb{m})\occm(\statlmb{m})(\mathrm{Id})$ is Lipschitz, see \Cref{sec:Slip}
\end{itemize}
The following technical lemma will be useful.
\begin{lemma}
    \label{lem:prod_lip_bound}
    Let $\alpha \in ]0, 1]$, $x \in \rsetpos$ and $\ell \in \nset$. Then for all $\lambda_{i} \in \rsetpos$, $i \in \intvect{0}{\ell}$, such that $\alpha \geq \prod_{i=0}^{\ell}(1 - \lambda_{i} x)$ it holds that $\alpha \geq 1 - x \sum_{i=0}^{\ell} \lambda_{i}$.
\end{lemma}
\begin{proof}
    Consider first the case where $x \lambda_i \leq 1$ for all $i \in \intvect{0}{\ell}$. We prove the result by induction. The case $\ell = 0$ is straightforward. Assume now that the result holds for some $r \in \intvect{0}{\ell - 1}$. Then, 
    \begin{multline*}
       \prod_{i=0}^{r+1}(1 - \lambda_{i} x) = (1 - \lambda_{r+1}x) \prod_{i=0}^{r}(1 - \lambda_{i} x) \geq (1 - \lambda_{r+1}x) (1 - x\sum_{i=0}^{r} \lambda_i ) \\
        = 1 - x\sum_{i=0}^{r+1} \lambda_i + x^2 \sum_{i=0}^{r} \lambda_i \lambda_{r+1}  \geq 1 - x\sum_{i=0}^{r+1} \lambda_i \eqsp .
    \end{multline*}
    Consider now the case where there is a index $j \in \intvect{0}{\ell}$ such that $x \lambda_j \geq 1$. Then $\alpha \geq 0 \geq  1 - (\sum_{i=0}^{\ell} \lambda_{i})x$.
\end{proof}
We begin with some important definitions.
Let $P$ and $Q$ be probability distributions on some common measurable space $(\xsp{}, \xfd{})$, and assume that these distributions admit densities $p$ and $q$ w.r.t some common reference measure $\lambda$. Let $\maxcoupling{P}{Q}$ denote a maximal coupling between $P$ and $Q$. As in \cite[Theorem 2]{lindholm2018}, it is possible to explicitly construct one such maximal coupling by
\begin{multline}
    \maxcoupling{P}{Q}(\rmd (x, y)) \eqdef \min\{p(x), g(x)\}\lambda(\rmd x) \delta_{x}(\rmd y) +\\
     \frac{\big[P(\rmd x) -  \min\{p(x), g(x)\}\lambda(\rmd x)\big] \big[Q(\rmd y) -  \min\{p(y), g(y)\}\lambda(\rmd y)\big]}{1 - \lambda\big(\min\{p, q\}\big)} \eqsp.
\end{multline}
From this definition it follows that for continuous and discrete dominating measures $\lambda$,
$$
    \int \indi{\{x = y\}}  \maxcoupling{P}{Q}\rmd (x, y) = \int \min\{p(x), g(x)\}\lambda(\rmd x) \,.
$$
Moreover, for two Markov transition kernels $K_1$ and $K_2$ on $(\xsp{}, \xfd{})$, which are assumed to admit transition densities with respect to some common dominating measure, we let, for $(x_1, x_2) \in \xsp{}^2$, $\maxcoupling{K_1}{K_2}((x_1, x_2), \cdot)$ denote the maximal coupling between the measures $K_1(x_1, \cdot)$ and $K_2(x_2, \cdot)$. Defined in this way, $\maxcoupling{K_1}{K_2}$ defines a Markov transition kernel on the product space $(\xsp{}^2, \xfd{}^{\varotimes 2})$
 
The following Lemma will be crucial in what follows. 

\begin{lemma}
    \label{lem:coupling:joint}
    \begin{itemize}
        \item[(i)] Let $(\mu_1, \mu_2)$ be two probability measures admitting a density with respect to a common dominating measure and let $(K_1, K_2)$ two Markov transition kernels also admitting transition densities with respect to some dominating measure. 
        Then the probability measure 
        \[ \maxcoupling{\mu_1}{\mu_2} \maxcoupling{K_1}{ K_2}(\rmd (x_1, x_2)) = \int \maxcoupling{\mu_1}{\mu_2}(\rmd (z_1, z_2))  \maxcoupling{K_1}{K_2}((z_1, z_2), \rmd (x_1, x_2)),
        \]
    is a coupling of $(\mu_1 K_1, \mu_2 K_2)$, and it holds that 
    \begin{multline*}
        \int \1_{x_1 = x_2} \maxcoupling{\mu_1 K_1}{\mu_2 K_2}(\rmd (x_1, x_2)) \\
        \geq \int \int \1_{z_1 = z_2} \1_{x_1 = x_2} \maxcoupling{\mu_1}{\mu_2}(\rmd (z_1, z_2)) \maxcoupling{K_1}{K_2}((z_1, z_2), \rmd (x_1, x_2)).
    \end{multline*}
    \item[(ii)] Let $(\mu_1, \cdots, \mu_n)$ and $(\nu_1, \cdots, \nu_n)$ be probability measures such that for all $i \in \intvect{1}{n}$, $\mu_i$ and $\nu_i$ admit densities with respect to the same dominating measure. Then $\bigotimes_{i = 1}^n \maxcoupling{\mu_i}{\nu_i}$ is a coupling of $\bigotimes_{i = 1}^n \mu_i$ and $\bigotimes_{i = 1}^n \nu_i$, and thus 
    \begin{multline*}
        \int \prod_{i = 1}^n \1_{x_i = y_i} \maxcoupling{\bigotimes_{i = 1}^n \mu_i}{\bigotimes_{i = 1}^n \nu_i}(\rmd (x_1, \ldots, x_n, y_1, \ldots, y_n)) \\
        \geq \int \prod_{i = 1}^n \1_{x_i = y_i} \bigotimes_{i = 1}^n \maxcoupling{\mu_i}{\nu_i}(\rmd (x_1, \ldots, x_n, y_1, \ldots, y_n)).
    \end{multline*}
    \end{itemize}
\end{lemma}
\begin{proof}
    It is enough to show that $\maxcoupling{\mu_1}{\mu_2} \maxcoupling{K_1}{K_2}$ admits $\mu_1K_1$ and $\mu_2 K_2$ as marginal distributions. This follows immediately from the fact that $\maxcoupling{\mu_1}{\mu_1}$ and $\maxcoupling{K_1}{K_2}$ admit the right marginal distributions; indeed, 
    \begin{align*} 
        \maxcoupling{\mu_1}{\mu_2}& \maxcoupling{K_1}{K_2}(\xsp{} \times A) \\
        & = \int \maxcoupling{\mu_1}{\mu_2}(\rmd z_1, \rmd _2) \, \maxcoupling{K_1}{K_2}(z_1, z_2, \rmd( x_1, x_2)) \, \1_{\xsp{} \times A}(x_1, x_2) \1_{\xsp{}^2}(z_1, z_2)\\
        & = \int \maxcoupling{\mu_1}{\mu_2}(\rmd z_1, \rmd _2) K_2(z_2, A) \\
        & = \int \mu_2(\rmd z_2) K_2(z_2, A) \\ 
        &= \mu_2 K_2(A).
    \end{align*}
    The derivation for the first marginal distribution follows similarly. For the second point, since $\maxcoupling{\mu_1}{\mu_2} \maxcoupling{K_1}{K_2}$ is a coupling of $(\mu_1 K_1, \mu_2 K_2)$ and $\maxcoupling{\mu_1 K_1}{\mu_2 K_2}$ is the maximal coupling, we have that
    \begin{align*}
        \int \1_{x_1 = x_2} & \maxcoupling{\mu_1 K_1}{\mu_2 K_2}(\rmd (x_1, x_2)) \\
        & \geq \iint \1_{x_1 = x_2} \maxcoupling{\mu_1}{\mu_2}(\rmd (z_1, z_2)) \, \maxcoupling{K_1}{K_2}(z_1, z_2;\rmd (x_1, x_2)) \\
        & \geq \iint \1_{x_1 = x_2} \1_{z_1 = z_2} \maxcoupling{\mu_1}{\mu_2}(\rmd (z_1, z_2)) \, \maxcoupling{K_1}{K_2}(z_1, z_2;\rmd (x_1, x_2)).
    \end{align*}
    The proof of the second item follows similarly.
\end{proof}
\subsubsection{$\paramopt{} \mapsto \mbjt{m, \paramopt{}}$ is Lipschitz.}
\label{sec:Clip}
We proceed by a coupling method that is inspired by \cite[Theorem 2]{lindholm2018}.
The coupling we consider is that where the \emph{selection} and \emph{mutation} steps of the particle filter are respectively coupled maximally.
\begin{algorithm}[H]
\KwData{$\paramopt{1}$, $\paramopt{2}$, $\zeta_{0:m}$}
\KwResult{$\xmb{0:m, 1}$, $\xmb{0:m, 1}$}
draw $\xmb{0, 1}, \xmb{0, 2} \sim \maxcoupling{\initmb[\zeta_0]}{\initmb[\zeta_0]}$\;

\For{$s \gets 1$ \KwTo $t$}
{
draw $(\xmb{s, 1}, \xmb{s, 2}) \sim \maxcoupling{\mkmb{s-1,\paramopt{1}}[\zeta_{s}](\xmb{s-1, 1}, \cdot)}{\mkmb{s-1, \paramopt{2}}[\zeta_{s}](\xmb{s-1, 2}, \cdot)}$\; \label{line:couplingPF}
}
\caption{Coupling $\mbjt{m, \paramopt{}}$} \label{alg:bootstrap:particlefilter}
\end{algorithm}

First, let us prove that the one step \emph{selection}--\emph{mutation} kernel is Lipschitz.
\begin{lemma}
    \label{lem:coupling:bootstrap}
    For all $t \in \nset$, $\xmb{t-1} \in \xspmb{t-1}$ and $(\paramopt{1}, \paramopt{2}) \in \paramspace^2$, 
\begin{equation}
    \int \indi{\{x_1 = x_2\}} \maxcoupling{\pd{t-1, \paramopt{1}}(\occm(\xmb{t-1}))}{\pd{t-1, \paramopt{2}}(\occm(\xmb{t-1}))}(\rmd (x_1, x_2))  
    \geq  1 - \frac{\sum_{i = 1}^N \lambda_t\big(\udlip{t-1}(\epartvar{t-1}{i}, \cdot) \big)}{\N\pothigh{n}}\|\paramopt{1} - \paramopt{2}\|.
\end{equation}
\end{lemma}
\begin{proof}
    By \Cref{assumption:strong_mixing}(i) and \Cref{assumption:lyapunov_smoothness}(iii),
    \begin{align*}
          \int \indi{\{x_1 = x_2\}} &\maxcoupling{\pd{t-1, \paramopt{1}}(\occm(\xmb{t-1}))}{\pd{t-1, \paramopt{2}}(\occm(\xmb{t-1}))}(\rmd (x_1, x_2)) \\
        & = \int  \min\left( \sum_{i = 1}^N \frac{\ud{t-1, \paramopt{1}}(\epartvar{t-1}{i}, x)}{\sum_{j = 1}^N \pot{t-1, \paramopt{1}}(\epartvar{t-1}{j})}, \sum_{i = 1}^N \frac{\ud{t-1, \paramopt{2}}(\epartvar{t-1}{i}, x)}{\sum_{j = 1}^N \pot{t-1,\paramopt{2}}(\epartvar{t-1}{j})}\right) \lambda_t(\rmd x) \\
        & \geq \sum_{j=1}^N \int \min\left(\frac{\ud{t-1, \paramopt{1}}(\epartvar{t-1}{i}, x)}{\sum_{j=1}^{N}\pot{t-1, \paramopt{1}}(\epartvar{t-1}{j})}, \frac{\ud{t-1, \paramopt{2}}(\epartvar{t-1}{i}, x)}{\sum_{j=1}^{N}\pot{t-1, \paramopt{2}}(\epartvar{t-1}{j})} \right) \lambda_t(\rmd x) \\
        & \geq \frac{1}{\sum_{j=1}^N \max\left(\pot{t-1, \paramopt{1}}(\epartvar{t-1}{j}), \pot{t-1, \paramopt{2}}(\epartvar{t-1}{j})\right)}\sum_{j=1}^N \int \min\left(\ud{t-1, \paramopt{1}}(\epartvar{t-1}{j}, x), \ud{t-1, \paramopt{2}}(\epartvar{t-1}{j}, x)\right) \lambda_t(\rmd x)\\
        &\geq \frac{\sum_{j=1}^N \max\left(\pot{t-1, \paramopt{1}}(\epartvar{t-1}{j}), \pot{t-1, \paramopt{2}}(\epartvar{t-1}{j})\right) -  \sum_{i = 1}^N \lambda_t \left( \udlip{t-1}(\epartvar{t-1}{i}, \cdot) \right) \|\paramopt{1} - \paramopt{2}\|}{\sum_{j=1}^N \max\left(\pot{t-1, \paramopt{1}}(\epartvar{t-1}{j}), \pot{t-1, \paramopt{2}}(\epartvar{t-1}{j})\right)} \\
        & \geq 1 - \frac{\sum_{i = 1}^N \lambda_t\big(\udlip{t-1}(\epartvar{t-1}{i}, \cdot)\big)}{\N\pothigh{n}}\|\paramopt{1} - \paramopt{2}\|,
    \end{align*}
    where we have used that 
    \begin{align*}
    \int \max(\ud{t-1, \paramopt{1}}(\epartvar{t-1}{i}, x), \ud{t-1, \paramopt{2}}(\epartvar{t-1}{i},x)) \lambda_t(\rmd x) & \geq \max\left(\int \ud{t-1, \paramopt{1}}(\epartvar{t-1}{i}, x) \lambda_t(\rmd x), \int \ud{t-1, \paramopt{2}}(\epartvar{t-1}{i}, x) \lambda_t(\rmd x) \right) \\
    & \geq \max(\pot{t-1, \paramopt{1}}(\epartvar{t-1}{i}), \pot{t-1, \paramopt{2}}(\epartvar{t-1}{i})).
    \end{align*}
\end{proof}
\begin{lemma}
    \label{lem:lip:mkmb}
    For all $t \in \nset$, $\xmb{t-1} \in \xspmb{t-1}$, $z \in \xsp{t}$ and $(\paramopt{1}, \paramopt{2}) \in \paramspace^2$,
    \begin{equation*}
        \tv{\mkmb{t-1, \paramopt{1}}[z](\xmb{t-1}, \cdot)}{\mkmb{t-1, \paramopt{2}}[z](\xmb{t-1}, \cdot)} \leq \mkmblip{t-1}(\xmb{t-1}) \|\paramopt{1} - \paramopt{2}\|
    \end{equation*}
    where $\mkmblip{t-1}(\xmb{t-1})= (1-N^{-1})\pothigh{t-1}^{-1}\sum_{i = 1}^N \lambda_t \left( \udlip{t-1}(\epartvar{t-1}{i}, \cdot)\right)
    $.
\end{lemma}
\begin{proof} Let us denote by $\mathrm{U}\intvect{1}{n}$ the uniform distribution on $\intvect{1}{n}$.
    By definition of the kernel $\mkmb{t-1, \paramopt{}}[z]$, we have that 
    \[ 
        \mkmb{t-1, \paramopt{}}[z](\xmb{t-1}, \rmd \xmb{t}) = \int \mathrm{U}\intvect{1}{n}(\rmd j) \big\{  \Phi_{t-1}(\occm(\xmb{t-1}))^{\tensprod j}
\tensprod \delta_{z} \tensprod \Phi_{t-1}(\occm(\xmb{t-1}))^{\tensprod (\N - j - 1)} \big\} (\rmd \xmb{t})
    \]
    and thus, applying the two items of Lemma~\ref{lem:coupling:joint} combined with the fact that $\maxcoupling{\mu}{\mu}\big(\rmd (x_1, x_2) \big) = \mu(\rmd x_1) \delta_{x_1}(\rmd x_2)$ for any probability measure $\mu$, we get that
    \begin{align*}
        \int \indi{\{\xmb{t, 1} = \xmb{t, 2}\}} &\maxcoupling{\mkmb{t-1, \paramopt{1}}[z](\xmb{t-1}, \cdot)}{\mkmb{t-1, \paramopt{2}}[z](\xmb{t-1}, \cdot)} \rmd (\xmb{t, 1}, \xmb{t, 2}) \\
        & \geq \int \1_{\xmb{t,1} = \xmb{t,2}, i_1 = i_2} \maxcoupling{\mathrm{U}\intvect{1}{n}}{\mathrm{U}\intvect{1}{n}}\big(\rmd (i_1, i_2)\big) \\
        & \hspace{2cm} \times \maxcoupling{\pd{t-1, \paramopt{1}}(\occm(\xmb{t-1}))}{\pd{t-1, \paramopt{2}}(\occm(\xmb{t-1}))}^{\otimes i_1} \otimes \maxcoupling{\delta_z}{\delta_z} \\
        & \hspace{2cm} \otimes \maxcoupling{\pd{t-1, \paramopt{1}}(\occm(\xmb{t-1}))}{\pd{t-1, \paramopt{2}}(\occm(\xmb{t-1}))}^{\otimes N - i_1 - 1} \rmd (\xmb{t,1}, \xmb{t,2}) \\
        & = \frac{1}{N} \sum_{i = 1}^N \int \prod_{k = 1, k \neq i}^n \1_{x^i _{t,1} = x^i _{t,2}} \maxcoupling{\pd{t-1, \paramopt{1}}(\occm(\xmb{t-1}))}{\pd{t-1, \paramopt{2}}(\occm(\xmb{t-1}))}\big(\rmd (x^i _{t,1}, x^i _{t,2}) \big)\\
        &\geq \left(1 - \frac{\sum_{i = 1}^N \lambda_t\big( \udlip{t-1}(\epartvar{t-1}{i},\cdot) \big)}{\N\pothigh{t-1}}\|\paramopt{1} - \paramopt{2}\|\right)^{N -1} \\
        &\geq 1 - \frac{N-1}{\pothigh{t-1}\N}\sum_{i = 1}^N \lambda_t\big(\udlip{t-1}(\epartvar{t-1}{i}, \cdot) \big)\|\paramopt{1} - \paramopt{2}\| \eqsp.\\
    \end{align*}
    where we have applied Lemma~\ref{lem:coupling:bootstrap}
in the penultimate line and Lemma~\ref{lem:prod_lip_bound} in the last one.   
\end{proof}
\begin{lemma}
    \label{lem:lip:mbjt}
    For every $t \in \nsetpos$, there exists $\mbjtlip{t} \in \meas(\xfd{0:t})$ such that
    \begin{equation}
        \tv{\mbjt{t, \paramopt{1}}(z_{0:t})}{\mbjt{t, \paramopt{2}}(z_{0:t})} \leq \mbjtlip{t}(z_{0:t}) \| \paramopt{1} - \paramopt{2} \| \eqsp,
    \end{equation}
    where $\mbjtlip{t}(z_{0:t}) = \sup_{\paramopt{}} \mbjt{t, \paramopt{}}{}\left[\sum_{i=0}^{t-1} \mkmblip{i}{}\right](z_{0:t})$.
    Under \cref{assumption:problem_lipschitz}(i), we obtain that $\|\mbjtlip{t}\|_{\infty} \leq (N - 1) \sum_{\ell=0}^{t-1}\pothigh{\ell}\|\udlip{\ell}\|_{\infty}$.
\end{lemma}
\begin{proof}
    This is a direct application of \cref{lem:mk:lip_prod}.
\end{proof}

\subsubsection{$\paramopt{} \mapsto \bdpart{t, \paramopt{}}(\xmb{0:t}, \cdot)$ is Lipschitz}
\label{sec:Blip}

We start by recalling the definition of $\bdpart{m}{}$
\begin{equation} \label{eq:def:bdpart}
\bdpart{t, \paramopt{}} : \xspmb{0:t} \times \xfd{0:t} \ni (\xmb{0:t}, A) \mapsto \idotsint  \1_A(x_{0:t}) \left( \prod_{s = 0}^{t - 1} \bk{s,\occm(\xmb{s})}{}(x_{s + 1}, \rmd x_s) \right) \occm(\xmb{t})(\rmd x_t) \eqsp.
\end{equation}

\begin{lemma}
    \label{lem:coupling:bkw}
    For all $s \in \intvect{0}{t}$, $x_{t+1} \in \xsp{t+1}$, $\xmb{t} \in \xspmb{t}$ and $(\paramopt{1}, \paramopt{2}) \in \paramspace^2$
\begin{equation}
    \tv{\bk{s,\occm(\xmb{s}),\paramopt{1}}(x_{s+1}, \cdot)}{\bk{s,\occm(\xmb{s}), \paramopt{2}}(x_{s+1}, \cdot)} \leq \bklip{s}(x_{s+1}, \xmb{s}) \|\paramopt{1} - \paramopt{2}\| \eqsp.
\end{equation}
with $\bklip{s}{}(x_{s+1}, \xmb{s}) =  (\N\pothigh{t}\mdhigh{s})^{-1}\sum_{i = 1}^N \udlip{s}(x_s^i, x_{s+1})$.
    Under \cref{assumption:problem_lipschitz}(i), we have  $\|\bklip{m}\|_{\infty} =  (\pothigh{m}\mdhigh{m})^{-1}\|\udlip{m}\|_{\infty}$.
\end{lemma}
\begin{proof}
    Note that $\bk{t,\occm(\xmb{t})}{}(x_{t+1}, \cdot) = \sum_{\ell = 1}^\N \frac{\ud{t}{}(x_t^\ell, x_{t + 1})}{\sum_{\ell' = 1}^\N \ud{t}{}(x_t^{\ell'}, x_{t + 1})} \delta_{x_t^\ell}$. Therefore, similarly to the proof of \Cref{lem:coupling:bootstrap},
    \begin{align*}
         \int \indi{\{x_{t, 1} = x_{t, 2}\}} &  \maxcoupling{\bk{t,\occm(\xmb{t}),\paramopt{1}}(x_{t+1}, \cdot)}
         {\bk{t,\occm(\xmb{t}), \paramopt{2}}(x_{t+1}, \cdot)} \rmd (x_{t, 1}, x_{t, 2}) \\
        &\geq \frac{\sum_{\ell=1}^{N}\max(\ud{t, \paramopt{1}}(x_t^{\ell}, x_{t + 1}), \ud{t, \paramopt{2}}(x_t^{\ell}, x_{t + 1})) - \udlip{t}(x_t^{\ell}, x_{t+1}) \|\paramopt{1} - \paramopt{2}\|}{\sum_{\ell=1}^{N}\max(\ud{t, \paramopt{1}}(x_t^{\ell}, x_{t + 1}), \ud{t, \paramopt{2}}(x_t^{\ell}, x_{t + 1}))} \\
        & \geq 1 - \frac{\sum_{\ell=1}^{N}\udlip{t}(x_t^{\ell}, x_{t+1})}{\N\pothigh{t}\mdhigh{t}}\|\paramopt{1} - \paramopt{2}\| \eqsp.
    \end{align*}
\end{proof}

\begin{lemma}
    \label{lem:lip:bdpart}
    For all $t \in \nset$, $\xmb{0:t} \in \xspmb{0:t}$ and $(\paramopt{1}, \paramopt{2}) \in \paramspace^2$
    \begin{equation}
        \tv{\bdpart{t, \paramopt{1}}(\xmb{0:t}, \cdot)}{\bdpart{t, \paramopt{2}}(\xmb{0:t}, \cdot)} \leq \bdpartlip{t}(\xmb{0:t}) \|\paramopt{1} - \paramopt{2}\|
    \end{equation}
    where $\bdpartlip{t}(\xmb{0:t}) = \sup_{\paramopt{}}\bdpart{t}{}\left[\sum_{i=0}^{t-1} \bklip{i}\right](\xmb{0:t})$.
    Under \cref{assumption:problem_lipschitz}(i), we have that $\|\bdpartlip{t}\|_{\infty} = \sum_{i=0}^{t-1}(\pothigh{i}\mdhigh{i})^{-1} \|\udlip{i}\|_{\infty}$.
\end{lemma}
\begin{proof}
    Apply \cref{lem:mk:lip_comp} and \cref{lem:coupling:bkw}.
\end{proof}

\subsubsection{ $\paramopt{} \mapsto \int\ckjt{t, \paramopt{}}(\xmb{0:t}, \rmd \statlmb{t})\occm(\statlmb{t})(\mathrm{id})$ is Lipschitz}
\label{sec:Slip}
Define the backward ancestors kernel
\begin{align}
    \bkwidx{t}{} : \xsp{t+1}  \times \xspmb{t} \times \sigma(\intvect{1}{N}) \label{eq:def:bkwidx} \mapsto \int \indi{A}(\tilde{j})\left(\sum_{\ell = 1}^\N \frac{\ud{t}{}(x^\ell _t, x_{t+1})}{\sum_{\ell' = 1}^\N\ud{t}{}(x^{\ell'} _t, x_{t+1})} \delta_{\ell}(\rmd \tilde{j})\right)  \eqsp . \nonumber
    \end{align}
\begin{lemma}($\bkwidx{t}{}$ is Lipschitz)
    For every $m \in \intvect{0}{t}$, there exists $\bkklip{m} \in \meas(\xfdmb{m:m+1})$ such that
    \begin{equation}
        \tv{\bkwidx{m}{1}(x_{m+1}, \xmb{m})}{\bkwidx{m}{2}(x_{m+1}, \xmb{m})} \leq \bklip{m}(x_{m+1}, \xmb{m}) \| \paramopt{1} - \paramopt{2} \| \eqsp,
    \end{equation}
    where $\bklip{s}$ is defined in \Cref{lem:coupling:bkw}
\end{lemma}
\begin{proof} $\bkwidx{s}{}$ is the index version of the kernel \eqref{eq:def:bdpart} and thus it is Lipschitz with the same constant.
\end{proof}
\begin{proposition}
    \label{prop:lip_cont_ppg}
    
    For every $m \in \intvect{0}{t}$, we have that 
    \begin{equation} 
        \big| \int \mbjt{m}{} \ckjt{m, \paramopt{}}(z_{0:m}, \rmd \statlmb{m})\occm(\statlmb{m})(\mathrm{Id}) \big| \leq \sum_{\ell = 0}^{m-1} s^\infty _\ell
    \end{equation}
    and 
    \begin{equation}
       \left| \int\ckjt{m, \paramopt{1}}(\xmb{0:m}, \rmd \statlmb{m})\occm(\statlmb{m})(\mathrm{Id})  - \int\ckjt{m, \paramopt{2}}(\xmb{0:m}, \rmd \statlmb{m})\occm(\statlmb{m})(\mathrm{Id}) \right| \leq \lipsmu{m}(\xmb{0:m}) \| \paramopt{1} - \paramopt{2} \| \eqsp.
    \end{equation}
    where $ \lipsmu{m}(\xmb{0:m}) = N^{-1} \sum_{i = 1}^N L^B _m(\epartvar{m}{k}, \xmb{0:m})$ and $L^B _m$ is defined recursively as 
    \begin{equation}
        L^B _{m+1}(\epartvar{m+1}{k}, \xmb{0:m}) = \bklip{m}(\epartvar{m+1}{k}, \xmb{m}) \sum_{\ell = 0}^{m} \supscore{\ell} + \int \bkwidx{m}{}(\epartvar{m+1}{k}, \xmb{m}, \rmd \idx{}{}) \left\{ \scorelip{m}(\epartvar{m}{\idx{}{}}, \epartvar{m+1}{k}) +  L^B _{m}(\epartvar{m}{\idx{}{}}, \xmb{0:m-1})\right\}.
    \end{equation}
    In particular, under \Cref{assumption:problem_lipschitz}, we have that $L^B _{m} \leq \sum_{j=1}^{m} \| \bklip{j} \| _\infty \left[\sum_{\ell=0}^{m-1}\supscore{\ell}\right] + \sum_{j=1}^{m}\| \scorelip{j}\|_\infty $.
\end{proposition}
\begin{proof}
Consider the following kernels,
\begin{align}
    \ckjtidx{m}{}(\xmb{0:m+1}, \rmd ( \idx{0}{i,j}, \dotsc, \idx{m}{i,j})_{i = 1, j =1}^{N, M}) &  \eqdef \prod_{\ell = 0}^m \prod_{k = 1}^N \ckidx{\ell}{}(\epartvar{\ell + 1}{k}, \xmb{\ell}, \rmd \big( \idx{\ell}{k, j}\big)_{j = 1}^M ) \eqsp, \\
    \ckidx{\ell}{}(\epartvar{\ell+1}{k}, \xmb{\ell}, \rmd ( \idx{\ell}{k, j})_{j = 1}^M)  & \eqdef \prod_{j = 1}^M \bkwidx{\ell}{}(\epartvar{\ell+1}{k}, \xmb{\ell}, \rmd \idx{\ell}{k,j}) \eqsp.
\end{align}
Define for all $k \in [1:N]$, $m \in \nset_{> 0}$,
\[
    B_{m+1, k} : \paramopt{} \mapsto \int \ckjtidx{m}{}(\xmb{0:m+1},  \rmd \big( \idx{0}{i,j}, \dotsc, \idx{m}{i,j}\big)_{i = 1, j =1}^{N, M}) \statvar{m+1}{k}\big(\xmb{0:m+1}, \big( \idx{0}{i,j}, \dotsc, \idx{m}{i,j}\big)_{i = 1, j =1}^{N, M}\big) \eqsp,
\]
where $b^k _{m+1}\big(\xmb{0:m+1}, \big( \idx{0}{i,j}, \dotsc, \idx{m}{i,j}\big)_{i = 1, j =1}^{N, M}\big)$ is defined recursively as
\[ 
b^k _{m+1}\big(\xmb{0:m+1}, \big( \idx{0}{i,j}, \dotsc, \idx{m}{i,j}\big)_{i = 1, j =1}^{N, M}\big) = M^{-1} \sum_{\ell = 1}^M \statvaridx{m}{k,\ell}\big(\xmb{0:m}, \big( \idx{0}{i,j}, \dotsc, \idx{m-1}{i,j}\big)_{i = 1, j =1}^{N, M}\big) + \score{}{m}(\epartvaridx{m}{k, \ell}, \epartvar{m+1}{k}).
\]
For notational convenience, we henceforth drop the arguments and simply write $b^k _{m+1}$.

We herebelow show that $B_{m+1,k}$ is Lipschitz with constant $L^B _{m}(\epartvar{m+1}{k}, \xmb{m})$ and bounded by $\sum_{\ell = 0}^{m-1} \supscore{\ell}$. For $m > 2$ and $k \in [1:N]$,
\begin{align*}
    B_{m+1,k}(\paramopt{}) & =  \int \ckjtidx{m}{}(\xmb{0:m+1},  \rmd ( \idx{0}{i,j}, \dotsc, \idx{m}{i,j})_{i = 1, j =1}^{N, M}) \statvar{m+1}{k} \\
    & = \int \cdots \int \ckjtidx{m-1}{}(\xmb{0:m}, \rmd ( \idx{0}{i,j}, \dotsc, \idx{m-1}{i,j})_{i = 1, j =1}^{N, M}) \ckidx{m}{}(\epartvar{m+1}{k}, \xmb{m}, \rmd (\idx{m}{k,j} )_{j = 1} ^M) \\
    & \hspace{2cm}  \times \left\{ M^{-1} \sum_{\ell = 1}^M \statvaridx{m}{k,\ell} + \score{}{m}(\epartvaridx{m}{k, \ell}, \epartvar{m+1}{k})\right\} \\
    & = \int \cdots \int \ckidx{m}{}(\epartvar{m+1}{k}, \xmb{m}, \rmd \{\idx{m}{k,j} \}_{j = 1} ^M) \bigg[  M^{-1} \sum_{\ell = 1}^M  \bigg\{ \score{}{m}(\epartvaridx{m}{k,\ell}, \epartvar{m+1}{k}) \\
    & \hspace{2cm} + \int \ckjtidx{m-1}{}(\xmb{0:m}, \rmd ( \idx{0}{i,j}, \dotsc, \idx{m-1}{i,j})_{i = 1, j =1}^{N, M}) \statvaridx{m}{k,\ell} \bigg\} \bigg] \\
    & = \int \cdots \int \ckidx{m}{}(\epartvar{m+1}{k}, \xmb{m}, \rmd (\idx{m}{k,j} )_{j = 1} ^M) \left[  M^{-1} \sum_{\ell = 1}^M  \left\{ \score{}{m}(\epartvaridx{m}{k,\ell}, \epartvar{m+1}{k}) + B_{m, \idx{m}{k, \ell}}(\paramopt{}) \right\} \right] \\
    & = \int \bkwidx{m}{}(\epartvar{m+1}{k}, \xmb{m}, \rmd \idx{}{}) \left\{ \score{}{m}(\epartvar{m}{\idx{}{}}, \epartvar{m+1}{k}) +  B_{m, \idx{}{}}(\paramopt{})\right\}
\end{align*}
Applying the induction hypothesis conditionally on $\idx{m}{k,\ell}$, $B_{m, \idx{m}{k, \ell}}$ is Lipschitz with constant $L^B _{m}(\epartvaridx{m}{k,\ell}, \xmb{0:m-1})$ and thus the Lipschitz constant of $B_{m+1,k}$ is
\begin{equation}
    L^B _{m+1}(\epartvar{m+1}{k}, \xmb{0:m}) = \bklip{m}(\epartvar{m+1}{k}, \xmb{m}) \sum_{\ell = 0}^{m} \supscore{\ell} + \int \bkwidx{m}{}(\epartvar{m+1}{k}, \xmb{m}, \rmd \idx{}{}) \left\{ \scorelip{m}(\epartvar{m}{\idx{}{}}, \epartvar{m+1}{k}) +  L^B _{m}(\epartvar{m}{\idx{}{}}, \xmb{0:m-1})\right\} \eqsp.
\end{equation}
where we have used the fact that $\bkwidx{m}{}$ and $\score{}{m}$ are also Lipschitz. Again by induction $B_{m+1,k}$ is bounded uniformly by $\sum_{\ell = 0}^{m} \supscore{\ell}$. The induction is concluded by noting that for the base case $m = 0$, $\beta^k _m = 0$ for all $k \in \nset$ and thus the result holds.

It now remains to check that for all $\paramopt{} \in \paramspace$, $m \in \intvect{0}{t}$ and $k \in [1:N]$,
\[
    B_{m,k}(\paramopt{}) = \int\ckjt{m}{}(\xmb{0:m}, \rmd \statlmb{m}) \statvar{m}{k} \eqsp.
\]
Again, we proceed by induction.
\begin{align*}
    & \int \ckjt{m}{}(\xmb{0:m}, \rmd \statlmb{m}) \statvar{m}{k} \\
    & = \int \cdots \int \ckjt{m-1}{}(\xmb{0:m-1}, \rmd \statlmb{m-1}) \ck{m}(\statlmb{m-1}, \xmb{m-1:m}, \rmd \statlmb{m}) \statvar{m}{k} \\
    & = \int \cdots \int \ckjt{m-1}{}(\xmb{0:m-1}, \rmd \statlmb{m-1}) \\
    & \hspace{1cm} \times \prod_{j = 1}^M \left( \sum_{p = 1}^N \frac{\ud{m-1}{}(\epartvar{m-1}{p}, \epartvar{m}{k})}{\sum_{\ell = 1}^N \ud{m-1}{}(\epartvar{m-1}{\ell}, \epartvar{m}{k})} \delta_{\epartvar{m-1}{p}, \statvar{m-1}{p}} \big(\rmd (\eparttdvar{m-1}{k,j}, \stattdvar{m-1}{k, j})\big)\right) \\
    & \hspace{1cm} \times \left[ M^{-1} \sum_{n = 1}^M \left\{ \stattdvar{m-1}{k, n} + \score{}{m}(\eparttdvar{m-1}{k,n}, \epartvar{m}{k}) \right\} \right] \\
    & = \int \cdots \int \ckjt{m-1}{}(\xmb{0:m-1}, \rmd \statlmb{m-1}) \\
    & \hspace{1cm} \times \prod_{j = 1}^M \left( \sum_{p = 1}^N \frac{\ud{m-1}{}(\epartvar{m-1}{p}, \epartvar{m}{k})}{\sum_{\ell = 1}^N \ud{m-1}{}(\epartvar{m-1}{\ell}, \epartvar{m}{k})} \delta_{p} (\rmd \idx{m-1}{k,j})\right) \left[ M^{-1} \sum_{n = 1}^M \left\{ \statvaridx{m-1}{k,n} + \score{}{m}(\epartvaridx{m-1}{k,n}, \epartvar{m}{k}) \right\}\right] \\
    & = \int \cdots \int \ckidx{m}{}(\epartvar{m-1}{k}, \xmb{\ell-1}, \rmd (\idx{\ell-1}{k,j} )_{j = 1} ^M) \\
    & \hspace{.5cm} \times  \left[  M^{-1} \sum_{\ell = 1}^M  \left\{ \score{}{m}(\epartvaridx{m-1}{k,\ell}, \epartvar{m}{k}) + \ckjt{m-1}{}(\xmb{0:m-1}, \rmd \statlmb{m-1}) \statvaridx{m-1}{k,\ell} \right\} \right] \\
    & = \int \cdots \int \ckidx{m}{}(\epartvar{m-1}{k}, \xmb{\ell-1}, \rmd (\idx{\ell-1}{k,j} )_{j = 1} ^M) \\
    & \hspace{.5cm} \times  \left[  M^{-1} \sum_{\ell = 1}^M  \left\{ \score{}{m}(\epartvaridx{m-1}{k,\ell}, \epartvar{m}{k}) + \int \ckjt{m-1}{}(\xmb{0:m-1}, \rmd \statlmb{m-1}) \statvaridx{m-1}{k,\ell} \right\} \right] \\
    & = \int \cdots \int \ckidx{m}{}(\epartvar{m-1}{k}, \xmb{\ell-1}, \rmd (\idx{\ell-1}{k,j} )_{j = 1} ^M) \left[  M^{-1} \sum_{\ell = 1}^M  \left\{ \score{}{m}(\epartvaridx{m-1}{k,\ell}, \epartvar{m}{k}) + B_{m-1, \idx{m-1}{k, \ell}}(\paramopt{}) \right\} \right] \\
    & = B_{m, k}(\paramopt{})
\end{align*}
The proof is finalized by noting that
\[
    \int\ckjt{m}{}(\xmb{0:m}, \rmd \statlmb{m})\occm(\statlmb{m})(\mathrm{Id}) = N^{-1} \sum_{k = 1}^N B_{m,k}(\paramopt{})
    \]
and thus it is Lipschitz with constant $ \lipsmu{m}(\xmb{0:m}) = N^{-1} \sum_{i = 1}^N L^B _m(\epartvar{m}{k}, \xmb{m-1})$.
\end{proof}

\subsection{Lipschitz properties of Markov Kernels}
\begin{lemma}[Composition of ergodic Lipschitz kernels is lipschitz]
    \label{lem:lip:ergodic}
    Let $P_{\paramopt{}}$ be a Markov kernel over $X \times \mathcal{Y}$ that is uniformly $\pi$-geometrically ergodic for any $\paramopt{}$ with contraction constant $\rho$ independent of $\paramopt{}$ and such that there exists $L_p > 0$ such that for every $x \in \xsp{}$
    \[ 
        \tv{P_{\paramopt{0}}(x, \cdot)}{P_{\paramopt{1}}(x, \cdot)} \leq L_P \| \paramopt{0} - \paramopt{1} \|.
    \]
    Then, for all $k > 0$
    \[ 
        \tv{P_{\paramopt{0}}^k(x, \cdot)}{P_{\paramopt{1}}^k(x, \cdot)} \leq \frac{L_P}{1 - \rho} \| \paramopt{0} - \paramopt{1} \|.
    \]
\end{lemma}
\begin{proof} 
    We use the following decomposition borrowed from \cite{Fort_2011}. For any $k \geq 1$,
    \[ 
    P_{\paramopt{0}}^k f - P_{\paramopt{1}}^k f = \sum_{j = 0}^{k-1} P_{\paramopt{0}}^j (P_{\paramopt{0}} - P_{\paramopt{1}}) \big( P_{\paramopt{1}}^{k - j -1} f - \pi f \big).
    \]
    Then, for any $f$ s.t. $\| f \|_\infty \leq 1$ and $x \in \xsp{}$,
    \begin{align*} 
        | P_{\paramopt{0}}^k f(x) - P_{\paramopt{1}}^k f(x) | & \leq \sum_{j = 0}^{k-1} \left| \int P^j _{\paramopt{0}}(x, \rmd y) \sup_{z \in \xsp{}} | P_{\paramopt{1}}^{k-j-1} f(z) - \pi f |\right| L_P \| \paramopt{0} - \paramopt{1} \| \\
        & \leq L_P \bigg( \sum_{j = 0}^{k-1} \rho^{k-j-1} \bigg) \| \paramopt{0} - \paramopt{1} \|\\
        & \leq \frac{L_P}{1 - \rho} \| \paramopt{0} - \paramopt{1} \|.
    \end{align*}
\end{proof}
\begin{lemma}[Composition of Lipschitz kernels is lipschitz]
    \label{lem:mk:lip_comp}
    Let $P_{\paramopt{}}, Q_{\paramopt{}}$ be two kernels defined over $X \times \mathcal{Y}$ and $Y \times \mathcal{Z}$ such that for ever $x \in X$, $y \in Y$ there are $L_{p} \in \meas(X)$, $L_q \in \meas(Y)$ that satisfy
    \begin{equation*}
        \tv{P_{\paramopt{0}}(x, \cdot)}{P_{\paramopt{1}}(x, \cdot)} \leq L_{p}(x) \|\paramopt{0} - \paramopt{1}\|
    \end{equation*}
    and
    \begin{equation*}
        \tv{Q_{\paramopt{0}}(y, \cdot)}{Q_{\paramopt{1}}(y, \cdot)} \leq L_{q}(y) \|\paramopt{0} - \paramopt{1}\| \eqsp.
    \end{equation*}
    Then
    \begin{equation*}
        \tv{P_{\paramopt{0}}Q_{\paramopt{0}}(x, \cdot)}{P_{\paramopt{1}}Q_{\paramopt{1}}(x, \cdot)} \leq L_{pq}   \|\paramopt{0} - \paramopt{1}\| \eqsp,
    \end{equation*}
    where $L_{pq} = (\sup_{\paramopt{}}P_{\paramopt{}}L_{q}(x) + L_{p}(x)\sup_{y} \sup_{\paramopt{}}Q_{\paramopt{}}(y, Z)) $.
\end{lemma}
\begin{proof}
    Let $f \in \meas{}$ such that $\|f\|_\infty \leq 1$.
    \begin{align*}
        \|P_{\paramopt{1}} Q_{\paramopt{1}}f - P_{\paramopt{2}} Q_{\paramopt{2}}f \| & \leq \|P_{\paramopt{1}} \left[Q_{\paramopt{1}}f - Q_{\paramopt{2}}f\right] \| + \| (P_{\paramopt{1}} - P_{\paramopt{2}}) Q_{\paramopt{2}}f \| \\
        & \leq (P_{\paramopt{1}}L_{q}(x) + L_{p}(x)\|Q_{\paramopt{2}}f\|_{\infty} )\| \paramopt{1} - \paramopt{2} \| \eqsp.
    \end{align*}
\end{proof}

\begin{corollary}
        \label{cor:mk:lip_comp}
    Let $P_{\paramopt{}}, Q_{\paramopt{}}$ be two \textbf{Markov} kernels defined over $X \times \mathcal{Y}$ and $Y \times \mathcal{Z}$ such that for ever $x \in X$, $y \in Y$ there are $L_{p} \in \meas(X)$, $L_q \in \meas(Y)$ that satisfy
    \begin{equation*}
        \tv{P_{\paramopt{0}}(x, \cdot)}{P_{\paramopt{1}}(x, \cdot)} \leq L_{p}(x) \|\paramopt{0} - \paramopt{1}\|
    \end{equation*}
    and
    \begin{equation*}
        \tv{Q_{\paramopt{0}}(y, \cdot)}{Q_{\paramopt{1}}(y, \cdot)} \leq L_{q}(y) \|\paramopt{0} - \paramopt{1}\| \eqsp.
    \end{equation*}
    Then
    \begin{equation*}
        \tv{P_{\paramopt{0}}Q_{\paramopt{0}}(x, \cdot)}{P_{\paramopt{1}}Q_{\paramopt{1}}(x, \cdot)} \leq L_{pq}   \|\paramopt{0} - \paramopt{1}\| \eqsp,
    \end{equation*}
    where $L_{pq} = (\sup_{\paramopt{}}P_{\paramopt{}}L_{q}(x) + L_{p}(x)) $.
\end{corollary}

\begin{lemma}[Product of Lipschitz kernels is lipschitz]
    \label{lem:mk:lip_prod}
    Let $P_{\paramopt{}}, Q_{\paramopt{}}$ be two markov kernels that are uniformly Lipschitz with constants $L_P, L_Q$.
    Then $P_{\paramopt{}} \tensprod Q_{\paramopt{}}$ is uniformly Lipschitz with constant $L_P + L_Q$.
\end{lemma}
\begin{proof}
    Let $h_{\paramopt{}} : y \mapsto \int Q_{\paramopt{}}(y, \rmd z) f(y,z)$. Then $(P_{\paramopt{i}} \otimes Q_{\paramopt{i}})(f) = P_{\paramopt{i}}( h_{\paramopt{i}})$ and the proof is similar to that of the previous Lemma since $h_{\paramopt{}}$ is Lipschitz with constant $L_Q$ and $\| h_{\paramopt{}} \|_\infty \leq 1$.
\end{proof}

\section{Additional numerical results}
\subsection{PPG}
\label{sec:appendix:numerics:ppg}
\begin{figure}[h]
    \centering
    \begin{subfigure}{0.45\textwidth}
        \includegraphics[width=\textwidth]{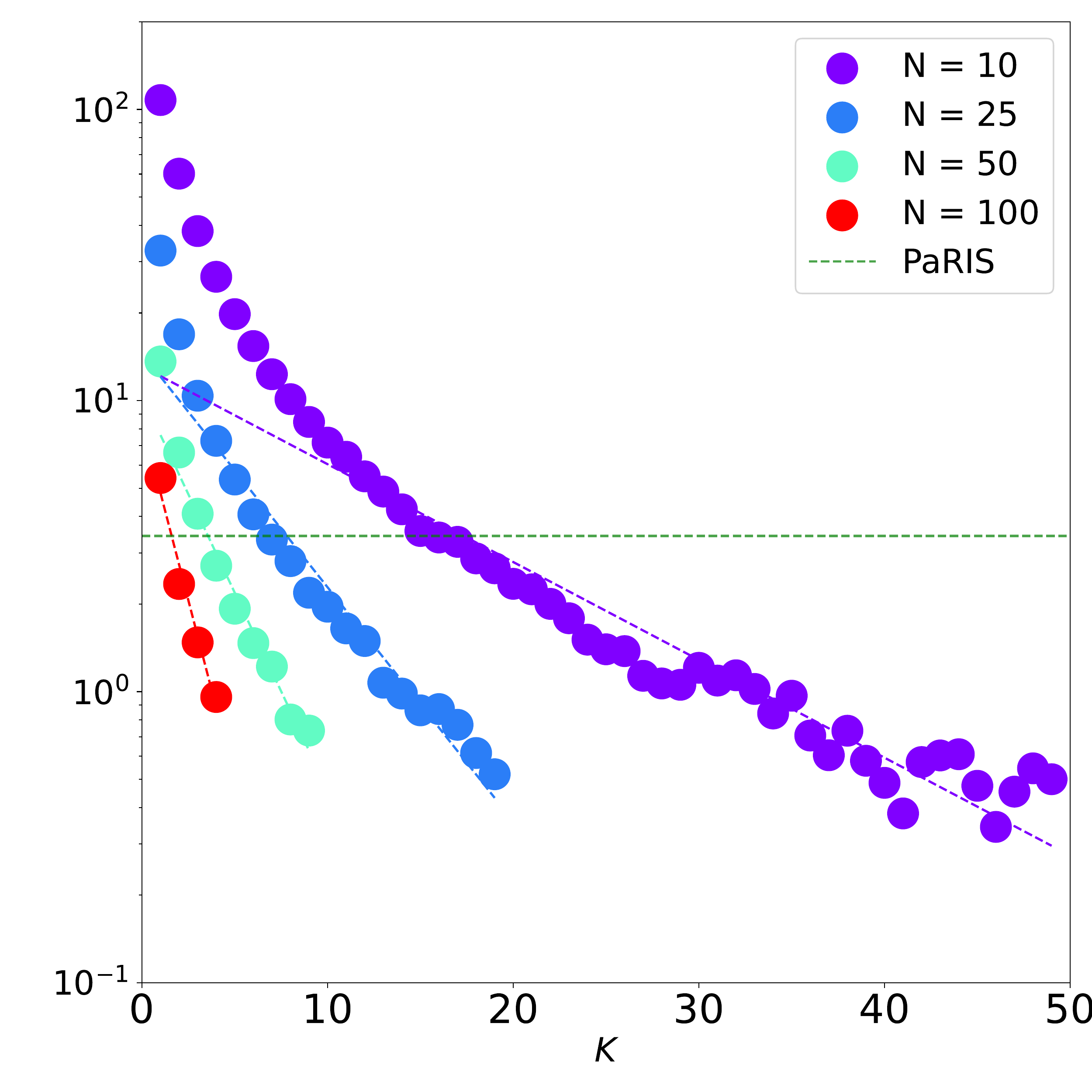}
        \label{fig:LGSSM:dim_1:bias_evolution_no_rollout}
    \end{subfigure}
    \begin{subfigure}{0.45\textwidth}
        \includegraphics[width=\textwidth]{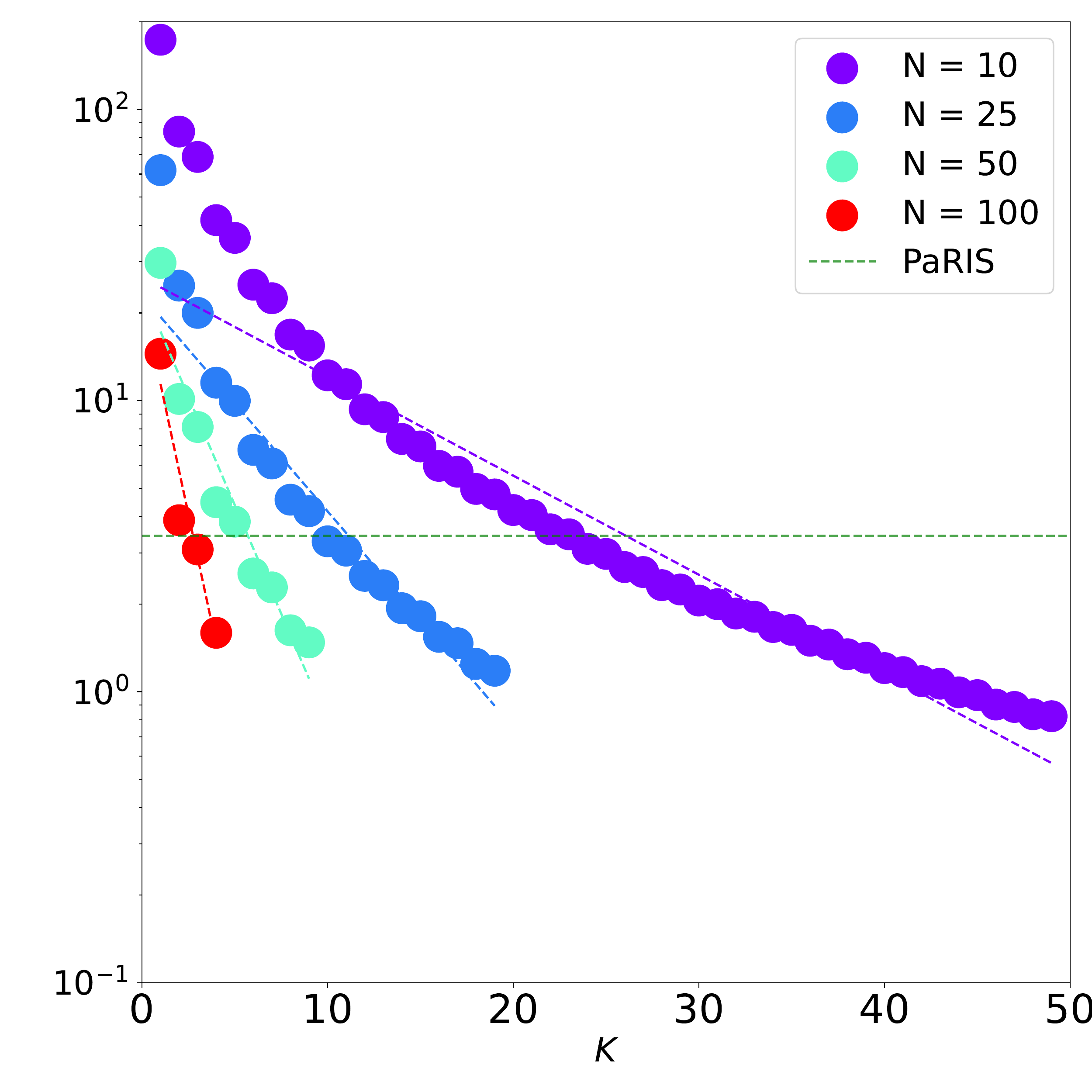}
        \label{fig:LGSSM:dim_1:bias_evolution_rollout_05}
    \end{subfigure}
    \caption{Output of the {\PPG} roll-out estimator for the LGSSM. The curves describe the evolution of the bias with increasing $\ki$ for different particle sample sizes $\N$. The left and right panels correspond to $\ki_0=\ki-1$ and $\ki_0=\lfloor\ki/2\rfloor$, respectively.}
    \label{fig:LGSSM:bias_evolution}
\end{figure}

\subsection{Learning}
\label{sec:appendix:numerics:learning}

For both experiments, all the parameters were initialized by sampling from a centered multivariate gaussian distribution with covariance matrix of $0.01 I$.
We have used the ADAM optimizer \cite{kingma:2014} with a learning rate decay of $1/\sqrt{\ell}$ where $\ell$ is the iteration index, with a starting learning rate of $0.2$.
We rescale the gradients by $T$.

\paragraph{LGSSM}

For LGSSM we evaluated for fixed number of particles ($N=64$) and number of gibbs iterations ($k=8$) the influence of the burn-in phase ($k_0$)
over the final distance obtained to the MLE estimator. \Cref{table:LGSSM_k_0} indicates that configurations with smaller $k_0$ perform better.
A possible interpretation of this phenomenon is that, since between two gradient ascent iterates the conditioning path is
being passed on, this conditioning path from a moment on makes the estimates less biased, so the importance
of having $k_0$ high to have less bias vanishes, but the effect of augmenting the variance with $k_0$ is still shown,
since the fact of having a conditioning particle from the right marginal does not affect the variance of the estimator, only it's bias.

\begin{filecontents*}{1_varying_k_0.csv}
Algorithm,Distance,NLL,lr,N,k,k_0,gradNorm
PPG,0.205 ± 0.013,3261.971 +/- 2.616,0.0,64,8,0,0.149 +- 0.005
PPG,0.213 ± 0.016,3263.749 +/- 3.890,1.0,64,8,1,0.154 +- 0.006
PPG,0.201 ± 0.010,3260.651 +/- 1.242,2.0,64,8,2,0.164 +- 0.006
PPG,0.201 ± 0.010,3260.772 +/- 1.242,3.0,64,8,3,0.171 +- 0.006
PPG,0.207 ± 0.012,3262.137 +/- 2.437,4.0,64,8,4,0.187 +- 0.007
PPG,0.212 ± 0.015,3263.599 +/- 3.850,5.0,64,8,5,0.205 +- 0.008
PPG,0.210 ± 0.017,3262.359 +/- 3.602,6.0,64,8,6,0.234 +- 0.008
PPG,0.211 ± 0.018,3263.885 +/- 4.237,7.0,64,8,7,0.288 +- 0.009
\end{filecontents*}
\DTLloaddb{lgss_k_0}{1_varying_k_0.csv}
\begin{table}
    \centering
    \resizebox{.45\textwidth}{!}{%
    \begin{tabular}{|c|c|c|c|c|}%
          \hline
          Algorithm & $N$ & $k_0$ & $k$ & $D_{\scriptsize{\mbox{{\it mle}}}}$\\
          \hline
          \DTLforeach*{lgss_k_0}{\Alg=Algorithm,\D=Distance, \kze=k_0, \k=k, \Npart=N, \NLL=NLL, \grad=gradNorm}{
            \Alg & \Npart & \kze & \k &\D \DTLiflastrow{}{\\
          }}
        \\\hline
    \end{tabular}
    }
    \caption{Distance to $\paramopt{\operatorname{MLE}}$ for each configuration in the LGSSM case.}
    \label{table:LGSSM_k_0}
\end{table}

\bibliographystyle{apalike}
\bibliography{motherofallbibs}

\begin{thebibliography}{}

\bibitem[Anderson and Qiu, 1997]{anderson:qiu:1997}
Anderson, G.~D. and Qiu, S.-L. (1997).
\newblock A monotonicity property of the gamma function.
\newblock {\em Proc. Amer. Math. Soc.}, 125(11):3355--3362.

\bibitem[Andrieu and Doucet, 2003]{andrieu:doucet:2003}
Andrieu, C. and Doucet, A. (2003).
\newblock Online {E}xpectation--{M}aximization type algorithms for parameter
  estimation in general state space models.
\newblock In {\em Proc. IEEE Int. Conf. Acoust., Speech, Signal Process.},
  volume~6, pages 69--72.

\bibitem[Andrieu et~al., 2010a]{andrieu2010particle}
Andrieu, C., Doucet, A., and Holenstein, R. (2010a).
\newblock Particle {M}arkov chain {M}onte {C}arlo methods.
\newblock {\em Journal of the Royal Statistical Society: Series B},
  72(3):269--342.

\bibitem[Andrieu et~al., 2010b]{andrieu:doucet:holenstein:2010}
Andrieu, C., Doucet, A., and Holenstein, R. (2010b).
\newblock Particle {M}arkov chain {M}onte {C}arlo methods (with discussion).
\newblock {\em J. Roy. Statist. Soc. B}, 72:269--342.

\bibitem[Andrieu et~al., 2018]{andrieu2018uniform}
Andrieu, C., Lee, A., and Vihola, M. (2018).
\newblock Uniform ergodicity of the iterated conditional {SMC} and geometric
  ergodicity of particle {G}ibbs samplers.
\newblock {\em Bernoulli}, 24(2):842--872.

\bibitem[Capp\'{e}, 2001]{cappe:2001}
Capp\'{e}, O. (2001).
\newblock Recursive computation of smoothed functionals of hidden {M}arkovian
  processes using a particle approximation.
\newblock {\em {M}onte {C}arlo Methods Appl.}, 7(1--2):81--92.

\bibitem[Capp\'{e}, 2011]{cappe:2009}
Capp\'{e}, O. (2011).
\newblock Online {EM} algorithm for hidden {M}arkov models.
\newblock {\em J. Comput. Graph. Statist.}, 20(3):728--749.

\bibitem[Capp\'{e} et~al., 2007]{cappe:godsill:moulines:2007}
Capp\'{e}, O., Godsill, S.~J., and Moulines, E. (2007).
\newblock An overview of existing methods and recent advances in sequential
  {M}onte {C}arlo.
\newblock {\em IEEE Proceedings}, 95(5):899--924.

\bibitem[Capp\'{e} and Moulines, 2005]{cappe:moulines:2005}
Capp\'{e}, O. and Moulines, E. (2005).
\newblock On the use of particle filtering for maximum likelihood parameter
  estimation.
\newblock In {\em European Signal Processing Conference (EUSIPCO)}, Antalya,
  Turkey.

\bibitem[Capp\'{e} et~al., 2005]{cappe:moulines:ryden:2005}
Capp\'{e}, O., Moulines, E., and Ryd\'{e}n, T. (2005).
\newblock {\em Inference in {H}idden {M}arkov Models}.
\newblock Springer.

\bibitem[Capp{\'e} et~al., 2009]{infhidden}
Capp{\'e}, O., Moulines, E., and Ryd{\'e}n, T. (2009).
\newblock Inference in hidden markov models.
\newblock In {\em Proceedings of EUSFLAT conference}, pages 14--16.

\bibitem[Chopin and Papaspiliopoulos, 2020]{chopin2020introduction}
Chopin, N. and Papaspiliopoulos, O. (2020).
\newblock {\em An Introduction to Sequential Monte Carlo}.
\newblock Springer.

\bibitem[Chopin and Singh, 2015a]{chopin:singh:2015}
Chopin, N. and Singh, S.~S. (2015a).
\newblock On particle {G}ibbs sampling.
\newblock {\em Bernoulli}, 21(3):1855--1883.

\bibitem[Chopin and Singh, 2015b]{chopin2015particle}
Chopin, N. and Singh, S.~S. (2015b).
\newblock On particle gibbs sampling.
\newblock {\em Bernoulli}, 21(3):1855--1883.

\bibitem[{Del Moral}, 2004]{delmoral:2004}
{Del Moral}, P. (2004).
\newblock {\em {F}eynman-Kac {F}ormulae. {G}enealogical and {I}nteracting
  {P}article {S}ystems with {A}pplications}.
\newblock Springer.

\bibitem[{D}el {M}oral, 2013]{delmoral:2013}
{D}el {M}oral, P. (2013).
\newblock {\em {M}ean {F}ield {S}imulation for {M}onte {C}arlo {I}ntegration}.
\newblock CRC Press.

\bibitem[{Del Moral} et~al., 2010]{delmoral:doucet:singh:2010}
{Del Moral}, P., Doucet, A., and Singh, S.~S. (2010).
\newblock A backward interpretation of {F}eynman--{K}ac formulae.
\newblock {\em ESAIM: Mathematical Modelling and Numerical Analysis},
  44:947--975.

\bibitem[Del~Moral and Jasra, 2018]{del2018sharp}
Del~Moral, P. and Jasra, A. (2018).
\newblock A sharp first order analysis of {F}eynman--{K}ac particle models,
  part {II}: Particle {G}ibbs samplers.
\newblock {\em Stoch. Proc. Appl.}, 128(1):354--371.

\bibitem[Del~Moral et~al., 2016]{del2016particle}
Del~Moral, P., Kohn, R., and Patras, F. (2016).
\newblock On particle {G}ibbs samplers.
\newblock {\em Ann. Inst. H. Poincar\'e Probab. Statist.}, 52(4):1687--1733.

\bibitem[Douc et~al., 2011]{douc:garivier:moulines:olsson:2009}
Douc, R., Garivier, A., Moulines, E., and Olsson, J. (2011).
\newblock Sequential {M}onte {C}arlo smoothing for general state space hidden
  {M}arkov models.
\newblock {\em Ann. Appl. Probab.}, 21(6):1201--2145.

\bibitem[Douc and Moulines, 2008]{douc:moulines:2008}
Douc, R. and Moulines, E. (2008).
\newblock Limit theorems for weighted samples with applications to sequential
  {M}onte {C}arlo methods.
\newblock {\em Ann. Statist.}, 36(5):2344--2376.

\bibitem[Douc et~al., 2018]{douc2018markov}
Douc, R., Moulines, E., Priouret, P., and Soulier, P. (2018).
\newblock {\em Markov Chains}.
\newblock Springer.

\bibitem[Douc et~al., 2014]{douc2014nonlinear}
Douc, R., Moulines, E., and Stoffer, D. (2014).
\newblock {\em Nonlinear time series: Theory, methods and applications with R
  examples}.
\newblock CRC press.

\bibitem[Dubarry and Le~Corff, 2013]{dubarry:lecorff:2013}
Dubarry, C. and Le~Corff, S. (2013).
\newblock {Non-asymptotic deviation inequalities for smoothed additive
  functionals in nonlinear state-space models}.
\newblock {\em Bernoulli}, 19(5B):2222 -- 2249.

\bibitem[Fort et~al., 2011]{Fort_2011}
Fort, G., Moulines, E., and Priouret, P. (2011).
\newblock Convergence of adaptive and interacting markov chain monte carlo
  algorithms.
\newblock {\em The Annals of Statistics}, 39(6).

\bibitem[Gloaguen et~al., 2022]{gloaguen:lecorff:olsson:2022}
Gloaguen, P., Le~Corff, S., and Olsson, J. (2022).
\newblock A pseudo-marginal sequential {M}onte {C}arlo online smoothing
  algorithm.
\newblock {\em Bernoulli}, 28(4):2606--2633.

\bibitem[Glynn and Rhee, 2014]{glynn_rhee_2014}
Glynn, P.~W. and Rhee, C.-H. (2014).
\newblock Exact estimation for markov chain equilibrium expectations.
\newblock {\em Journal of Applied Probability}, 51(A):377–389.

\bibitem[Godsill et~al., 2004]{godsill:doucet:west:2004}
Godsill, S.~J., Doucet, A., and West, M. (2004).
\newblock Monte {C}arlo smoothing for non-linear time series.
\newblock {\em J. Am. Statist. Assoc.}, 50:438--449.

\bibitem[Jacob et~al., 2020]{https://doi.org/10.1111/rssb.12336}
Jacob, P.~E., O’Leary, J., and Atchadé, Y.~F. (2020).
\newblock Unbiased markov chain monte carlo methods with couplings.
\newblock {\em Journal of the Royal Statistical Society: Series B (Statistical
  Methodology)}, 82(3):543--600.

\bibitem[Karimi et~al., 2019]{pmlr-v99-karimi19a}
Karimi, B., Miasojedow, B., Moulines, E., and Wai, H.-T. (2019).
\newblock Non-asymptotic analysis of biased stochastic approximation scheme.
\newblock In Beygelzimer, A. and Hsu, D., editors, {\em Proceedings of the
  Thirty-Second Conference on Learning Theory}, volume~99 of {\em Proceedings
  of Machine Learning Research}, pages 1944--1974. PMLR.

\bibitem[Kingma and Ba, 2014]{kingma:2014}
Kingma, D.~P. and Ba, J. (2014).
\newblock Adam: A method for stochastic optimization.

\bibitem[Lee et~al., 2020]{lee2020coupled}
Lee, A., Singh, S.~S., and Vihola, M. (2020).
\newblock Coupled conditional backward sampling particle filter.
\newblock {\em The Annals of Statistics}, 48(5):3066--3089.

\bibitem[Lindholm and Lindsten, 2018]{lindholm2018}
Lindholm, A. and Lindsten, F. (2018).
\newblock Learning dynamical systems with particle stochastic approximation em.

\bibitem[Lindsten et~al., 2014a]{lindsten:jordan:schoen:2014}
Lindsten, F., Jordan, M.~I., and Sch\"{o}n, T.~B. (2014a).
\newblock Particle {G}ibbs with ancestor sampling.
\newblock {\em J. Mach. Learn. Res.}, 15(1):2145--2184.

\bibitem[Lindsten et~al., 2014b]{lindsten14a}
Lindsten, F., Jordan, M.~I., and Sch{{\"o}}n, T.~B. (2014b).
\newblock Particle gibbs with ancestor sampling.
\newblock {\em Journal of Machine Learning Research}, 15(63):2145--2184.

\bibitem[Naesseth et~al., 2020]{naesseth:2O2O}
Naesseth, C.~A., Lindsten, F., and Blei, D. (2020).
\newblock Markovian score climbing: Variational inference with kl(p||q).

\bibitem[Olsson and Westerborn, 2017]{olsson:westerborn:2017}
Olsson, J. and Westerborn, J. (2017).
\newblock Efficient particle-based online smoothing in general hidden {M}arkov
  models: The {P}a{RIS} algorithm.
\newblock {\em Bernoulli}, 23(3):1951--1996.

\bibitem[Poyiadjis et~al., 2005]{poyiadjis:doucet:singh:2005}
Poyiadjis, G., Doucet, A., and Singh, S.~S. (2005).
\newblock Particle methods for optimal filter derivative: application to
  parameter estimation.
\newblock In {\em Proc. IEEE Int. Conf. Acoust., Speech, Signal Process.},
  pages v/925--v/928.

\bibitem[Poyiadjis et~al., 2011]{poyiadjis:doucet:singh:2011}
Poyiadjis, G., Doucet, A., and Singh, S.~S. (2011).
\newblock Particle approximations of the score and observed information matrix
  in state space models with application to parameter estimation.
\newblock {\em Biometrika}, 98(1):65--80.

\bibitem[S{\"a}rkk{\"a}, 2013]{sarkka2013bayesian}
S{\"a}rkk{\"a}, S. (2013).
\newblock {\em Bayesian Filtering and Smoothing}.
\newblock Cambridge University Press.

\bibitem[Singh et~al., 2017]{singh2017blocking}
Singh, S.~S., Lindsten, F., and Moulines, E. (2017).
\newblock Blocking strategies and stability of particle gibbs samplers.
\newblock {\em Biometrika}, 104(4):953--969.

\bibitem[Whiteley, 2010]{Whiteley:2010}
Whiteley, N. (2010).
\newblock Discussion on particle markov chain monte carlo methods.
\newblock pages 306--307.

\bibitem[Zhao et~al., 2021]{zhao2021streaming}
Zhao, Y., Nassar, J., Jordan, I., Bugallo, M., and Park, I.~M. (2021).
\newblock Streaming variational monte carlo.

\end{thebibliography}
\end{document}